\documentclass[letterpaper,12pt]{article}
\usepackage[margin=1in,letterpaper]{geometry} % decreases margins
\usepackage{setspace}
\usepackage{amsmath}
\usepackage{amssymb}
\usepackage{cite}
\usepackage{graphicx} % takes care of graphic including machinery
\usepackage{hyperref}
\usepackage[dvipsnames]{xcolor}
\usepackage{tabularx} % extra features for tabular environment
\usepackage{multirow}
\newcommand{\QY}{\text{QY}}
\newcommand{\QYfixedlen}[2]{\% \Delta \text{QY}(W:#1 \rightarrow #2)}
\newcommand{\QYfixednoise}[2]{\% \Delta \text{QY}(N: #1 \rightarrow #2)}
\newcommand{\noise}[1]{W=#1 \text{ cm}^{-1}}
\newcommand{\refpanel}[1]{\textbf{(#1)}}

\begin{document}

\section*{ }
\begin{minipage}{\textwidth}
    \LARGE
    \centering
    Quantum-enhanced photoprotection in neuroprotein architectures emerges from collective light-matter interactions
    \vspace{10mm}
\end{minipage}

\begin{minipage}{\textwidth}
    \large
    \centering
    Hamza Patwa\,$^{1}$, Nathan S. Babcock\,$^{1}$, and Philip Kurian\,$^{1,*}$
    \vspace{10mm}
\end{minipage}

\begin{minipage}{\textwidth}
    \centering
    $^{1}$Quantum Biology Laboratory, Howard University, Washington, D.C. 20060, USA\\
    \url{https://www.quantumbiolab.com}
    \vspace{10mm}
\end{minipage}

\begin{minipage}{\textwidth}
    \centering
    *E-mail: pkurian@howard.edu
    \vspace{5mm}
\end{minipage}

\begin{abstract}
\textbf{Background}: Superradiance is the phenomenon of many identical quantum systems absorbing and/or emitting photons collectively at a higher rate than any one system can individually. This phenomenon has been studied analytically in idealized distributions of electronic two-level systems (TLSs), each with a ground and excited state, as well as numerically in realistic photosynthetic nanotubes and cytoskeletal architectures. \textbf{Methods}: Superradiant effects are studied here in realistic biological mega-networks of tryptophan (Trp) molecules, which are strongly fluorescent amino acids found in many proteins. Each Trp molecule acts as a chromophore absorbing in the ultraviolet spectrum and can be treated approximately as a TLS, with its $1L_a$ excited singlet state; thus, organized Trp networks can exhibit superradiance. Such networks are found, for example, in microtubules, actin filaments, and amyloid fibrils. Microtubules and actin filaments are spiral-cylindrical protein polymers that play significant biological roles as primary constituents of the eukaryotic cytoskeleton, while amyloid fibrils have been targeted in a variety of neurodegenerative diseases. We treat these proteinaceous Trp networks as open quantum systems, using a non-Hermitian Hamiltonian to describe interactions of the chromophore network with the electromagnetic field. We numerically diagonalize the Hamiltonian to obtain its complex eigenvalues, where the real part is the collective energy and the imaginary part is its associated enhancement rate. \textbf{Results}: We obtained the energies and enhancement rates for realistic microtubules, actin filament bundles, and amyloid fibrils of differing lengths, and we used these values to calculate the fluorescence quantum yield, which is the ratio of the number of photons emitted to the number of photons absorbed. We find that all three of these structures exhibit highly superradiant states near the low-energy portion of the spectrum, which enhances the magnitude and robustness of the quantum yield even in the presence of static disorder and thermal noise. \textbf{Conclusions:} The high quantum yield and stable superradiant states in these biological architectures may play a photoprotective role \textit{in vivo}, downconverting highly energetic ultraviolet photons emitted from reactive free radical species and thereby mitigating biochemical stress and photophysical damage. Contrary to conventional assumptions that quantum effects cannot survive in large biosystems at high temperatures, our results suggest that macropolymeric collectives of TLSs in microtubules, actin filaments, and amyloid fibrils exhibit increasingly observable and robust effects with increasing length, at least up to the micron scale, due to quantum coherent interactions in the single-photon limit. Superradiant enhancement and high quantum yield exhibited in neuroprotein polymers could thus play a crucial role in information processing in the brain, the development of neurodegenerative diseases such as Alzheimer's and related dementias, and a wide array of other pathologies characterized by anomalous protein aggregates.

\end{abstract}

\section{Introduction}

Superradiance is a quantum coherent phenomenon first explored in detail by Robert Dicke \cite{Dicke1954} in 1954. Superradiance arises from the interaction of a collective of quantum systems with the external electromagnetic field. Thus, the theoretical formalism that describes superradiance is given frequently in the language of open quantum systems. In collectives of quantum systems with discrete energy levels, collective superradiant states are characterized by the collective decay rate $\Gamma$ of the system being much larger than the single-system decay rate $\gamma$. An eigenstate with a larger decay is more short-lived than an eigenstate with a small decay rate. In other words, an absorbed photon in an eigenstate with a larger decay rate will be very quickly re-emitted into the environment. The reason that the decay rate is larger for a collective of quantum systems than one system is that in a collective, the excitation is delocalized across the collective, rather than being incoherently concentrated on a single system. 

Superradiant effects in the ultraviolet region of the electromagnetic spectrum have been studied for biosystems \cite{Celardo2019, Babcock2023, kurian2017oxidative} and emerge largely due to collective light-matter interactions involving tryptophan (Trp), which is a strongly fluorescent amino acid found in many proteins. It has many notable photophysical properties, such as its strong ultraviolet absorption, significant absorption-emission Stokes shift, and large transition dipole moment. Trp can be modeled as a two-level system (TLS), which has a ground and an excited state \cite{Callis1997Trp1La1Lb}. Other amino acids such as tyrosine, phenylalanine, and cysteine also absorb in the ultraviolet, but much more weakly than Trp. The fact that Trp networks absorb in the ultraviolet means that the excitation wavelengths are frequently shorter than the characteristic length scales of the biological scaffolds in which such networks lie ($\lambda \lesssim L$), a sharp distinction from the longer visible wavelengths that excite smaller photosynthetic light-harvesting complexes. This implies that long-range interactions in the ultraviolet-excited system will play a more prominent role in the light-matter dynamics.

Coherent quantum phenomena arising from organized networks of chromophores in protein scaffolds have been shown to play a role in the efficiency of photosynthetic complexes \cite{engel2007evidence, panitchayangkoon2010long,collini2010coherently,strumpfer2012quantum,jang2018delocalized} and of other light-harvesting structures (see \cite{Celardo2019,doria2018photochemical,mattiotti2020thermal,mattiotti2022efficient,werren2023light} and references therein). More recently, superradiant states have been experimentally confirmed in tryptophan networks of microtubules (MTs) and theoretically predicted in centrioles\footnote{Centrioles are cylindrically symmetric organelles formed from nine triplets of microtubules exhibiting a pinwheel-like structure (see \cite{Winey2014CentrioleStructure} for more specifics on their geometry). They are highly conserved in most eukaryotic cells, but notably absent in yeast and higher plants, among others \cite{Winey2014CentrioleStructure}. Centrioles play an important role in forming the spindle complex in cell division, where they help ensure that the correct number of chromosomes are present in each daughter cell after replication \cite{firat2014centriole,rodrigues2008centriole}. They have also been shown, in several studies by Guenter Albrecht-Buehler \cite{AlbrechtBuehler1992CellularVision, AlbrechtBuehler1994InfraredDetector, AlbrechtBuehler1997Autoflourescence,AlbrechtBuehler1977ActinTubulin}, to aid orientation of the cell to an external light stimulus.} and neuronal axon bundles \cite{Babcock2023}. In this work, we study the role of superradiance in a wider class of neuroprotein polymers, including cytoskeletal filaments and pathological aggregates, thereby demonstrating the generalizability of our prior experimental results and theoretical predictions for a novel group of chromophore architectures with significant implications for a host of neurodegenerative and other complex diseases.

\section{Background}

\begin{figure*}[tbhp]
\includegraphics[width=0.86\linewidth]{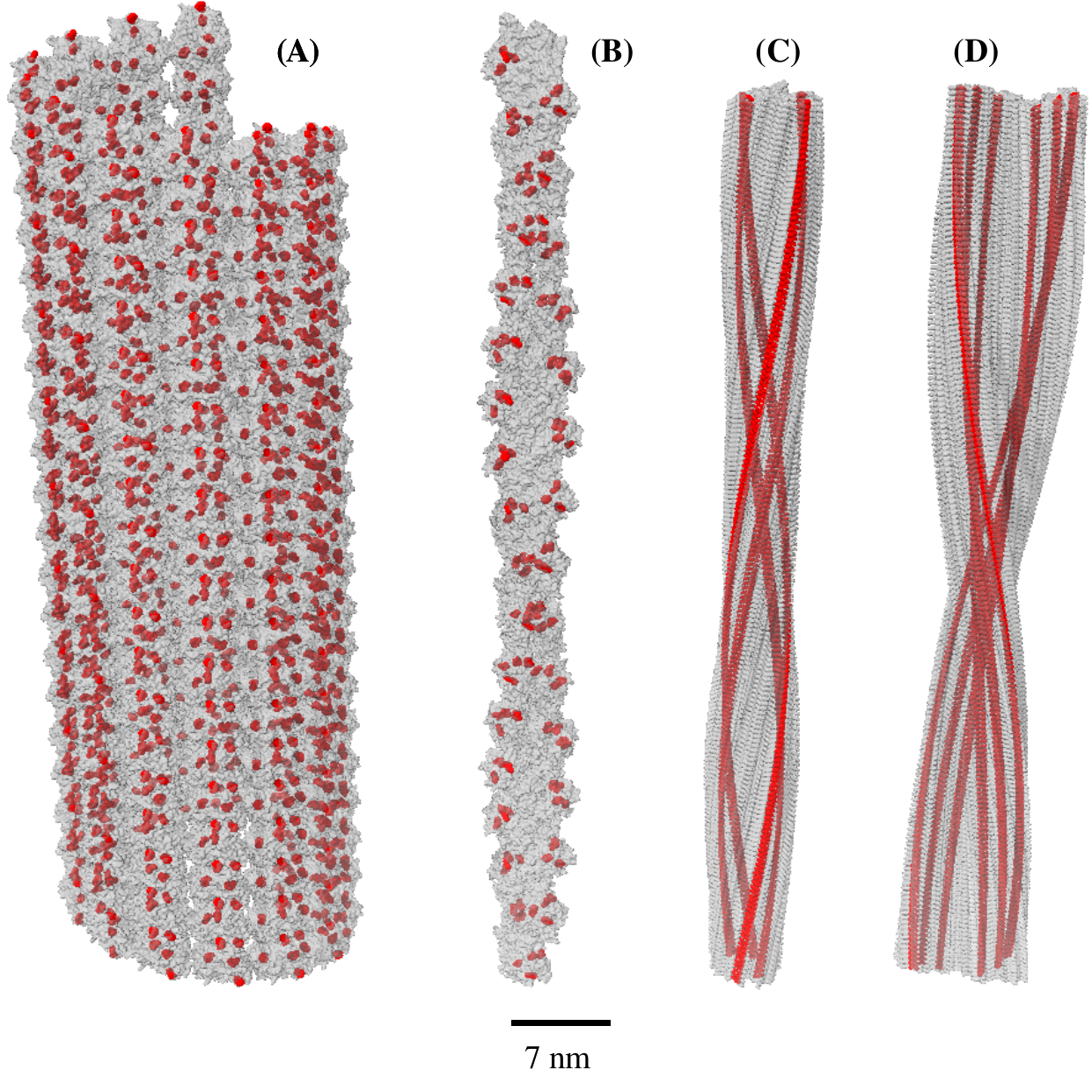}

\caption{\label{fig:trp_positions}Tryptophan (Trp) network geometries in a model \refpanel{A} $80$-nm microtubule (built from tubulin dimer PDB entry 1JFF), \refpanel{B} $90$-nm actin filament (built from actin subunit PDB entry 6BNO), \refpanel{C} $86.4$-nm human amyloid fibril (built from amyloid subunit PDB entry 6MST), and \refpanel{D} $86.4$-nm mouse amyloid fibril (built from amyloid subunit PDB entry 6DSO). The Trp molecules are colored in red and have been enlarged for ease of viewing, within each gray protein lattice. Scale bar is valid for the entire figure.}
\end{figure*}

\begin{figure*}[tbhp]
\centering
\refpanel{A}\\\vspace{5mm}
\includegraphics[width=0.75\linewidth]{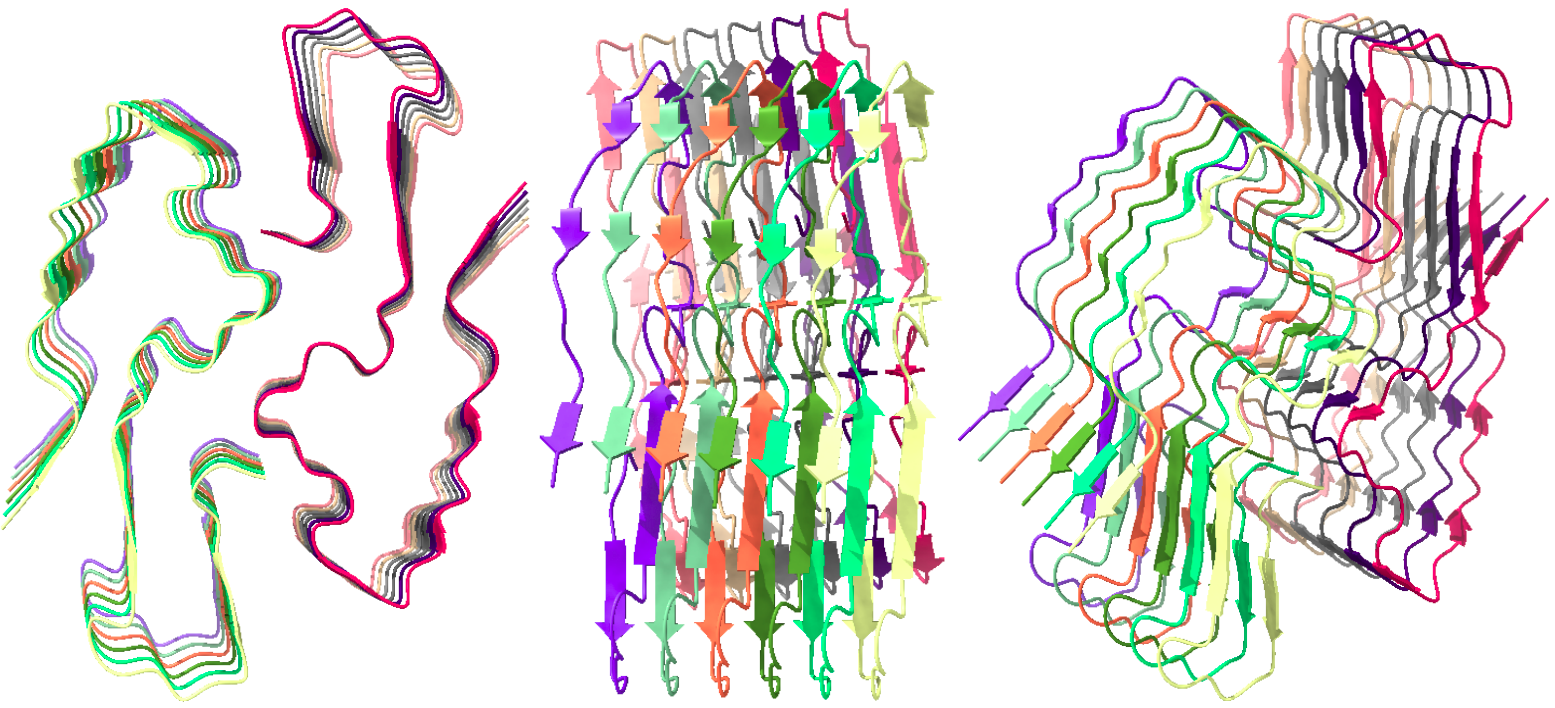}\\\vspace{10mm}
\refpanel{B}\\
\includegraphics[width=0.75\linewidth]{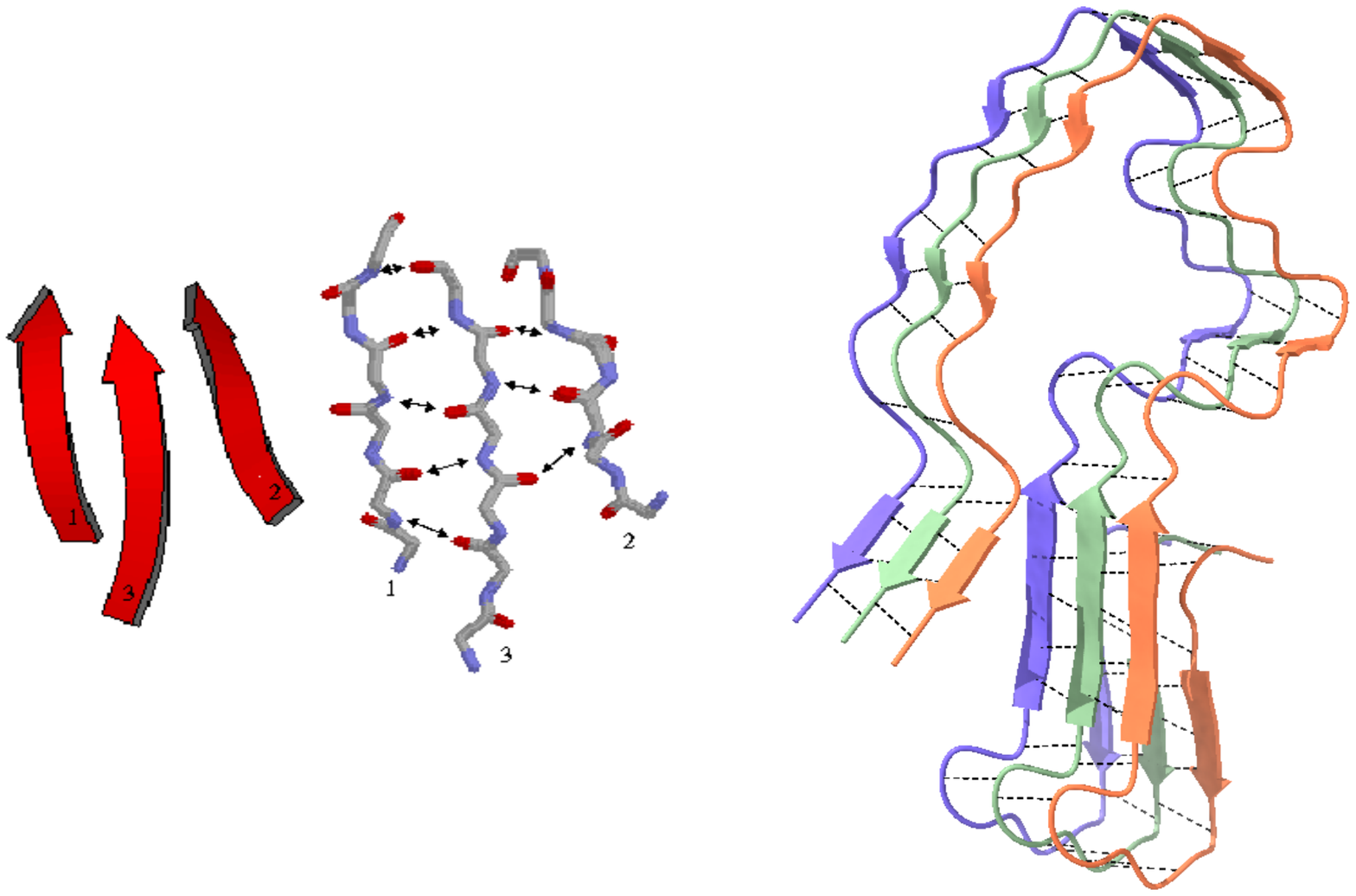}
\caption{\label{fig:amyloid_beta_sheet}Structure of a single human amyloid subunit and its distinctive parallel $\beta$-sheet structure. \refpanel{A} The structure of a human amyloid subunit (PDB entry 6MST) in cartoon/ribbon from a top-down view (left), side-on view (middle), and angled view (right). \refpanel{B} Visual of parallel $\beta$-sheet structure of the human amyloid subunit. The left image is the structure of a parallel $\beta$-sheet, reproduced from \cite{Everse2014BetaStrandsImage}. The right image is three chains from a human amyloid subunit with the hydrogen bonds highlighted in black dashes, showing the similarity with the left image.}
\end{figure*}

\subsection{Cytoskeleton: Microtubules and Actin Filaments}
The cytoskeleton is a dynamic structure in eukaryotic cells that provides structural support and acts as a transport mechanism for molecules inside the cell. It has three main components: microtubules, actin filaments, and intermediate filaments. In this work, we specifically focus on microtubules and actin filaments. 

Microtubules (pictured in Fig. \ref{fig:trp_positions}\refpanel{A}), spiral-cylindrical structures made of tubulin dimers, play a role in cell communication and mitosis. They are a dynamic part of the cytoskeleton, being able to quickly grow and shrink via polymerization or depolymerization, respectively. Microtubules also make up the internal structure of external appendages such as flagella and cilia, which are important for locomotion and movement. Intra-cell communication is also facilitated by microtubules, as well as the whole cytoskeleton in general. Microtubules are a primary constituent of axons in the brain. They have been shown to play an important role in transport along axonal processes in neurons \cite{Ahmad2006}, so disruption of microtubule transport processes in neurons has been linked to several neurodegenerative diseases.

Actin filaments (pictured in Fig. \ref{fig:trp_positions}\refpanel{B}) are strandlike structures that play an important role in the cytoskeleton. Within the context of the cytoskeleton, they are known as microfilaments, reflecting their small diameter generally less than one-third that of microtubules (see Fig. \ref{fig:trp_positions} for comparison). Actin filaments can bundle together to form hexagonal arrangements \cite{Matsudaira1983ActinHexagonal}, which we also analyze in this work. They provide contractile and protrusive forces to stabilize the cytoskeleton and assist with the mobility of the cell. Like microtubules, they assist in transport from outside the cell to the inside. Actin also plays a role in the contractile apparatus of muscle cells, in the form of so-called thin filaments, and helps to maintain the structure of dendritic spines, tiny protrusions from dendrites that form functional contacts with neighboring axons of other neurons in the brain. Dendritic spines play a significant role in plasticity and processing of memory. Therefore, the role of actin has been investigated in synaptic failure in neurodegenerative diseases such as Alzheimer's \cite{Pelucchi2020DendriticSpines}.

\subsection{Pathological Aggregates: Amyloid Fibrils}
Amyloid fibrils (Figs. \ref{fig:trp_positions}\refpanel{C} and \ref{fig:trp_positions}\refpanel{D}) are helical aggregates of amyloid proteins. Amyloids, the building blocks of amyloid fibrils, are a class of self-assembling proteins that fold in a $\beta$-sheet structure. The $\beta$-sheet structure, originally discovered by Herman Branson, Linus Pauling, and collaborators \cite{Pauling1951BetaSheet}, consists of so-called $\beta$-strands, each of which forms a zig-zag pattern, and which are connected laterally to each other via hydrogen bonding to form a pleated sheet. The $\beta$-sheet has a twist (i.e., the zig-zag sheet is not confined to undulations in a single plane). Multiple $\beta$-sheets stacked on top of one another form amyloid fibrils, which are also called $\beta$-helices. An image of a human amyloid subunit (PDB entry 6MST) is pictured in Fig. \ref{fig:amyloid_beta_sheet}\refpanel{A}. In Fig. \ref{fig:amyloid_beta_sheet}\refpanel{B}, we can see how a human amyloid subunit forms a $\beta$-sheet structure. Typical amyloid fibrils can grow up to several micrometers in length \cite{Eves2021AmyloidLength,Toyama2011AmyloidLength,SegersNolten2011AmyloidLength}. Many different proteins can form amyloids, such as amyloid-beta (A$\beta$) \cite{Findeis1999Amyloid}, islet amyloid polypeptide (IAPP) \cite{Akter2016Amylin}, lysozyme \cite{Kyle2001Lysozyme}, and insulin \cite{Nielsen2001Insulin}, and they are all associated with different pathological diseases. A$\beta$ is associated with Alzheimer's disease, while IAPP, lysozyme, and insulin are associated with type II diabetes, lysozyme amyloidosis, and injection-localized amyloidosis, respectively \cite{Chiti2017AmyloidSummaryOfProgress}. There are three main models that were proposed for how amyloids are created from the original protein fold: the refolding, natively disordered, and gain-of-interaction models \cite{Nelson2006AmyloidStructure}. The most well-known of these is the refolding model, in which the protein folds from its native state to an amyloid state. Through such a mechanism, amyloids form amyloid fibrils, which can further aggregate and form clumps known as amyloid plaques. Amyloid, amyloid fibrils, and amyloid plaques are a hallmark of neurodegenerative diseases, such as Alzheimer's and related dementias. 

Another indicator of Alzheimer's disease is the formation of neurofibrillary tangles, abnormal aggregates of the tau protein. The tau protein aids in structural support of microtubules in the brain \cite{Brion1991}, which start to disintegrate in Alzheimer's disease. The tau proteins fall off and undergo hyperphosphorylation, which causes them to transition from an unfolded state to a folded state capable of aggregating into threadlike structures inside neurons, called tangles \cite{Binder2005}. Tangles block transport and inhibit communication between neurons. The population density of tangles is strongly linked to the severity of cognitive decline in Alzheimer's disease \cite{Ow2014BriefOverview, Binder2005}.

\subsection{\label{sect:toy-models}Toy models of cylindrical geometries of ultraviolet-excited transition dipoles}

\begin{figure*}[tbhp]
\centering
\includegraphics[width=0.9\linewidth]{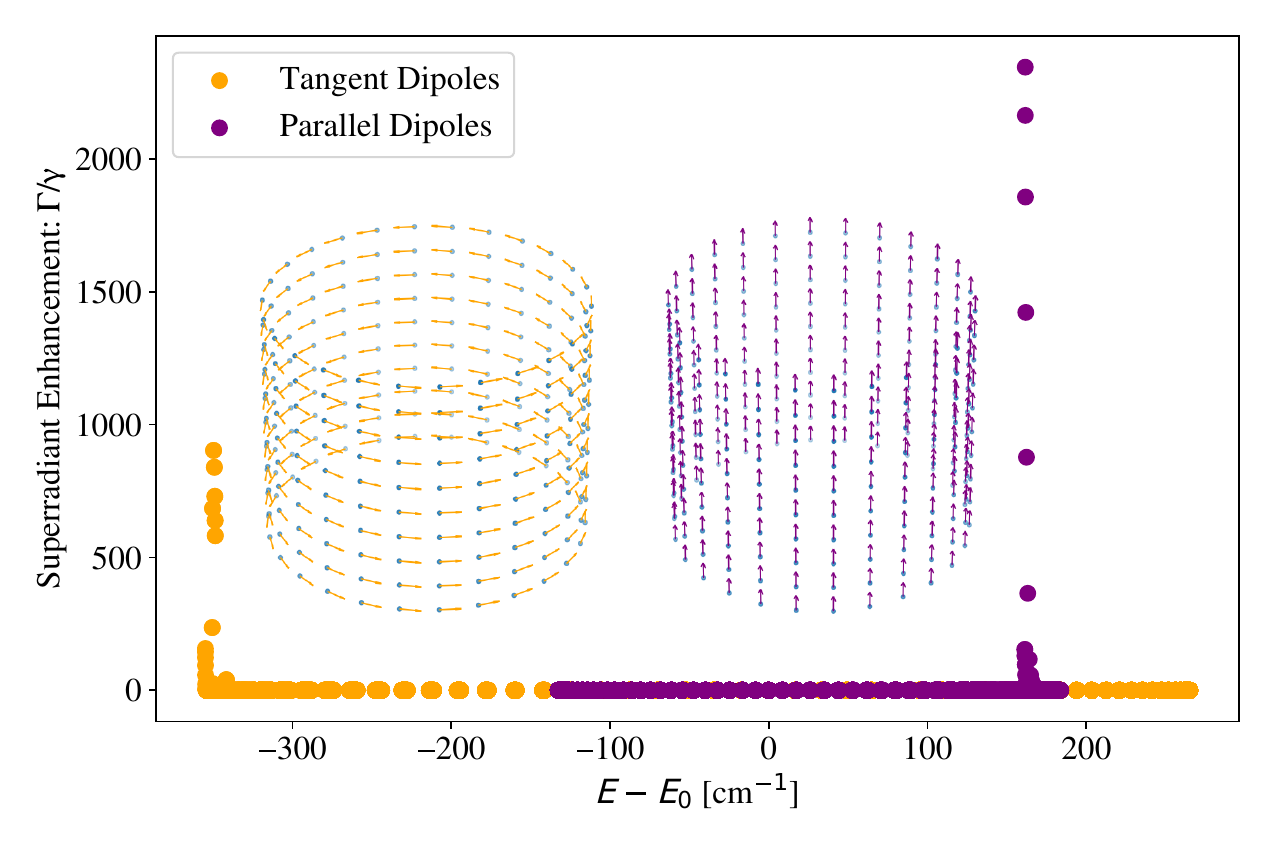}
\caption{\label{fig:perfect_rings_spectrum}Plot of the eigenvalue spectrum (superradiant enhancement rate vs. energy) of idealized ring structures, with the transition dipole vectors of each cylinder plotted inset in their corresponding colors. Each transition dipole has the photophysical parameters of a tryptophan (Trp) chromophore: an excitation wavelength of 280 nm and a fluorescent decay rate of $\sim2.73 \times 10^{-3}\text{ cm}^{-1}$. Each ring has a diameter of 22.4 nm, corresponding to the inner diameter of a microtubule. We see that, for these idealized structures, the superradiant states arise at only a very few specific energies.}
\end{figure*}

First, we present a preliminary analysis of some toy models, in order to gain physical intuition and insight on how transition dipole vector orientations affect the photophysical properties of prototypical biological structures with cylindrical symmetry. We consider two idealized architectures of molecules based on the ones studied in \cite{Gulli2019}. Each molecule has the photophysical parameters of Trp (excitation wavelength of $280$ nm and decay rate of $\sim2.73\times10^{-3} \text{ cm}^{-1}$), but different transition dipole geometries. The architectures consist of multiple rings parallel to the $x$-$y$ plane, each stacked on top of one another and separated by a distance $L$ in the $z$-direction. The first dipole vector arrangement is the case in which all vectors point in the $+z$ direction. We call this the parallel dipole (PD) arrangement. The next arrangement is where the dipoles are all pointing in the $x$-$y$ plane tangent to the ring, the so-called tangent dipole (TD) arrangement. See the insets from Fig. \ref{fig:perfect_rings_spectrum} for a visual representation of the structures. We solve for the the eigenstates of the PD and TD arrangements under the effective Hamiltonian \eqref{eq:non-Hermitian-Hamiltonian-simplified-form}.

Fig. \ref{fig:perfect_rings_spectrum} shows the eigenspectrum of the TD and PD arrangements. An interesting feature of the spectrum is that both the TD and PD structures have superradiant states at a few specific energies, rather than being distributed across many energies. This feature arises from a specific property of the transition dipole vector arrangements: each vector's orientation is only slightly deformed from its nearest neighbors (or, in the case of the PD arrangement, not modified at all). This symmetrical geometry of the transition dipole vectors creates a selection for a very small range of energies that contain superradiant states.

The spectrum having this unique property is significant because it influences the quantum yield (QY), defined as the ratio of the number of photons emitted to the number of photons absorbed, as well as its thermal average ($\langle\text{QY}\rangle_{th}$; see the Methods in Section \ref{sec:methods} for more details on the quantum yield). A spectrum with the majority of superradiant states lying at the lower end of the energy spectrum will have a higher $\langle\text{QY}\rangle_{th}$, since lower energies are weighted higher in a thermal Gibbs distribution, while a spectrum with superradiant states near the high-energy portion will have a lower $\langle\text{QY}\rangle_{th}$. This means that structures that are similar to a TD structure are likely to have a higher $\langle\text{QY}\rangle_{th}$, while structures similar to PD arrangements may have a lower $\langle\text{QY}\rangle_{th}$ value. This will be discussed further in relevant biostructures, in the ensuing sections.

This analysis is different from the analysis done in \cite{Gulli2019} in significant ways. Our analysis uses the parameters for the Trp chromophore, which has an absorption peak at $\sim280$ nm and a decay rate of $\sim2.73\times 10^{-3} \text{ cm}^{-1}$, while in \cite{Gulli2019}, the photosynthetic chromophores absorb in the visible ($\sim650$ nm) and have a smaller decay rate of $\sim1.821\times 10^{-4} \text{ cm}^{-1}$. The physics changes in a critical way when the excitation wavelength changes from the visible to the ultraviolet: the biosystem sizes considered generally become comparable to or larger than the excitation wavelength, as can be the case for characteristic microtubules, actin filaments, and amyloid fibrils in the brain. Thus, we employ a widely used effective Hamiltonian for the light-matter interactions that couples the Trp chromophores at long range due to their collective interactions with the electromagnetic field. This Hamiltonian is non-Hermitian because the large number of degrees of freedom of the electromagnetic field are traced out to give an effective description of the (collective) open quantum system, whose probability amplitude decays to the field with time. For further details on the non-Hermitian formalism, please see \cite{Moiseyev2011NonHermitian, Breuer2007OpenQuantumSystems, Manzano2020LindbladOverview, Lindblad1976Original}.

\section{Results}

\begin{figure*}[tbhp]
\centering
\includegraphics[width=0.75\linewidth]{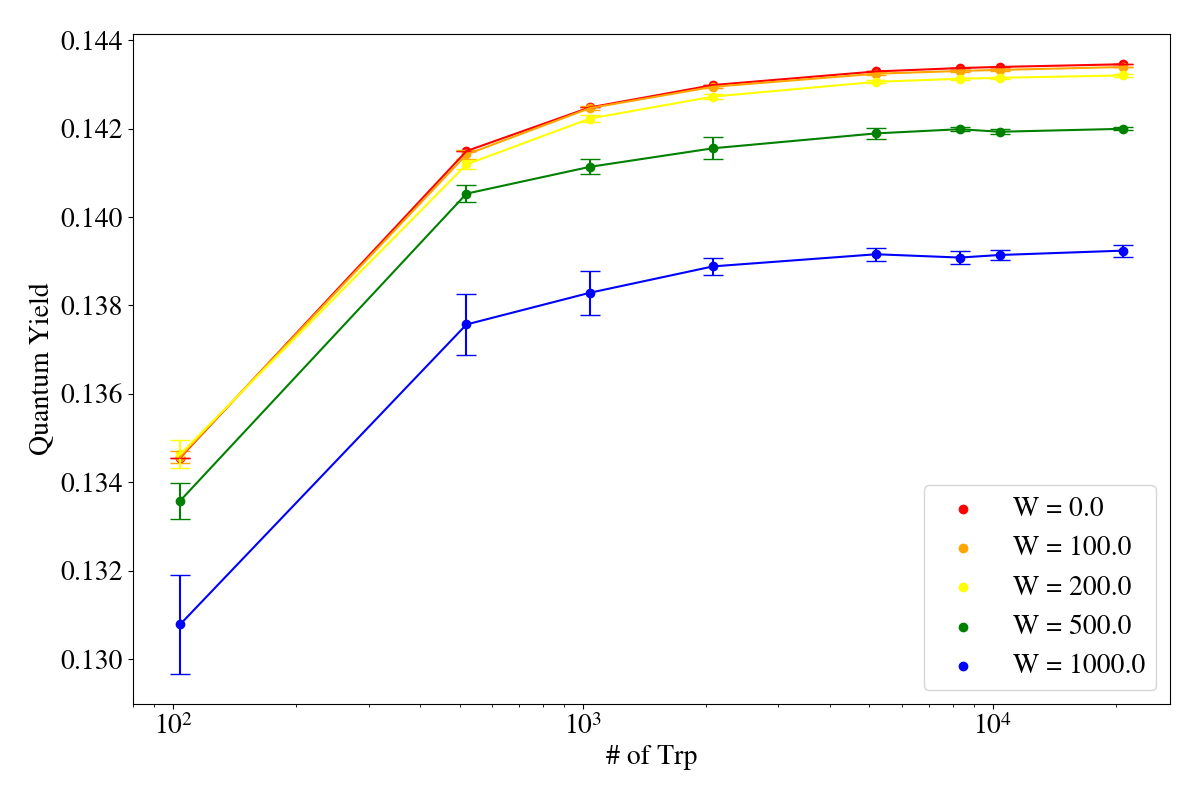}\hspace{15mm}
\caption{\label{fig:quantum_yield_microtubules}Thermal average of the quantum yield (QY) vs. number of tryptophan (Trp) molecules for varying static disorder strengths for single microtubules (pictured in Fig. \ref{fig:trp_positions}\refpanel{A}). $W$ (in units of $\text{cm}^{-1}$) represents the strength of static disorder applied to \eqref{eq:non-Hermitian-Hamiltonian-simplified-form}, where each diagonal element of the effective Hamiltonian $H_{\text{eff},\,ii}$ is replaced by a random value in the range $[H_{\text{eff},\,ii} - W/2, H_{\text{eff},\,ii} + W/2]$. Then the Hamiltonian is diagonalized to find the eigenvalues, and from the eigenvalues the thermal average of the quantum yield is obtained. This is repeated ten times, and the mean is taken to obtain a data point. The error bars on each point represent one standard deviation of the ten QY values calculated for that point.}
\end{figure*}

We present the eigensolutions of the effective Hamiltonian given in Eq. \eqref{eq:non-Hermitian-Hamiltonian-simplified-form} for microtubules, actin filaments, and amyloid fibrils of varying lengths (for details on the geometry of these structures, see the Methods in Section \ref{sec:methods}, and for a visual image, see Fig. \ref{fig:trp_positions}). We also consider the thermal average of the quantum yield (see Section \ref{sec:methods}: Methods for further details), its dependence on system size, and its robustness to static disorder. The effective Hamiltonian is non-Hermitian, and can be written as the sum of a Hermitian part and a non-Hermitian part
\begin{equation}
    \label{eq:non-Hermitian-Hamiltonian-simplified-form}
    H_\text{eff} = H_0 + \Delta - \frac{i}{2}G,
\end{equation}
where $H_0$, $\Delta$, and $G$ are real matrices. Because of the non-Hermitian part $-\frac{i}{2}G$, the eigenvalues of this matrix are complex numbers. Assuming that the dimension of the matrix is $N$, then the (right) eigenvectors $|\mathcal{E}_j\rangle$ and their associated eigenvalues $\mathcal{E}_j$ are
\begin{equation}
    \label{eq:eigenvalues-of-non-Hermitian-H}
    H_{\text{eff}}|\mathcal{E}_j\rangle = \mathcal{E}_j |\mathcal{E}_j\rangle;\,\,\,\,\mathcal{E}_j = E_j - \frac{i}{2}\Gamma_j,
\end{equation}
where $E_j$ is the energy (real part of eigenvalue) and $\Gamma_j$ is the decay rate ($-2$ times imaginary part of eigenvalue) of the eigenvector $|\mathcal{E}_j\rangle$. For further details, please see the Methods in Section \ref{sec:methods}.
\subsection{Microtubules}
\label{sec:ResultsMTs}

We study single microtubules of varying length constructed as spiral-cylindrical collectives of tubulin dimers (PDB entry 1JFF). One tubulin dimer contains 8 Trp molecules, and one spiral of the microtubule contains 13 tubulin dimers (so one spiral contains $8\times13=104$ Trp molecules). For example, a 100-spiral microtubule contains $104\times100=10400$ Trp molecules.

Shown in Fig. \ref{fig:quantum_yield_microtubules} is the thermal average of the quantum yield (QY) as a function of the length of the microtubule, reported as the number of Trp molecules. Each color represents a differing static disorder strength, with red being the smallest ($\noise{0}$) and blue being the largest ($\noise{1000}$), which is approximately five times larger than static disorder in a room-temperature environment ($\noise{200}$). Let us now define $\QYfixedlen{W_i}{W_f}$ as the percent change in quantum yield when the static disorder goes from $W_i \text{ cm}^{-1}$ to $W_f \text{ cm}^{-1}$ for a set system size. Using this notation for a system size of 20800 Trp molecules (200 spirals; the rightmost data points in Fig. \ref{fig:quantum_yield_microtubules}), $\QYfixedlen{0}{1000} = -3.08\%$. The quantum yield dampening by only $3.08\%$ when the static disorder is five times that of room temperature demonstrates its robustness. Such quantum yield robustness to static disorder has recently been experimentally confirmed for microtubules at room temperature \cite{Babcock2023} (also see Figs. S6 and S7 of \cite{Babcock2023} for theoretical predictions of the enormous superradiant enhancements for axonal microtubule bundles, and of the quantum yield robustness for centrioles, respectively). This suggests that  quantum yield robustness can be observed for similar biological structures, once realized experimentally.

The robustness of the quantum yield for microtubules (and their bundled architectures) is explainable from the shape of the spectrum of eigenvalues of the non-Hermitian Hamiltonian \eqref{eq:non-Hermitian-Hamiltonian-simplified-form}. The spectrum of single microtubules has been studied in \cite{Celardo2019}. Specifically, in Fig. 2c) of \cite{Celardo2019}, the spectrum of a 100-spiral microtubule (10400 Trp molecules) is shown. It can be seen that the most superradiant states lie in the low-energy portion of the spectrum. Examining Eq. \eqref{eq:thermal_avg_of_decay_rate}, if a large $\Gamma_j$ is associated with a smaller $E_j$, then the term $\Gamma_j \exp(-\beta E_j)$ in the Gibbs thermal ensemble will be weighted more strongly, thereby augmenting the quantum yield.

In the case of microtubules, the dependence of the thermal average of the quantum yield on system size also highlights that collective light-matter interactions can enhance quantum effects beyond the length scales normally associated with quantum behavior. Let us define $\QYfixednoise{N_i}{N_f}$ to be the percent change in the quantum yield at a fixed static disorder strength, when the structure goes from $N_i$ Trp molecules to $N_f$ Trp molecules. The QY has been determined experimentally to be $0.124$ for Trp alone in BRB80 aqueous buffer solution \cite{Babcock2023}. For microtubules, when $\noise{0}$, $\QYfixednoise{1}{20800} = 15.76\%$. With a static disorder of $\noise{1000}$, $\QYfixednoise{1}{20800} = 12.31\%$. So, even with extremely large static disorder strengths, at thermal equilibrium, the quantum yield for microtubules is enhanced as the system size grows.

We also study the effects of mechanical/vibrational degrees of freedom on the superradiant states of microtubules (see Figs. S1, S2 in the Supplementary Material), within the Born-Oppenheimer approximation. As seen in these figures, the superradiance is dynamically altered by nuclear geometry and changes when microtubules are mechanically deformed in different ways. Purely longitudinal modes (only deforming along the microtubule main axis) have higher superradiance than modes with purely twisting motions around the microtubule main axis, which have higher superradiance than bending motions off the microtubule main axis. Mode 15 has purely longitudinal stretching/contracting motions, which still preserve a $\Gamma/\gamma$ factor of greater than $27$. This mode has the highest superradiance of any mode in Figs. S1 and S2. The mode with the next highest superradiance is mode 9, with a 180-degree twisting motion along the microtubule axis. Some of the large bending motions displayed in modes 12, 13, and 17 dampen the superradiance down to a $\Gamma/\gamma$ factor of less than $10$. These results show that the superradiance is modified depending on not only the biological type of structure, but its intrinsic mechanical modes. Microtubules form long, straight, packed bundles in neuronal axons, which would mostly have longitudinal stretching/contracting modes. From this analysis, we show that longitudinal vibrational modes would not dampen superradiance as much as other modes, suggesting that highly stable structures such as axons may actively be exploiting quantum coherent effects based on their architecture and which mechanical modes are allowed. Although we only conduct this analysis for microtubules, the modulation of superradiant effects with vibrational state can be extended to other structures.

The vibrational modes studied have a mechanical frequency in the low gigahertz range \cite{havelka2017deformation}, corresponding to a timescale on the order of nanoseconds. As a microtubule oscillates mechanically, any superradiant states supported by the given atomic/nuclear configuration can vary from their enhancement factors reported in both the left and right columns of Figs. S1 and S2, to enhancement factors one or two orders of magnitude higher near the amplitude node of each vibration (see middle panels), where the structure is closest to a ``straight'' longitudinal configuration. From Table \ref{table:superradiant-subradiant-values}, we can see that the lifetimes for the superradiant states of microtubules and of many other structures are on the order of picoseconds.
Thus, the photophysical effect of superradiance is operating on a timescale at least three orders of magnitude faster than the mechanical motion of the microtubule, which can be considered more or less static in this ultrafast regime. However, even though our predictions have all been calculated within the Born-Oppenheimer approximation, it is clear from Table \ref{table:superradiant-subradiant-values} that the most subradiant states---and even a few of the superradiant states---supported by these neuroprotein architectures are extremely long-lived, suggesting potential influence and interaction across electronic and nuclear degrees of freedom in these structures. In many of the middle columns, we can see high exciton probabilities near the ends of the structures, despite the mode being symmetric. This could be a biological manifestation of topological edge states, which have been previously studied in paradigmatic non-Hermitian systems \cite{Yao2018Topological, Song2019Topological}.

\begin{figure*}[tbhp]
\hspace{5mm} \refpanel{A} \hspace{83mm} \refpanel{B} \\
\includegraphics[width=0.50\linewidth,height=0.22\paperheight]{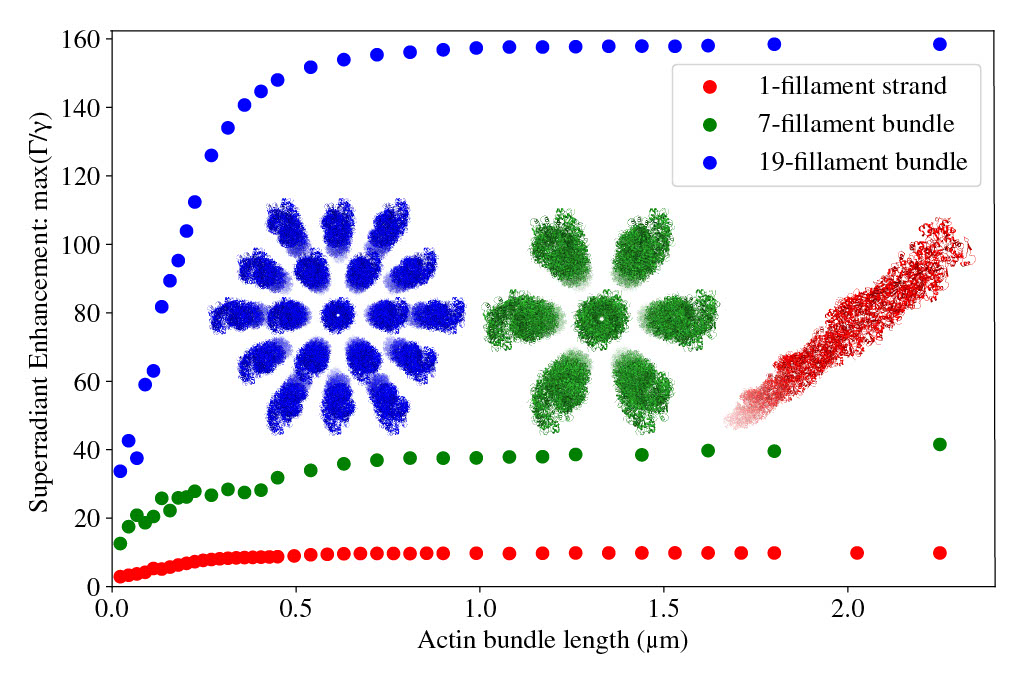}
\includegraphics[width=0.50\linewidth,height=0.22\paperheight]{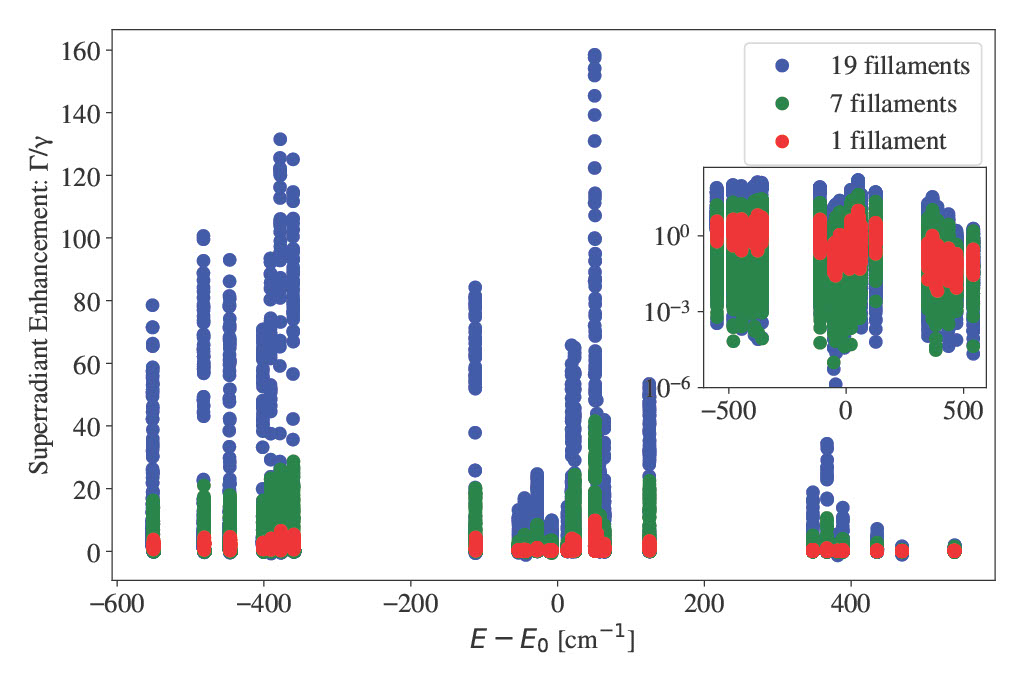}
\caption{\label{fig:actin_max_decay_rate_and_spectrum}\refpanel{A} Plot of the maximum superradiance $\max(\Gamma_j/\gamma)$ vs. structure length for model actin filament structures. Filaments have diameters of $\sim7$ nm, and the hexagonal bundles have filaments spaced $12$ nm from each other center-to-center. Sample images of 1-filament (3200 tryptophan), 7-filament (22400 tryptophan), and 19-filament (60800 tryptophan) actin structures are shown inset to the plot with their corresponding colors. \refpanel{B} The eigenvalue spectrum ($\Gamma_j/\gamma$ \text{vs} $E-E_0$) of $2.25$-$\mu\text{m}$ actin structures with 1 filament, 7 filaments, and 19 filaments in their corresponding colors. Inset is the same spectrum plotted with the $y$ axis on a semi-log scale.}
\end{figure*}

\begin{figure*}[tbhp]
\centering
\includegraphics[width=0.75\linewidth]{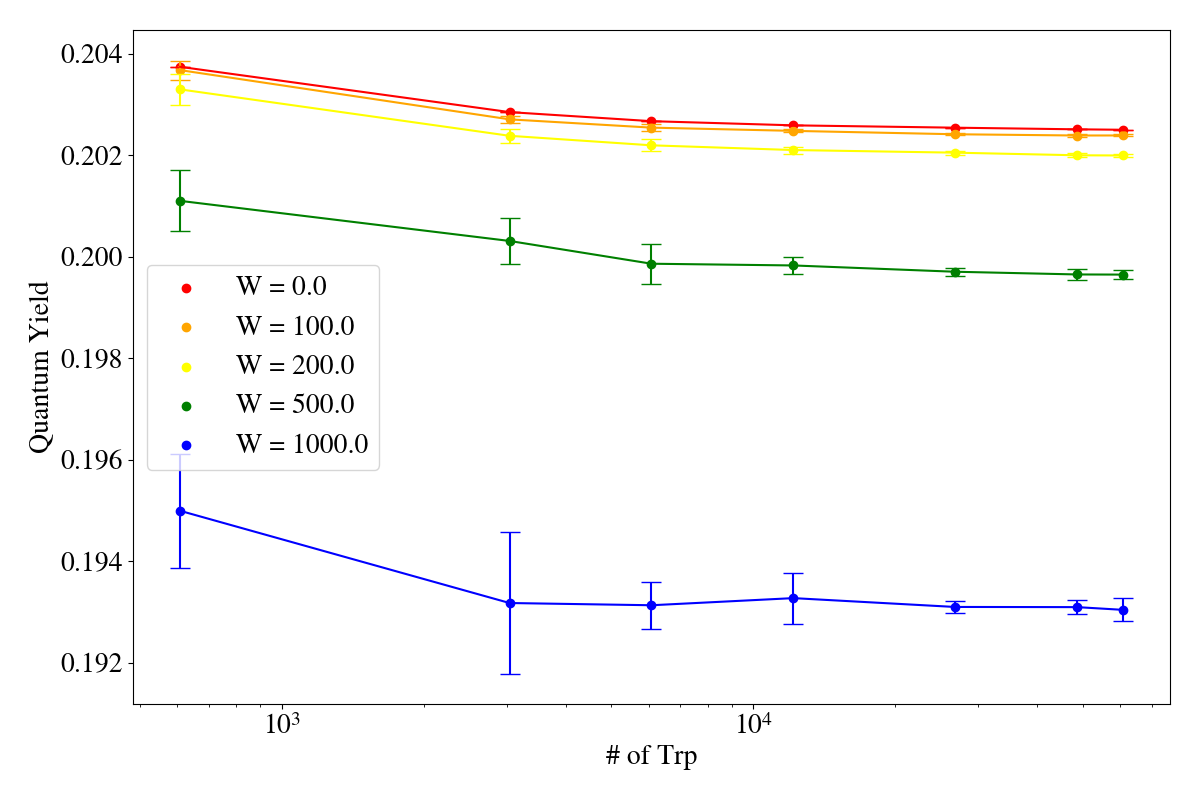}\hspace{15mm}
\caption{\label{fig:quantum_yield_actin6BNO_19_MT}Thermal average of the quantum yield (QY) vs. number of tryptophan (Trp) molecules for varying static disorder strengths for 19-filament bundles of actin (one actin filament is pictured in Fig. \ref{fig:trp_positions}\refpanel{B}, and a 19-filament bundle is pictured inset in blue in Fig. \ref{fig:actin_max_decay_rate_and_spectrum}\refpanel{A}). $W$ (in units of $\text{cm}^{-1}$) represents the strength of static disorder applied to \eqref{eq:non-Hermitian-Hamiltonian-simplified-form}, where each diagonal element of the effective Hamiltonian $H_{\text{eff},\,ii}$ is replaced by a random value in the range $[H_{\text{eff},\,ii} - W/2, H_{\text{eff},\,ii} + W/2]$. Then the Hamiltonian is diagonalized to find the eigenvalues, and from the eigenvalues the thermal average of the quantum yield is obtained. This is repeated ten times, and the mean is taken to obtain a data point. The error bars on each point represent one standard deviation of the ten QY values calculated for that point.}
\end{figure*}

\subsection{Actin filaments}

We study two different types of actin collectives: actin filaments and actin bundles. An actin filament is made from a concatenation of bare actin subunits (PDB entry 6BNO), each of which is a homo-octamer (protein consisting of eight identical chains) about 22.5 nm long. We then study two sizes of actin bundles corresponding to the smallest hexagonal configurations: 7-filament bundles, and 19-filament bundles (a top-down view of both of these are pictured in the inset of Fig. \ref{fig:actin_max_decay_rate_and_spectrum}\refpanel{A}). One bare actin subunit contains 32 Trp molecules. So, for example, a 19-filament bundle comprised of single-subunit actin filaments contains $32\times19=608$ Trp molecules. Also, the Trp network in actin filaments forms a helical structure, which repeats approximately every $40$ nm.

In Fig. \ref{fig:actin_max_decay_rate_and_spectrum}, panel \refpanel{A} shows the enhancement rate of the maximally superradiant state against the length of an actin filament or of a bundle of actin filaments. Similarly to the case of microtubules, the maximum superradiant enhancement rate increases with length at first, and then eventually saturates when the length approaches or exceeds the excitation wavelength. This feature is most pronounced in the 19-filament actin bundles. In the single filament and the 7-filament bundle, saturation of the maximum enhancement rate starts to occur when the length of the actin structure is around the length scale of excitation (280 nm). However, in the 19-filament bundle, saturation occurs at about twice that length. This is an interesting difference between microtubules/microtubule bundles \cite{Babcock2023,Celardo2019}, which saturate in their maximum superradiant enhancement (in the single-photon limit) at about three or more times the excitation wavelength, and actin filaments/bundles.

In panel \refpanel{B} of Fig. \ref{fig:actin_max_decay_rate_and_spectrum}, we can see that the maximally superradiant states of 2.25 $\mu$m-long actin structures are not close to the lowest exciton state. This impacts the quantum yield of actin bundles and filaments, as seen in Fig. \ref{fig:quantum_yield_actin6BNO_19_MT}. With zero static disorder, $\QYfixednoise{1}{60800} = 62.31\%$ and with $\noise{1000}$, $\QYfixednoise{1}{60800} = 55.14\%$. Even though the QY of actin bundles is enhanced for large structures with respect to Trp alone in solution, it can be seen in Fig. \ref{fig:quantum_yield_actin6BNO_19_MT} that all curves show a very slight decrease in the QY. Specifically, $\QYfixednoise{608}{60800} = -1.11\%$ for $\noise{0}$. For room-temperature static disorder of $\noise{200}$, such a decrease is still present, although for $\noise{1000}$, the change is within the error bars of the static disorder. Despite this decrease being small, its existence means that once a 19-filament bundle of single-subunit actin filaments is created, increasing the filament length further does not enhance the QY at all, contrary to the microtubule case. For the dependence of the QY of actin bundles with static disorder, we calculated $\QYfixedlen{0}{1000} = -4.17\%$ for a 60800-Trp actin bundle ($60800 \text{ Trp} = 2.25 \,\mu\text{m}$). This means that, at thermal equilibrium, the QY for large actin structures is dampened only slightly more than the QY for single microtubules.

Comparing Fig. \ref{fig:actin_max_decay_rate_and_spectrum} to the analogous spectrum for microtubules (see Fig. 2 of \cite{Celardo2019}), and by examining the entries of Table \ref{table:superradiant-subradiant-values}, we see that microtubules have brighter superradiant states than those for all actin structures. However, 19-filament actin bundles still have higher predicted QY values (see Fig. \ref{fig:quantum_yield_actin6BNO_19_MT}) than those for microtubules (see Fig. \ref{fig:quantum_yield_microtubules}). We can understand this by revisiting Eq. \eqref{eq:thermal_avg_of_decay_rate}. Each decay rate $\Gamma_j$ is weighted by a Bolztmann factor $\exp(-\beta E_j)$. If the energy $E_j$ is much smaller than the single-Trp excitation energy (our ``zero'' reference) and has a relatively large absolute value compared to this collective Lamb shift, this can compensate for the decay rate being small. Fig. \ref{fig:actin_max_decay_rate_and_spectrum} shows that the lowest energy states are shifted about $-600\text{ cm}^{-1}$, while for microtubules the lowest energy states are shifted only about $-100\text{ cm}^{-1}$. This is due to the Trp-Trp interactions in actin filaments being much larger than the Trp-Trp interactions in microtubules. For a 35-spiral microtubule ($280$-nm length), the average Trp-Trp interaction is $0.0311\text{ cm}^{-1}$, with a standard deviation of $0.898\text{ cm}^{-1}$. The nearest-neighbor Trp-Trp interaction is $62.82\text{ cm}^{-1}$, which is small compared to room temperature ($k_BT\approx 200\text{ cm}^{-1}$). An actin filament of 13 subunits ($292.5\text{ nm}$), on the other hand, has an average Trp-Trp interaction of $0.683\text{ cm}^{-1}$, with a standard deviation of $14.06\text{ cm}^{-1}$. The nearest-neighbor Trp-Trp interaction for an actin filament is $537.2\text{ cm}^{-1}$, much larger than room temperature.
The strength of the Trp-Trp couplings in actin compared to microtubules explains the larger collective Lamb shift for lower-energy states in actin, thereby explaining its high quantum yield despite having dimmer superradiant states than microtubules.

These results show that although the absolute values of the QYs for 19-filament actin bundles are larger than that of single microtubules, these QYs for actin bundles decrease with system size after even a single twist (608 Trp). However, 19-filament actin bundles are comparable to microtubules in their QY robustness to static disorder. Although such actin bundles do exhibit observable and important superradiant effects via the QY, our results imply that their role in cytoskeletal dynamics may be restricted, more so than microtubules, to their conventional mechanical roles rather than having significant photophysical enhancements at long length scales. Experiments \textit{in vitro} to detect superradiant QY enhancements (from Trp in solution) in actin bundles would be warranted, as has been demonstrated with microtubules \cite{Babcock2023}.

\begin{figure*}[tbhp]
\hspace{7mm} \refpanel{A} \hspace{83mm} \refpanel{B} \\
\includegraphics[width=0.50\linewidth,height=0.22\paperheight]{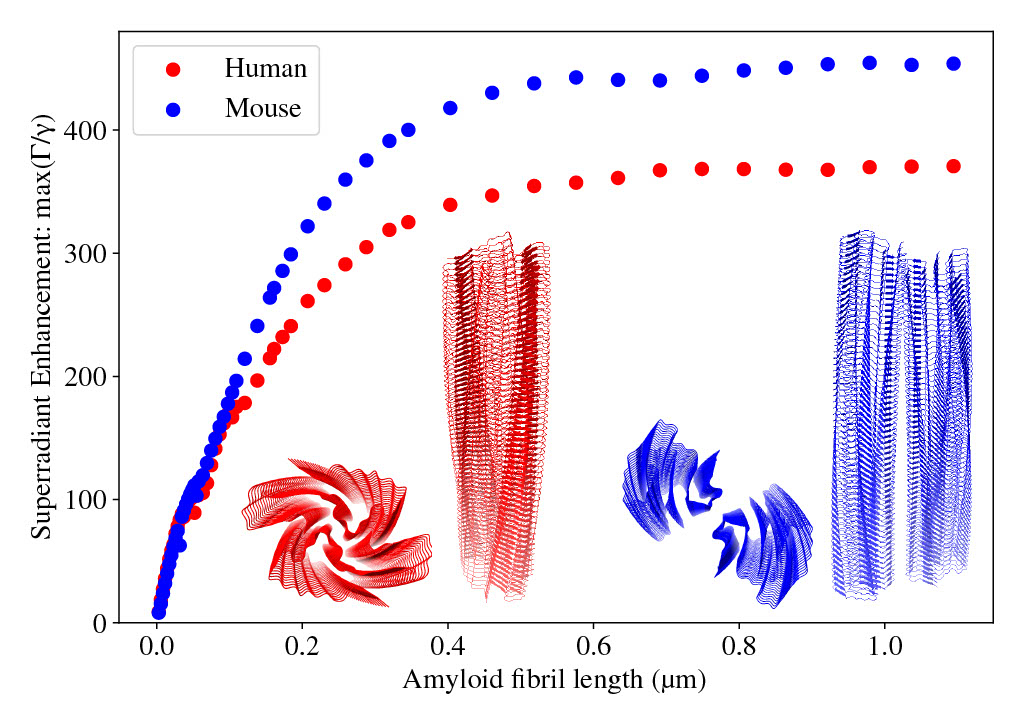}
\includegraphics[width=0.50\linewidth,height=0.22\paperheight]{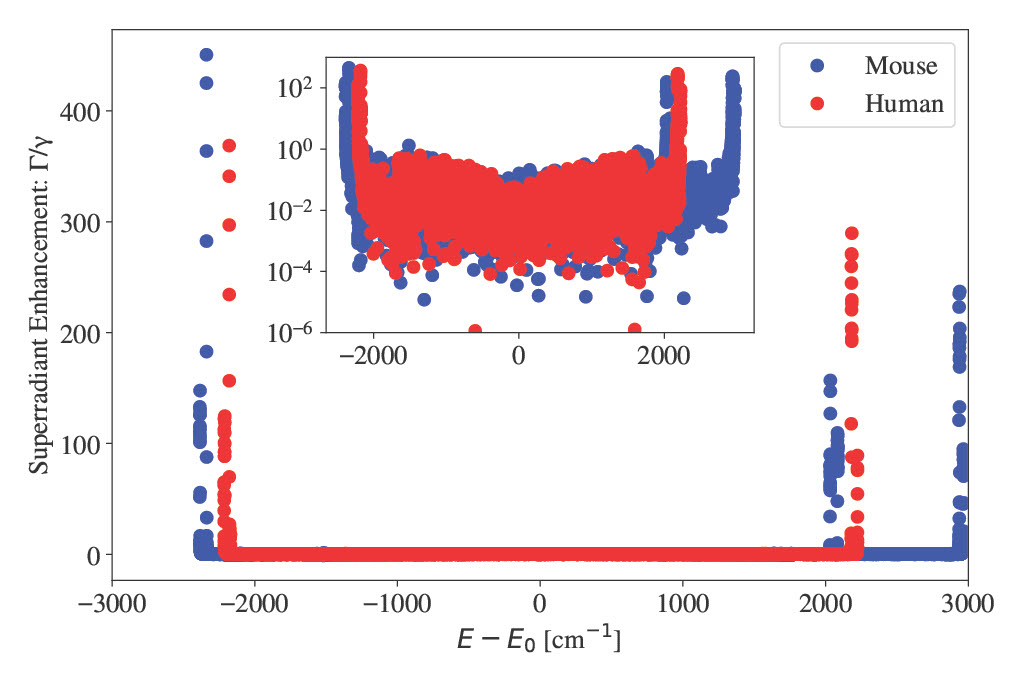}
\caption{\label{fig:amyloid_max_decay_rate_and_spectrum}\refpanel{A} Plot of the maximum superradiance $\max(\Gamma_j/\gamma)$ vs. structure length for model amyloid fibrils. Sample images of amyloid structures built from PDB files 6MST (\textit{Homo sapiens}) and 6DSO (\textit{Mus musculus}) are shown inset to the plot with their corresponding colors, both in a cross-sectional view and a longitudinal view. \refpanel{B} The eigenvalue spectrum ($\Gamma_j/\gamma$ \textit{vs} $E-E_0$) of $864$-$\text{nm}$ 6MST (7200 tryptophan) and 6DSO (10800 tryptophan) amyloid fibrils. Inset is the same spectrum plotted with the $y$ axis on a semi-log scale.}
\end{figure*}

\begin{figure*}[tbhp]
\centering
\includegraphics[width=0.75\linewidth]{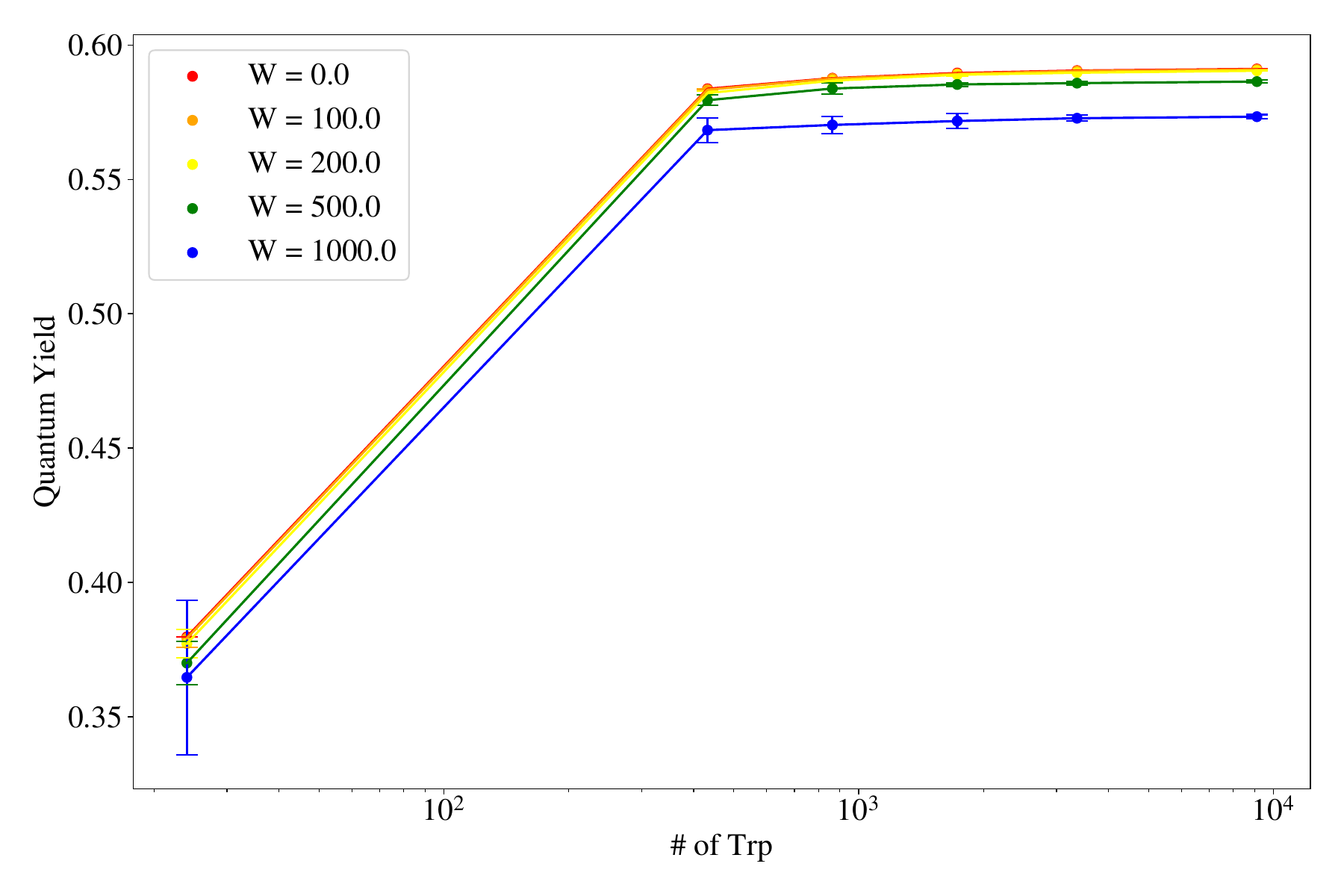}\hspace{15mm}
\caption{\label{fig:quantum_yield_amyloid6MST}Thermal average of the quantum yield (QY) vs. number of tryptophan (Trp) molecules for varying static disorder strengths for human amyloid fibrils (pictured in Fig. \ref{fig:trp_positions}\refpanel{C}). $W$ (in units of $\text{cm}^{-1}$) represents the strength of static disorder applied to \eqref{eq:non-Hermitian-Hamiltonian-simplified-form}, where each diagonal element of the effective Hamiltonian $H_{\text{eff},\,ii}$ is replaced by a random value in the range $[H_{\text{eff},\,ii} - W/2, H_{\text{eff},\,ii} + W/2]$. Then the Hamiltonian is diagonalized to find the eigenvalues, and from the eigenvalues the thermal average of the quantum yield is obtained. This is repeated ten times, and the mean is taken to obtain a data point. The error bars on each point represent one standard deviation of the ten QY values calculated for that point.}
\end{figure*}

\begin{figure*}[tbhp]
\centering
\includegraphics[width=0.75\linewidth]{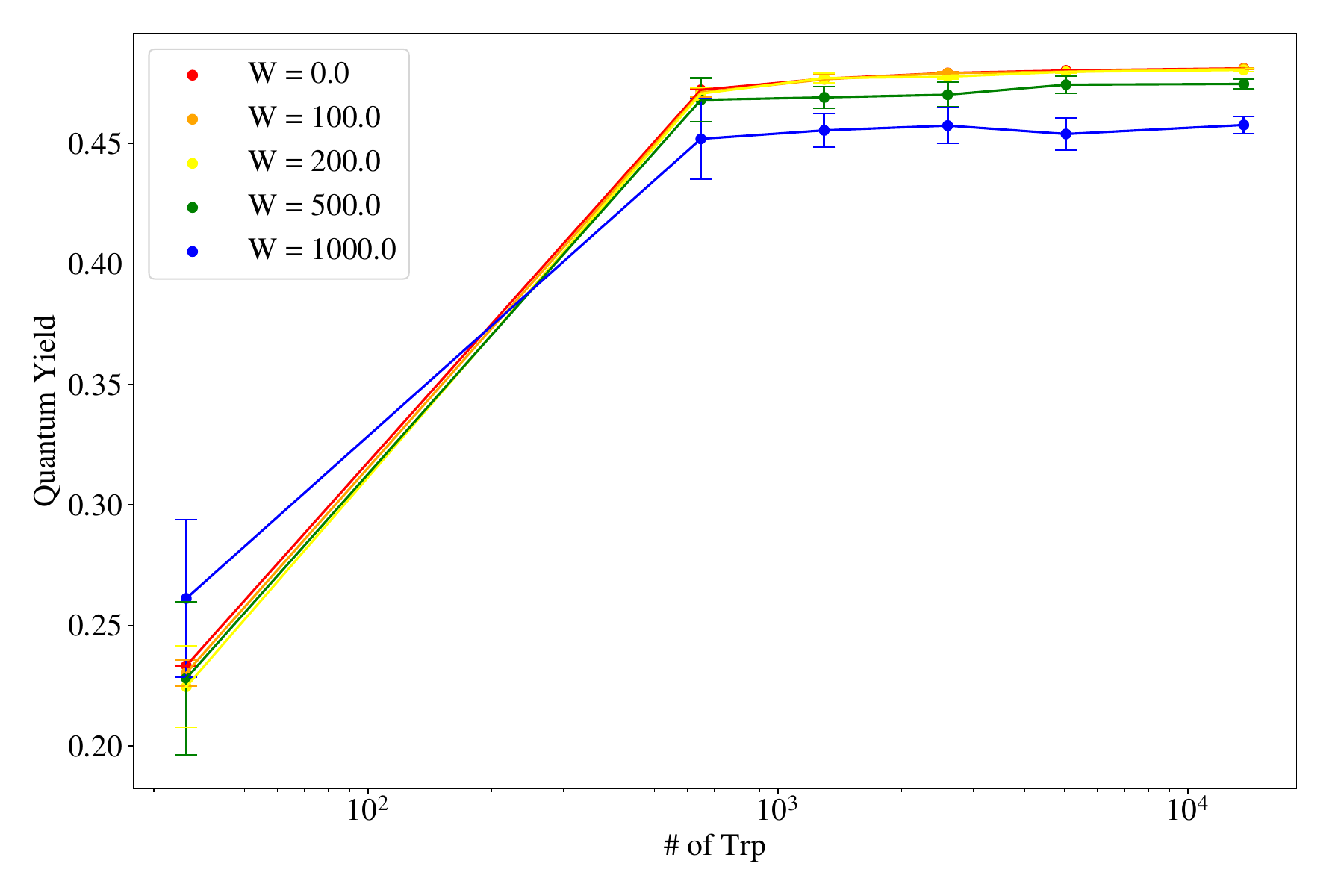}\hspace{15mm}
\caption{\label{fig:quantum_yield_amyloid6DSO}Thermal average of the quantum yield (QY) vs. number of tryptophan (Trp) molecules for varying static disorder strengths for murine amyloid fibrils (pictured in Fig. \ref{fig:trp_positions}\refpanel{D}). $W$ (in units of $\text{cm}^{-1}$) represents the strength of static disorder applied to \eqref{eq:non-Hermitian-Hamiltonian-simplified-form}, where each diagonal element of the effective Hamiltonian $H_{\text{eff},\,ii}$ is replaced by a random value in the range $[H_{\text{eff},\,ii} - W/2, H_{\text{eff},\,ii} + W/2]$. Then the Hamiltonian is diagonalized to find the eigenvalues, and from the eigenvalues the thermal average of the quantum yield is obtained. This is repeated ten times, and the mean is taken to obtain a data point. The error bars on each point represent one standard deviation of the ten QY values calculated for that point.}
\end{figure*}

\begin{figure*}[tbhp]
\centering
\refpanel{A} \hspace{69mm} \refpanel{B} \\
\includegraphics[width=0.25\linewidth]{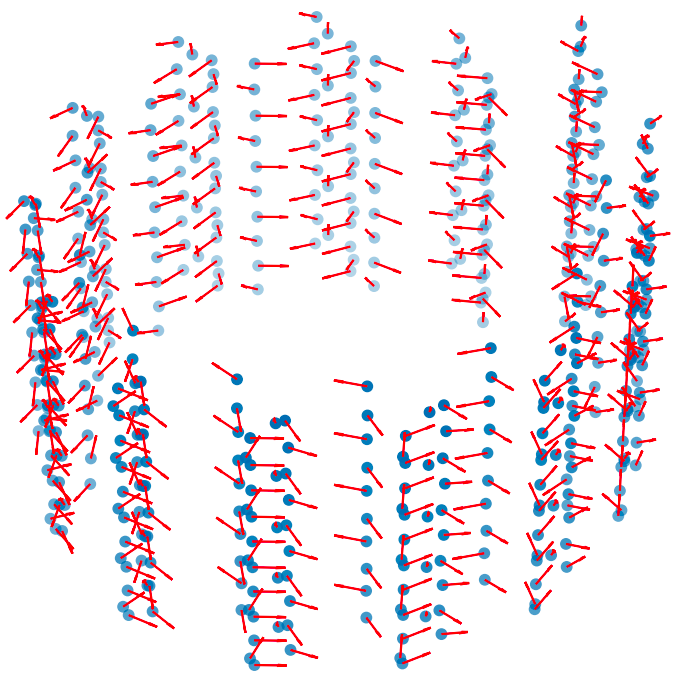}
\hspace{30mm}
\includegraphics[width=0.25\linewidth]{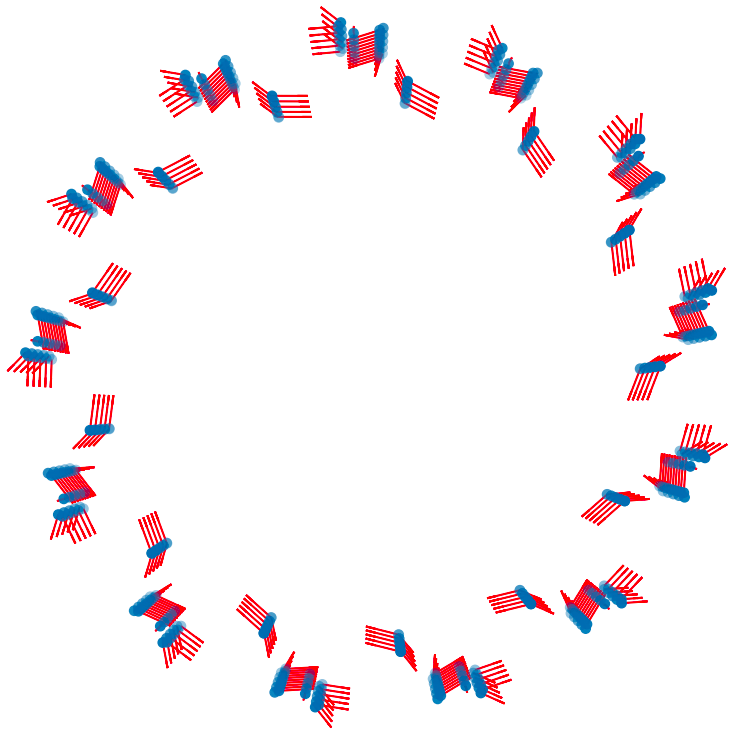}\\\,
\refpanel{C} \hspace{69mm} \refpanel{D} \\
\includegraphics[width=0.25\linewidth]{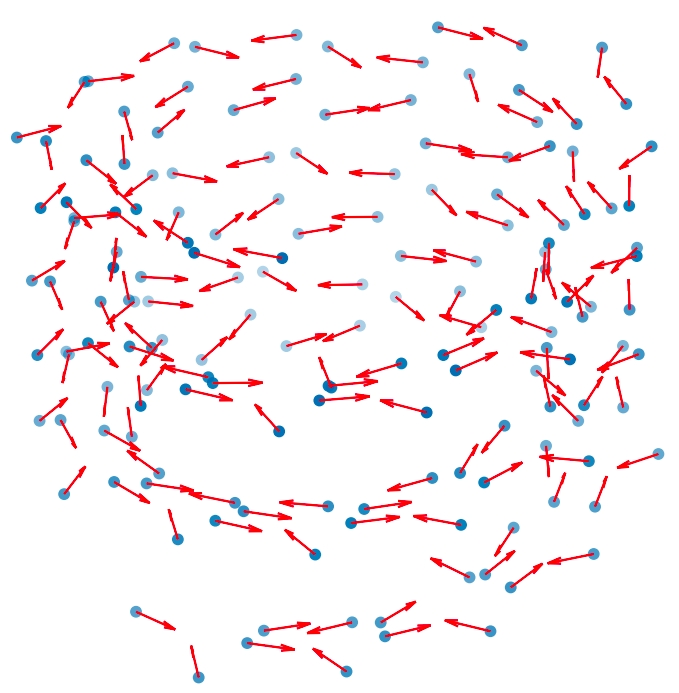}
\hspace{30mm}
\includegraphics[width=0.25\linewidth]{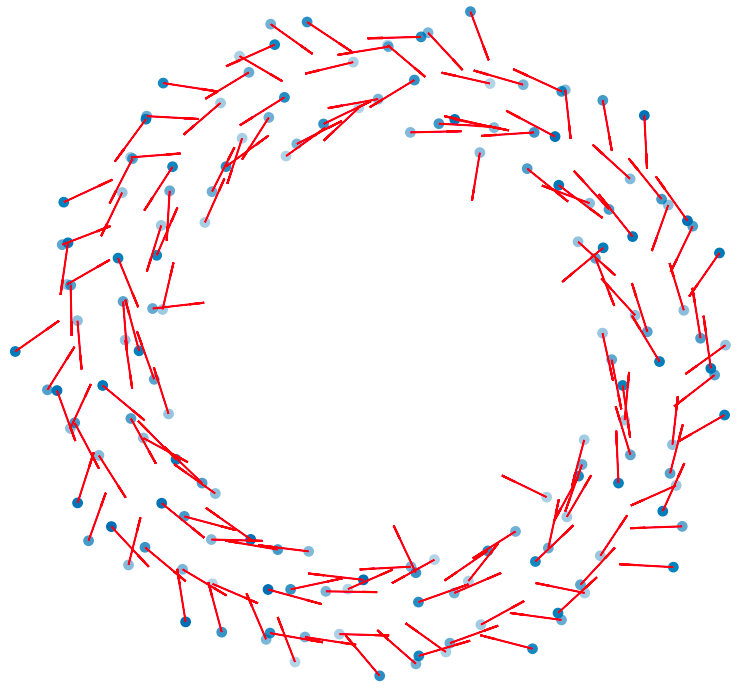}\\\,
\refpanel{E} \hspace{69mm} \refpanel{F} \\
\includegraphics[width=0.25\linewidth]{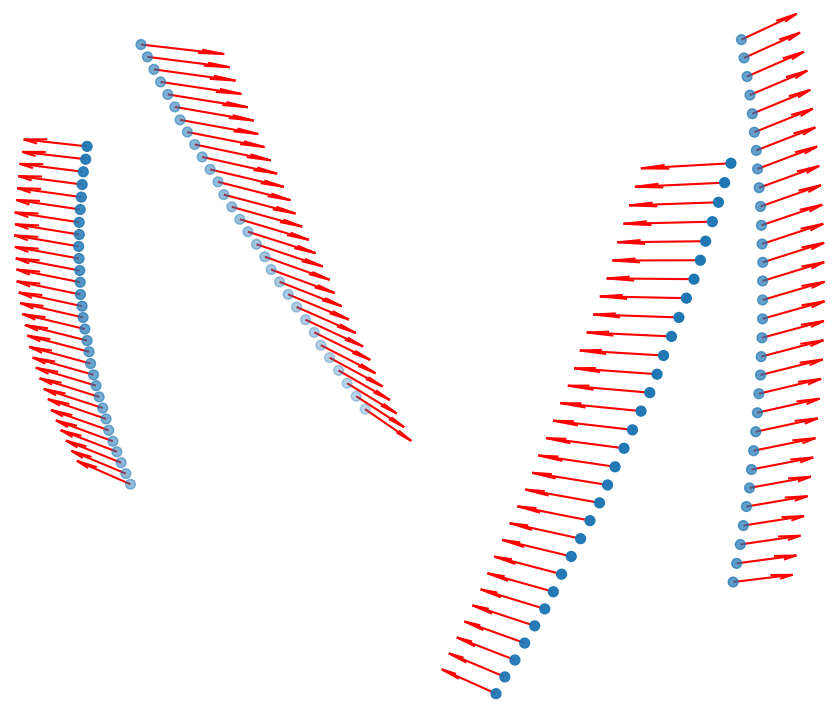}
\hspace{30mm}
\includegraphics[width=0.25\linewidth]{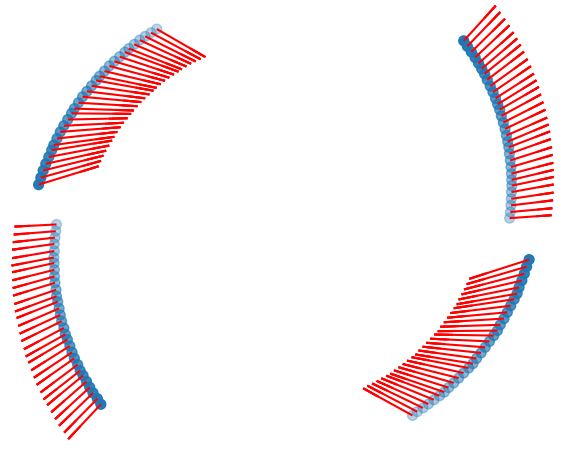}\\\,
\refpanel{G} \hspace{69mm} \refpanel{H} \\
\includegraphics[width=0.25\linewidth]{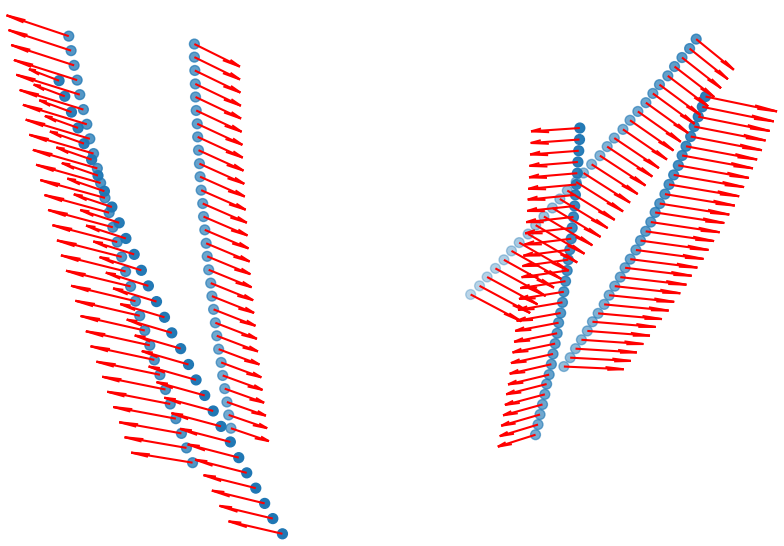}
\hspace{30mm}
\includegraphics[width=0.25\linewidth]{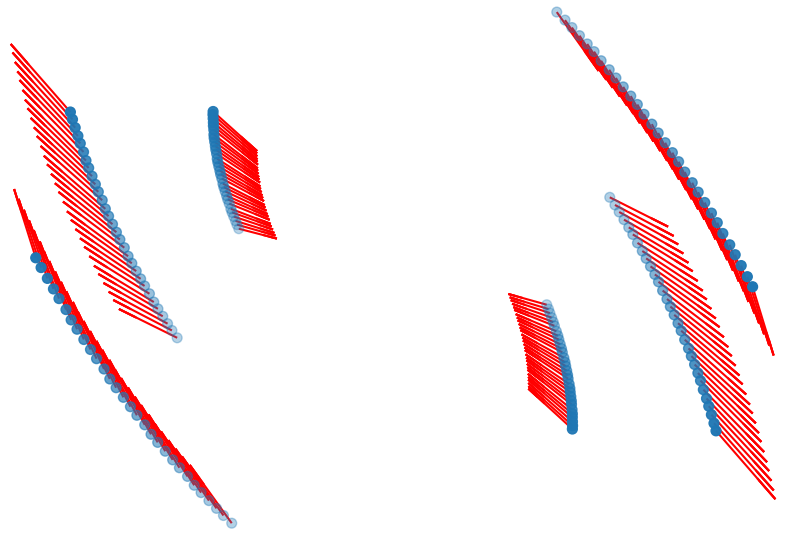}
\caption{\label{fig:dipole_vectors}Transition dipole vector geometries of tryptophan (Trp) in the realistic biological structures considered in this paper. The blue dots represent the Trp molecules. Relative size of structures with respect to one another is not to scale. The size of vectors has been enlarged for viewing. \refpanel{A} Angled longitudinal view of $45$ nm microtubule (built from tubulin dimer PDB entry 1JFF). \refpanel{B} Cross-sectional view of the microtubule. \refpanel{C} Angled longitudinal view of $112.5$ nm actin filament (built from actin subunit PDB entry 6BNO). \refpanel{D} Cross-sectional view of the actin filament. \refpanel{E} Angled longitudinal view of $20.16$ nm human amyloid fibril (built from amyloid subunit PDB entry 6MST). \refpanel{F} Cross-sectional view of human amyloid fibril. \refpanel{G} Angled longitudinal view of $20.16$ nm mouse amyloid fibril (built from amyloid subunit PDB entry 6DSO). \refpanel{H} Cross-sectional view of mouse amyloid fibril.}
\end{figure*}

\subsection{Amyloid fibrils}
Fig. \ref{fig:amyloid_max_decay_rate_and_spectrum} presents the same data as in Fig. \ref{fig:actin_max_decay_rate_and_spectrum}, but for single amyloid fibrils in human and in mouse. We will refer to the length of an amyloid fibril by its absolute length (in $\mu$m), and/or the number of subunits it is made of. The subunit of human (mouse) amyloid is given by the PDB entry 6MST (6DSO), contains 24 (36) Trp molecules, and is $2.88$ nm in length.  The Trp network of amyloid forms a helical pattern, which repeats approximately every 40 subunits (every $115.2$ nm).

In panel \ref{fig:amyloid_max_decay_rate_and_spectrum}\refpanel{A}, it can be seen that for both human and mouse amyloid fibrils, the maximum superradiant enhancement \(\text{max}(\Gamma/\gamma)\) reaches a saturating value at large lengths. For $1.09$-$\mu$m human (mouse) amyloid fibrils, $\text{max}(\Gamma/\gamma) = 371.21 \,(453.14)$, far surpassing that of even a $2.2$-$\mu$m 19-filament actin bundle, which has $\text{max}(\Gamma/\gamma) = 160.93$. This is despite the Trp network in the amyloid fibrils being comprised of significantly fewer chromophores than that in actin. In fact, if one considers a human amyloid fibril and a 19-filament actin bundle that have the same number of Trp molecules (e.g., picking 1824 Trp molecules), the actin bundle has $\text{max}(\Gamma/\gamma) = 37.70$, while the human amyloid fibril has $\text{max}(\Gamma/\gamma) = 270$. This shows that the density of Trp chromophores within a unit volume and the transition dipole orientations of amyloid are much more suited for maintaining bright superradiant states than the Trp density and transition dipole orientations of actin bundles.

In panel \ref{fig:amyloid_max_decay_rate_and_spectrum}\refpanel{B}, we can see that superradiant states emerge at very specific bands in both the low- and high-energy portions of the spectrum, and at all other energies, the superradiant enhancement rate is very close to 0. For the human amyloid fibril (built from PDB file 6MST), the superradiant states are only present near the smallest and largest energies, and every other state is subradiant ($\Gamma_j < \gamma$, close to zero enhancement rate). The emergence of superradiant states only at a few energies arises due to the structure of the Trp networks in question, as discussed with the toy models in Section \ref{sect:toy-models}. Both amyloid fibril structures have dipole vector orientations that vary more smoothly from one dipole to its nearest neighbor, as compared with microtubules and actin filaments, which don't exhibit this feature (Fig. \ref{fig:dipole_vectors}). The presence of a large proportion of superradiant states in the low-energy portion of the amyloid fibril spectrum gives it a very large QY, as seen in Figs. \ref{fig:quantum_yield_amyloid6MST} and \ref{fig:quantum_yield_amyloid6DSO}. Specifically, for system sizes of 864 Trp or above, the human (mouse) amyloid has a quantum yield between 0.55 and 0.60 (0.44 and 0.49) for all considered values of static disorder up to 1000 cm$^{-1}$; these quantum yields are about two to three times that of actin filaments, and more than three to four times that of microtubules. 

For the dependence of the amyloid fibril QY on static disorder, for $864$nm structures, $\QYfixedlen{0}{1000} = -3.05\%$ for human, and $\QYfixedlen{0}{1000} = -5.01\%$ for mouse, making these fibrils at least as robust to static disorder as microtubules and actin filament bundles, and potentially more so. The QY of amyloid is also strongly enhanced with system size, with $\QYfixednoise{1}{9120} = 130.65\%$ for human and $\QYfixednoise{1}{13680} = 118.05\%$ for mouse, at zero static disorder. For $\noise{1000}$, $\QYfixednoise{1}{9120} = 128.88\%$ for human and $\QYfixednoise{1}{13680} = 114.73\%$ for mouse. Thus, amyloid displays a very high QY that increases with system size, up to a certain point at which it begins to saturate: at zero static disorder, $\QYfixednoise{432}{9120} = 1.26\%$ for human amyloid fibrils and $\QYfixednoise{648}{13680} = 1.87\%$ for mouse amyloid fibrils, a clear indication of flattening of the monotonically increasing QY, as compared with the initial more-than-doubling and more-than-tripling of the QY from Trp alone in solution. For a human (mouse) amyloid fibril with greater than or equal to 432 (648) Trp molecules, the QY stays constant when $\noise{1000}$ (any variation is within the error bars, i.e., random fluctuations caused by the static disorder).

It should be noted that for both amyloid fibrils, some inter-Trp distances are as small as $\sim 5 \text{\AA}$. In the Hamiltonian \eqref{eq:non-Hermitian-Hamiltonian-simplified-form}, the point dipole approximation is made. But, in reality, the Trp molecule extends over space: the distance from the oxygen atom to the CZ2 atom in Trp is $\sim7.7\text{\AA}$. The inter-Trp distances being smaller than the Trp molecules themselves means that orbitals of different Trp molecules may overlap, and this can lead to the formation of charge-transfer states, which are intermediate between an exciton and an electron donor-acceptor complex. Such charge-transfer states in biomolecular complexes with closely spaced chromophores have previously been described in DNA \cite{Bauer2022DNA, Renaud2013DNA, Bittner2006DNA}, proteins \cite{Spata2014Oligonucleotide, Jong2019Proteins, Oakley2006Proteins}, and photosynthetic complexes \cite{AndreaRozzi2013LightHarvesting}. In the case of our amyloid fibrils, strictly speaking, the point dipole approximation and the approximation of Trp as a two-level system would both break down. However, this only applies to the nearest-neighbor Trp-Trp interactions. For Trp molecules that are not as closely spaced (the majority of Trp-Trp pairs), the long-range terms that go as $r^{-1}$ in Eqns. \eqref{eq:non-hermitian-hamiltonian-Delta-full} and \eqref{eq:non-hermitian-hamiltonian-G-full} are dominant over the $r^{-2}$ and $r^{-3}$ terms. Quantum coherent effects such as superradiance are greatly enhanced by this type of long-range interaction, and for these Trp molecules, the aforementioned approximations remain valid.

Due to the close Trp-Trp spacings in amyloid fibrils, we would expect the Trp-Trp couplings to be very high, and indeed they are. The average Trp-Trp coupling strength for a 100-subunit human (mouse) amyloid fibril, which has a length of $288$ nm, is $0.971\text{ cm}^{-1}$ ($0.741\text{ cm}^{-1}$) with a standard deviation of $26.6\text{ cm}^{-1}$ ($24.6\text{ cm}^{-1}$). The nearest-neighbor Trp-Trp coupling for human (mouse) amyloid fibril is $1012\text{ cm}^{-1}$ ($1306\text{ cm}^{-1}$). As we expected, this leads to energies with much larger collective Lamb shifts for the lowest exciton states of amyloid fibrils, which can be seen in Fig. \ref{fig:amyloid_max_decay_rate_and_spectrum}\refpanel{B} at about $-2500\text{ cm}^{-1}$. This explains the very high quantum yield for amyloid fibrils (see Figs. \ref{fig:quantum_yield_amyloid6MST} and \ref{fig:quantum_yield_amyloid6DSO}): they have their brightest superradiant states at large negative shifts from the single-Trp excitation energy, thereby increasing the weight of these states in the thermal ensemble beyond those in either microtubules or actin bundles.

\subsection{Energy gaps in the complex plane, and thermal robustness}

\begin{figure*}[tbhp]
    \centering
    \refpanel{A} \hspace{85mm} \refpanel{B}\\
    \includegraphics[width=0.49\linewidth]{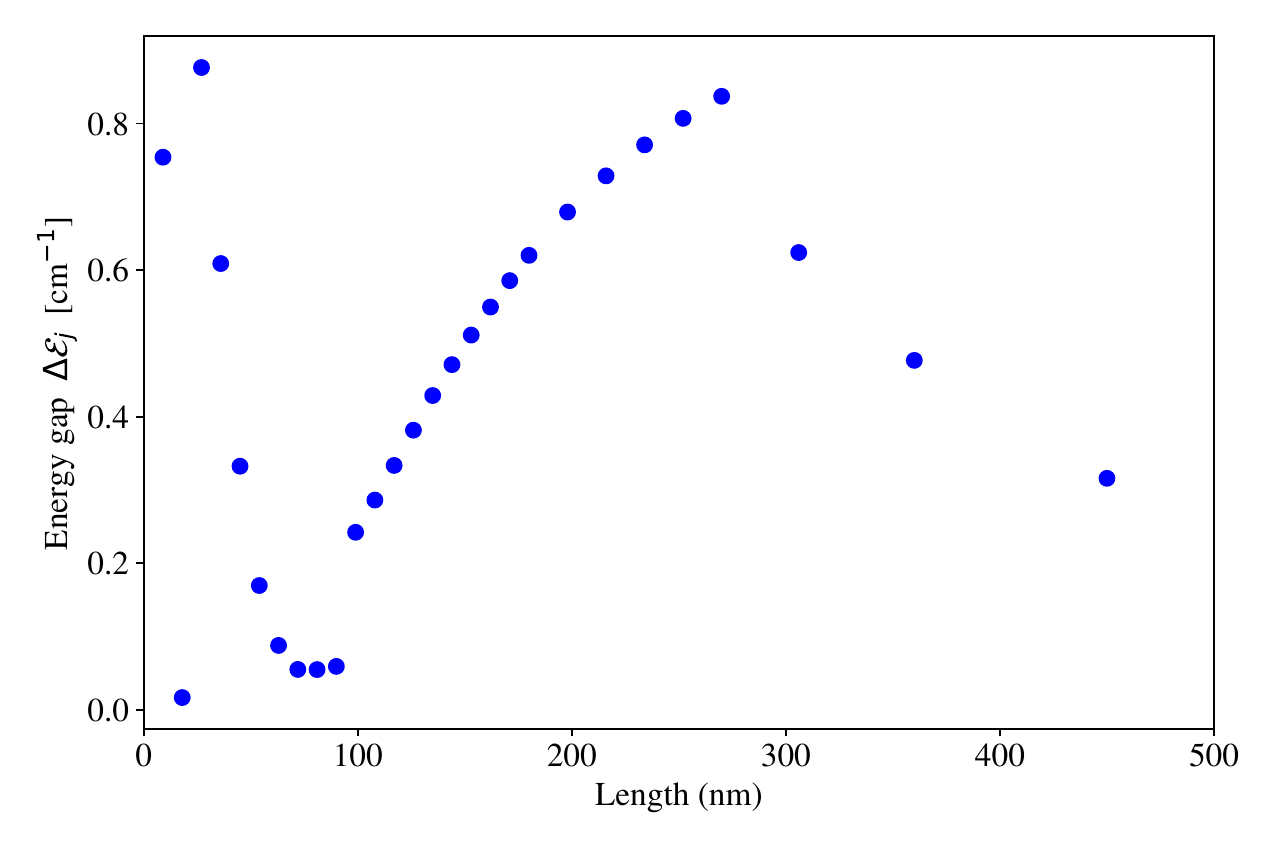}
    \includegraphics[width=0.49\linewidth]{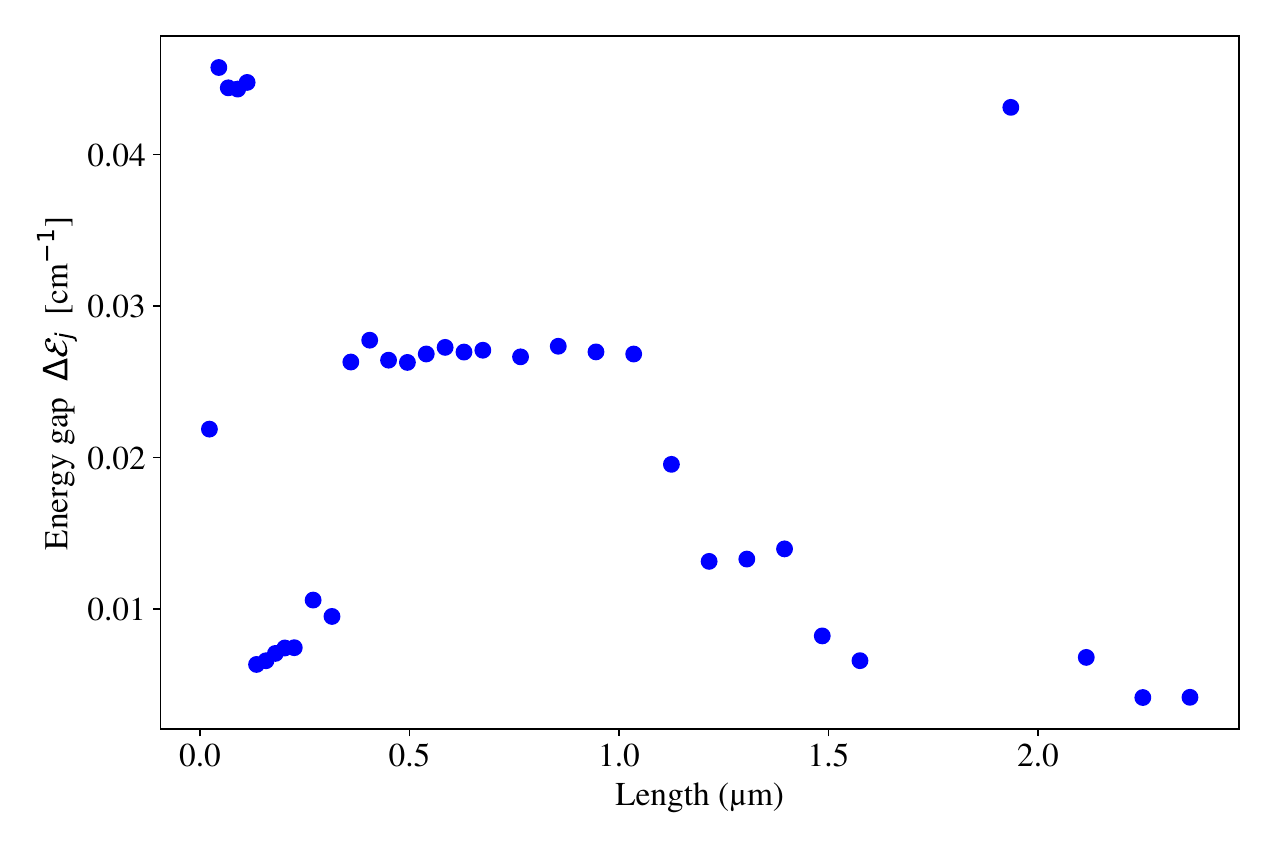}\\
    \refpanel{C} \hspace{85mm} \refpanel{D}\\
    \includegraphics[width=0.49\linewidth]{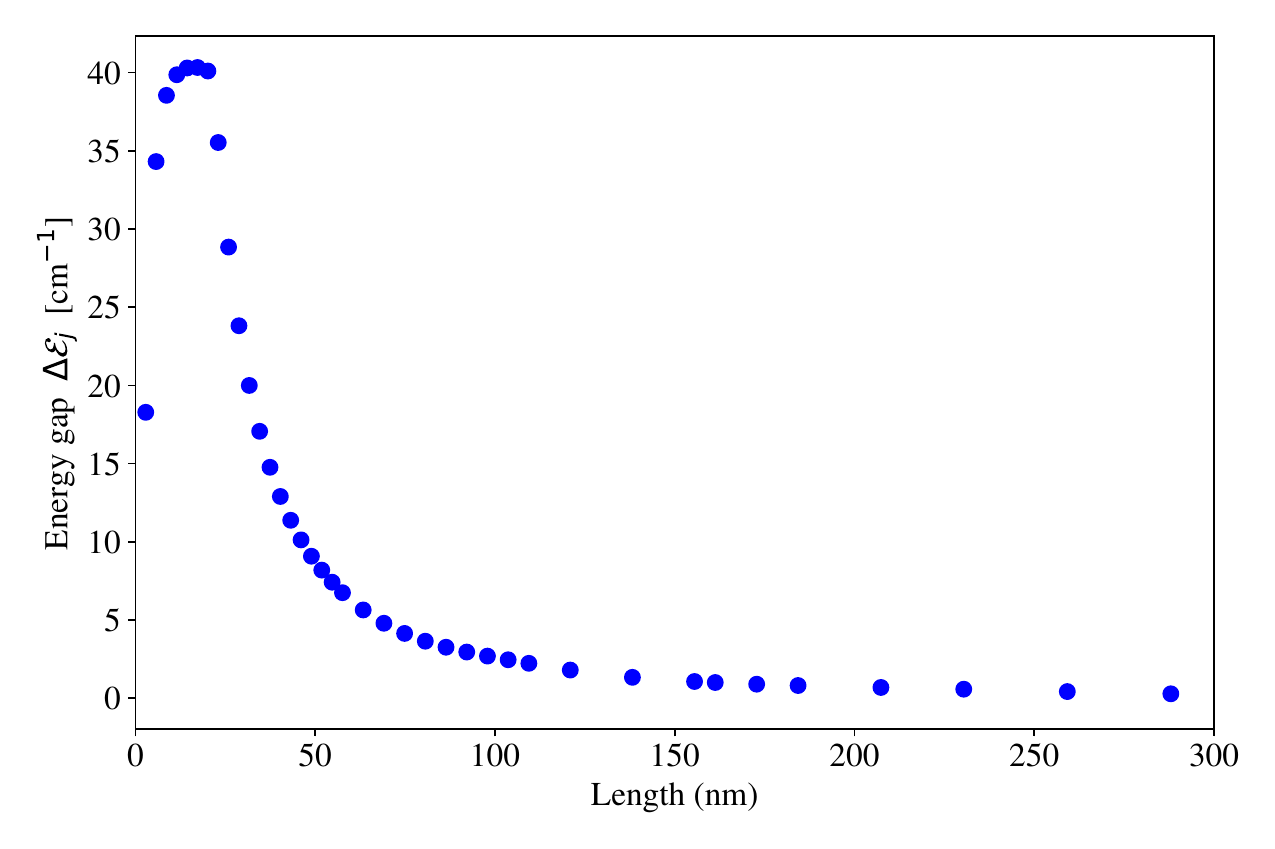}
    \includegraphics[width=0.49\linewidth]{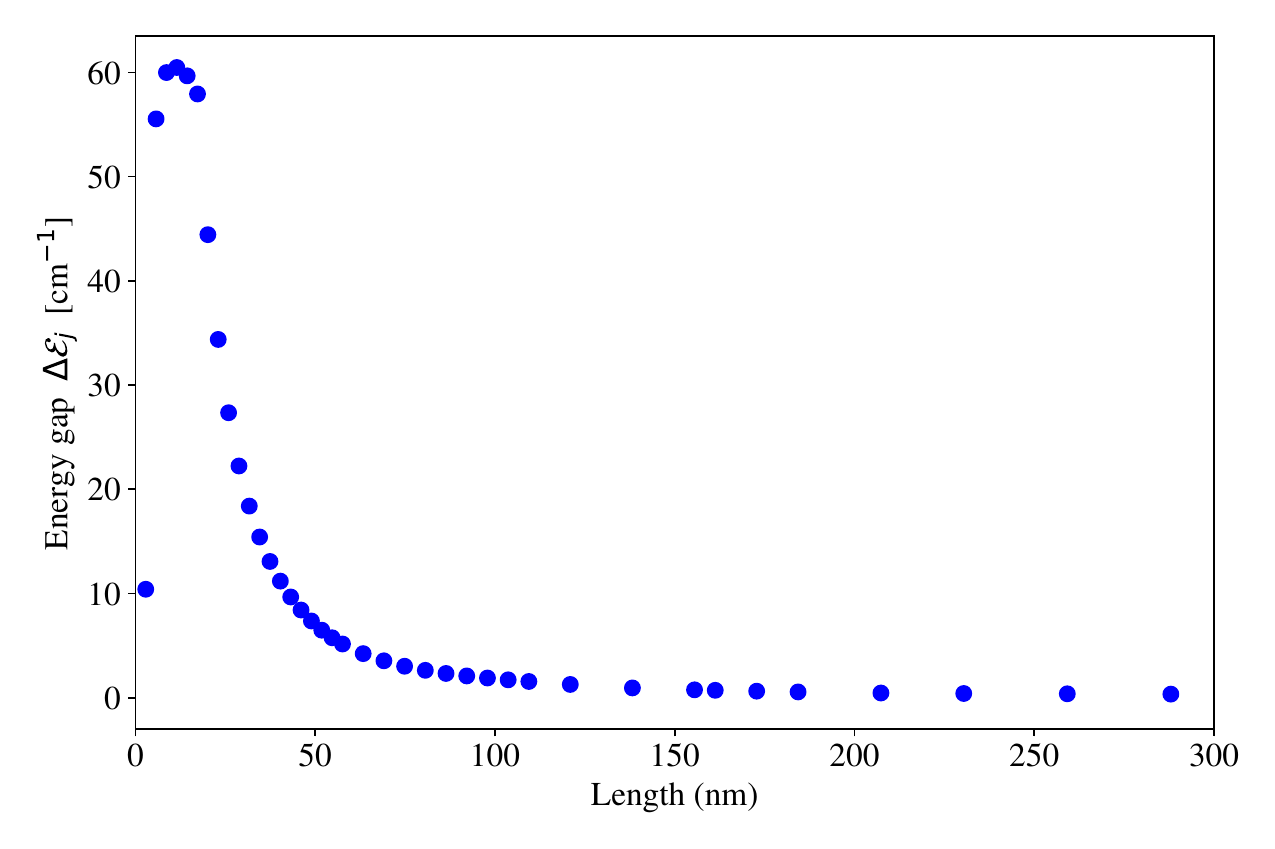}
    \caption{Plots of the energy gap $\Delta \mathcal{E}$ in the complex plane, given by Eq. \eqref{eq:energy_gap_in_complex_plane}, vs. length. The energy gap is calculated using the eigenspectrum obtained by diagonalizing the non-Hermitian Hamiltonian \eqref{eq:non-Hermitian-Hamiltonian-simplified-form} for the tryptophan networks in biological structures. \refpanel{A} Energy gap plot for single microtubules (pictured in Fig. \ref{fig:trp_positions}\refpanel{A}). \refpanel{B} Energy gap plot for 19-filament actin bundles. A top-down view of a 19-filament actin bundle is pictured inset in Fig. \ref{fig:actin_max_decay_rate_and_spectrum}\refpanel{A}, and a single filament is pictured in Fig. \ref{fig:trp_positions}\refpanel{B}. \refpanel{C} Energy gap plot for human amyloid fibrils (pictured in Fig. \ref{fig:trp_positions}\refpanel{C}). \refpanel{D} Energy gap plot for murine amyloid fibrils (pictured in Fig. \ref{fig:trp_positions}\refpanel{D}).}\label{fig:energy_gaps}
\end{figure*}

We create plots of the energy gap ($\Delta \mathcal{E}$) in the complex plane for all the biological structures we study in this work. Let each complex eigenvalue of the non-Hermitian Hamiltonian \eqref{eq:non-Hermitian-Hamiltonian-simplified-form} be denoted by $\mathcal{E}_j$, where the real and imaginary parts of $\mathcal{E}_j$ are $E_j$ and $-\frac{i}{2}\Gamma_j$, respectively. The energy gap in the complex plane is defined as
\begin{equation}
    \label{eq:energy_gap_in_complex_plane}
    \Delta \mathcal{E} \equiv | \mathcal{E}_1 - \mathcal{E}_0 |,
\end{equation}
where $|\cdot|$ represents the complex norm, $\mathcal{E}_0$ is the state with the lowest energy and $\mathcal{E}_1$ is the state with the second lowest energy (i.e., $E_0 < E_j \,\,\,\, \forall \,j\neq0$ and $E_1 < E_j\,\,\,\, \forall \,j\neq0,1$). This energy gap can be considered a measure of how ``quantum mechanical" the system is and has been associated with robustness to thermal noise and supertransfer processes \cite{Celardo2019, Babcock2023, Gulli2019}. If the classical limit is taken for a quantum system (the limit $\hbar/S_{\text{sys}} \rightarrow 0$, where $S_{\text{sys}}$ is the characteristic scale of the action of the system), then the spacing between the energy levels will approach 0. Therefore, as the system size increases, it would be conventionally expected that the system would become more classical, and thus exhibit a decrease in the energy gap. 

In Fig. \ref{fig:energy_gaps}, $\Delta \mathcal{E}$ is plotted against structure length for single microtubules, 19-filament actin bundles, human amyloid fibrils, and mouse amyloid fibrils. In microtubules, we can observe an \textit{increase} in the energy gap from around $100$ nm to $280$ nm. Since $280$ nm is the excitation wavelength of Trp, this means that the system increases its energy gap up to this characteristic length scale induced by matter interacting with the electromagnetic field. This observation has been made in \cite{Celardo2019}. Interestingly, we do not see this behavior for any of the other structures in Fig. \ref{fig:energy_gaps}. For both amyloids (panels \refpanel{C} and \refpanel{D} in Fig. \ref{fig:energy_gaps}), there is a huge peak of the energy gap around $10$-$25$ nm, and then a smooth dropoff that tends to 0. The maximum energy gap of human (mouse) amyloid is $40.32 \text{ cm}^{-1}$ ($60.47 \text{ cm}^{-1}$), which is about 47 (71) times larger than the maximum energy gap of microtubules around $280$ nm, which is $0.85 \text{ cm}^{-1}$. Even though the energy gap of amyloid is decreasing at the excitation wavelength of Trp (280 nm), its value for human (mouse) amyloid fibrils at this length is $0.278\text{ cm}^{-1}$ ($0.344\text{ cm}^{-1}$), which is on the same order as the energy gap of microtubules. This is consistent with the robustness of amyloid fibrils compared with microtubules.

The energy gap behavior of 19-filament actin bundles (panel \refpanel{B} from Fig. \ref{fig:energy_gaps}) is vastly different from all the other structures. It has a sporadic behavior that does not trend upward with increasing system size. The maximum energy gap for 19-filament actin bundles is at $\sim0.045\, \mu$m, and it is $0.046 \text{ cm}^{-1}$, which is a whole order of magnitude below the maximum energy gap of microtubules, and three orders of magnitude less than the maximum energy gap of amyloid. The sporadic behavior and low values for the energy gap for actin may reflect the more strictly mechanical nature of actin bundles compared with the other structures studied in this work. However, a more comprehensive analysis of these energy gaps averaged over multiple realizations of different static disorder strengths is warranted, to understand better how the transition to classicality in these structures is affected by a thermal environment.

\section{Discussion}

\begin{table}[tbhp!]
    \begin{tabularx}{\linewidth}{|X|ccc|cc|}
    \hline
    $\displaystyle\mathrm{Protein\, Structure,}\atop\displaystyle\mathrm{Length\, in\, nm}$ & $\displaystyle\frac{{\max}(\Gamma_{j})}{N\gamma}$ & $\tau_\text{super}$ (ps) & $\mathcal{P}_\text{super}$ ($\mu$W) & $\displaystyle{\frac{\min(\Gamma_{j})}{\gamma}}$ & $\tau_\text{sub}$ (s) \vspace{1mm} \\
    \hline
    91-MT Axon, 320 (fit) & 0.012  & 0.428 & 1.65 & --- & --- \\
    61-MT Axon, 320 (fit) & 0.016 & 0.479 & 1.47 & --- & --- \\
    Centriole, 400 & 0.028 & 0.495 & 1.42 & $4.6\times 10^{-8}$ & 0.042 \\
    61-MT Axon, 224 & 0.020 & 0.547 & 1.30 & $3.6\times 10^{-10}$ & 5.4 \\
    37-MT Axon, 320 & 0.026 & 0.602 & 1.19 & $2.3\times 10^{-10}$ & 8.5\\
    91-MT Axon, 152 & 0.017 & 0.636 & 1.13 & $2.6 \times 10^{-10}$ & 7.5 \\
    Axoneme (1JFF), 320 & 0.031 & 0.754 & 0.93 & $ 2.8 \times 10^{-10}$ & 6.9\\
    19-MT Axon, 320 & 0.032 & 0.769 & 0.92 & $9.9\times 10^{-10}$ & 2.0 \\
    7-MT Axon, 640 & 0.039 & 0.856 & 0.81 & $1.4\times 10^{-10}$ & 13.9 \\
    7-MT Axon, 320 & 0.071 & 0.941 & 0.75 & $2.8\times 10^{-9}$ & 0.69 \\
    Axoneme (6U42), 320 & 0.010 & 2.64 & 0.26 & $1.0\times 10^{-8}$ & 0.19 \\
    1 Microtubule, 320 & 0.120 & 3.89 & 0.18 & $1.36\times 10^{-6}$ & 0.001\\
    Murine amyloid, 1094 & 0.033 & 4.29 & 0.15 & $8.2\times 10^{-8}$ & 0.024\\
    Murine amyloid, 346 & 0.093 & 4.85 & 0.14 & $3.27\times 10^{-7}$ & 0.006\\
    Human amyloid, 1094 & 0.041 & 5.24 & 0.13 & $6.41\times 10^{-8}$ & 0.030\\
    Human amyloid, 346 & 0.113 & 5.98 & 0.11 & $1.47\times 10^{-7}$ & 0.013\\
    19-F actin bundle, 2250 & 0.003 & 12.1 & 0.059 & $5.00\times 10^{-6}$ & $3.89 \times 10^{-4}$ \\
    19-F actin bundle, 450 & 0.012 & 13.1 & 0.054 & $5.01\times 10^{-6}$ & $3.88 \times 10^{-4}$ \\
    7-F actin bundle, 2700 & 0.002 & 46.0 & 0.015 & $5.18\times10^{-6}$ & $3.75\times10^{-4}$ \\
    7-F actin bundle, 450 & 0.007 & 61.0 & 0.012 & $5.18\times10^{-6}$ & $3.75\times10^{-4}$ \\
    1-F actin, 2700 & 0.003 & $197$ & 0.004 & 0.006 & $3.16\times10^{-7}$ \\
    1-F actin, 450 & 0.014 & $222$ & 0.003 & 0.007 &  $2.81\times10^{-7}$ \\
    \hline
    \end{tabularx}
    \caption{Values for the extremal superradiant and subradiant decay rate values calculated by diagonalizing the non-Hermitian Hamiltonian in Eq. \eqref{eq:non-Hermitian-Hamiltonian-simplified-form} for the tryptophan (Trp) network in each protein structure. ``MT" stands for microtubule; the ``-F" stands for "filament." Images of a microtubule, 1-F actin, human amyloid, and murine amyloid, are shown in Fig. \ref{fig:trp_positions}\refpanel{A}-\refpanel{D}, respectively. For the first eleven rows with structures not analyzed in this paper, the values were taken from \cite{Babcock2023}. For the last two columns of the first two rows, the ``---" indicate that an analytical fit was taken of the equivalent of Figs. \ref{fig:actin_max_decay_rate_and_spectrum} and \ref{fig:amyloid_max_decay_rate_and_spectrum} for those structures. So, subradiant data was not available for these structures: only superradiance data was available. See \cite{Babcock2023} for more details. The column $\text{max}(\Gamma_j)/N\gamma$ represents the value of the maximum decay rate from the eigenspectrum (the enhancement rate of the maximally superradiant state). It is normalized by the single-Trp decay rate $\gamma\approx2.73 \times 10^{-3} \text{ cm}^{-1}$ and the number of emitters $N$, which varies for each structure. The column $\tau_\text{super}=(2\pi c \text{ max}(\Gamma_j))^{-1}$ is the lifetime of the maximally superradiant state in picoseconds. The column $\mathcal{P}_\text{super}$ represents the power output $E_\text{super}/\tau_\text{super}$ of the superradiant state. In the next portion of the table, information about the long-lasting subradiant states is listed. The column $\text{min}(\Gamma_j)/\gamma$ is the maximally subradiant state (the one with the smallest $\Gamma_j$). The column $\tau_\text{sub}=(2\pi c\text{ min}(\Gamma_j))^{-1}$ is the lifetime of the maximally subradiant state in seconds.}\label{table:superradiant-subradiant-values}
\end{table}

We discuss here the implications of our findings of high quantum yield and robustness for microtubules, actin bundles, and amyloid fibrils. In Alzheimer's disease and related dementias, the prevailing theory is that amyloid plaque formation is a direct cause of the onset of the disease \cite{Ow2014BriefOverview}. Based on our results we propose an alternative hypothesis to these previous observations on amyloid. By definition, if the QY of a structure was 1, then every photon that it absorbed would be re-emitted into the environment. We have predicted here that the QY of even small human amyloid fibrils $<1\mu$m is over 0.5. Given the significant absorption-emission Stokes shift of Trp, every photon that is re-emitted will be redshifted to a much lower energy than the UV photon that was absorbed, indicating that amyloid may serve a photoprotective role that downconverts dangerous UV photons in the brain to lower-energy photons which can be safely managed. Lower QY structures like microtubules and actin may assist with this process as well. 

Table \ref{table:superradiant-subradiant-values} shows the decay rate values of the most superradiant and most subradiant states, along with other key observables associated with these diverse neuroprotein architectures. One trend we can see in Table \ref{table:superradiant-subradiant-values}---and according to which it is organized---is that the maximum superradiance of the axon bundles is the highest ($\tau_\text{super} \propto 1/\text{max}(\Gamma_j)$), and as we go down the table, this value decreases for the other protein structures, with the smallest maximum superradiance (longest superradiant lifetime) belonging to the single actin filament. Large bundles of long, stably configured, straight microtubules in axons can exhibit enhanced, robust superradiance several thousands of times the single-Trp decay rate \cite{Babcock2023}, suggesting possible routes for quantum information processing in the brain that would be at least nine orders of magnitude faster than Hodgkin-Huxley chemical signaling at the millisecond scale in spiking neurons.

After the microtubule-based structures, amyloid fibrils in both humans and mice have the next largest superradiant power outputs ($E_\text{super}/\tau_\text{super}$), both due to their high-energy clustering of superradiant states and their picosecond-scale superradiant lifetimes. Interestingly, among the protein structures we consider in Table \ref{table:superradiant-subradiant-values}, the maximum superradiant density ($\text{max}(\Gamma_j)/N\gamma$) is the highest for a single microtubule of 320 nm (0.120) and a human amyloid fibril of 346 nm (0.113). Protein aggregates supporting such superradiant states with high power outputs could dissipate high-energy UV photons in an intensely oxidative cellular environment more quickly, and mitigate any potential damage. This is especially important in neuropathological conditions such as Alzheimer's, but also in a host of other complex diseases characterized by high allostatic load and oxidative stress, where high-energy photons can be produced due to metabolic photon emissions. Our prediction of high power outputs for amyloid fibrils supports the hypothesis that amyloid fibrils and plaques could actually serve as photoprotective entities in the brain. 

We also see in Table \ref{table:superradiant-subradiant-values} the general trend of highly subradiant long-lived states correlated with highly superradiant short-lived states. ($\tau_\text{sub}$ generally decreases down the table, while $\tau_\text{super}$ strictly increases down the table.) Subradiant states have been predicted with high excitonic occupation probabilities on the inner lumen surface of microtubules, while superradiant states have more delocalized occupation probabilities but with a preference for the external microtubule surface (see Fig. 4 of \cite{Celardo2019}). The lifetime of such superradiant states is faster than thermal noise from the aqueous environment surrounding the microtubule, whereas the lumen surface of the microtubule is exposed to a more ordered, gel-like matrix, and is thus subjected to far less thermal fluctuation. Such a locally protected environment could enable the potential exploitation of subradiant states (which are not particularly robust to noise) by living systems, possibly serving as a collective quantum mechanism for synchronizing behavior and information processing over long periods. ($\tau_\text{sub}$ values in Table \ref{table:superradiant-subradiant-values} are on the order of microseconds to tens of seconds.) In quantum information applications, the long lifetime of subradiant states in qubit architectures has been used to create very long-lived quantum memories \cite{AsenjoGarcia2017Subradiance}, implement mechanisms for lossless transport of photons \cite{GutirrezJuregui2022Subradiance}, and generate phase-imprinting for potential quantum storage of multiphotonic qubits in two-level systems \cite{Jen2017PhaseImprinting}.

Also, the quantum yield of amyloid may be even larger than we have predicted here. In the $\beta$-sheets displayed in Fig. \ref{fig:amyloid_beta_sheet}\refpanel{B}, we can see that there are so-called short hydrogen bonds (SHBs) 
that connect the $\beta$-strands side-by-side (labeled by dashed black lines). Even in the absence of aromatic chromophores like Trp, networks of these SHBs have been observed to absorb strongly in the UV band through proton transfer events, and emit in the visible band \cite{Grisanti2017ShortHydrogenBonds, Stephens2021ShortHydrogenBonds}. This means that, like Trp, networks of SHBs may aid in downconverting UV photons to lower energies. Networks of SHBs in amyloid fibrils may therefore exhibit their own bright superradiant states and enhance the QY of amyloid even beyond what we predict in this work. Due to the incredibly high QY of amyloid fibrils and its potential for photoprotection, rather than being a cause of pathological conditions, amyloid fibrils could be a \textit{response} to them, and to the highly oxidative environments that characterize them. Therapies that target amyloid in the brain for elimination could therefore exacerbate such diseases rather than ameliorating them.

There are at least 37 known proteins that form pathological amyloids \cite{Chiti2017AmyloidSummaryOfProgress}. We have found that the amyloids 6MST and 6DSO, which are associated with systemic AA amyloiosis \cite{liberta2019cryo}, exhibit extremely robust QY. Since amyloids are a geometric class of protein architectures characterized by helical superstructures made of $\beta$-sheets, it is likely that amyloids formed from other proteins (such as lysozyme, insulin, and IAPP) will have similar transition dipole networks as discussed in Section \ref{sect:toy-models}, and thus may also exhibit the high QYs that would play a strongly photoprotective role in the pathological cellular environment.

Furthermore, the formation of structures called cofilin-actin rods from pools of actin and the cofilin protein have also been studied \cite{Jang2005CofilinActin,Bamburg2016CofilinActin,Nichols2023JournalofMultiScaleNeuro}, with recent suggestions of potentially quantum behavior being disrupted in Alzheimer's pathogenesis \cite{Nichols2023JournalofMultiScaleNeuro}. Cofilin-actin rods do have a helical structure: every subunit that is added to the rod comes with an approximate $5^{\circ}$ twist \cite{Bamburg2016CofilinActin}. Therefore, without having conducted any of the detailed analyses presented here for microtubules, actin bundles, and amyloid fibrils, we would hypothesize that cofilin-actin rods exhibit significant superradiance that may translate into robust, observable quantum yield effects based on the symmetry and interactions of their helical Trp networks. This is yet another instance in which a cylindrically or helically symmetric structure is created in the context of neurodegeneration, further stressing the importance of chromophore network geometry in protein lattices as the source of these robust superradiant effects. 

As the microtubule results in the Supplementary Material and in Section \ref{sec:ResultsMTs} attest, the morphology and mechanical deformations of protein structures are crucial to understanding the modulation of superradiant effects. For example, amyloid fibrils are known to aggregate into macroscopic structures called amyloid plaques, together with glial and neuritic debris \cite{Bazan2012Plaques}. They are found in the grey matter of the brain in the areas associated with memory and cognition. Amyloid plaques can form spherically symmetric aggregates of amyloid fibrils with very dense cores \cite{Merz1983PlaqueStructure}. Plaques have also been observed to form symmetric superarchitectures such as bundles, as well as mesh-like and star-like geometries \cite{Han2017PlaqueStructure}. Given these observations, amyloid plaques may exhibit even higher, more robust superradiance and quantum yield values than those of single fibrils, which would strengthen the argument of amyloid's photoprotective role and of the mitigating effect of plaque formation in neurodegenerative pathology.

Our predictions of robust, observable increases in the QY for Trp networks in large protein polymeric architectures has implications for many other diseases outside the neurodegenerative context. For example, sickle-cell anemia results from a $\text{Glu}\rightarrow\text{Val}$ mutation of the amino acid at the sixth position on the beta chain of normal hemoglobin (HbA). The resulting deoxygenated hemoglobin (HbS) is known to aggregate in erythrocytes (red blood cells without nuclei) \cite{Mu1998Hemoglobin, Henry2020Hemoglobin} and form helical structures. These helical hemoglobin strands, commonly known as Wishner-Love helices, would then manifest helical Trp networks. We have found that helical Trp networks exhibit superradiance in three distinct cases (in microtubules, actin filaments, and amyloid fibrils), so hemoglobin's helical Trp network may also exhibit significant superradiance and/or quantum yield. If these hemoglobin aggregates are indeed found to exhibit superradiance and robust increases in QY with increasing size, then quantum-enhanced photoprotection may also play an important role in the onset, progression, and treatment of hemoglobinopathies like sickle-cell, which are also associated with intensely oxidative and damaging cellular environments.

Our results pose opportunities for a paradigm shift in the theory of neuronal information processing and signaling. The role of microtubules in information processing in the brain has been studied extensively \cite{Ahmad2006, craddock2012cytoskeletal, craddock2014feasibility, CraddockKurian2019Chapter}. Also, classical energy scalings cannot account for the sub-neuronal information processing capacity of the brain \cite{Penrose1996SoTM}, given its extremely low input power of around 20 W. There must be another physical mechanism that enables the human brain to achieve the computational efficiency that it does, at orders of magnitude lower power consumptions than high-performance hardware systems. A tantalizing possibility is that extended protein architectures, such as those described here and elsewhere \cite{Celardo2019, Babcock2023}, including axons in the brain, may form a highly interconnected, ultrafast quantum-optical network that gives rise to incredibly efficient transfer and processing of information. This mechanism would be much faster than chemical Hodgkin-Huxley-type transport based on neuronal sodium-potassium gradients firing at the millisecond timescale, which is currently used as a standard paradigm in neuroscience.

\section{Conclusion and Future Plans}
We analyze the interaction of the electromagnetic field with networks of tryptophan chromophores. Geometrical information on these networks and the orientations of the tryptophan transition dipoles are extracted from realistic simulations of three types of neuroprotein architectural elements: microtubules, actin filaments/bundles, and amyloid fibrils. The tryptophan chromophores are modeled as two-level systems and exhibit superradiant behavior as a collective when coherently superposed in the single-excitation limit. We see this by diagonalizing the non-Hermitian Hamiltonian used to describe the collective light-matter interaction of such a weakly photoexcited system. All three structures were found to exhibit bright superradiant states due to symmetry and long-range couplings, which support robustness of the quantum yield as a figure of merit with increasing static disorder even up to five times room-temperature energy. In the case of microtubules and amyloid fibrils, the brightest superradiant states are clustered near the lowest-energy portion of their spectra, and these photophysical properties result in a large quantum yield that counterintuitively increases with system size and has been experimentally confirmed for microtubules \cite{Babcock2023}.

Our results display the observable and important consequences that quantum coherent effects have on neuroprotein architectures. These analyses could strengthen our understanding of the etiology of neurodegenerative and other complex diseases, which are frequently characterized by anomalous protein polymers. Furthermore, our investigations of superradiance and subradiance in these neuroproteins are revealing an ultrafast mechanism that our brains may use to process information, which is paired with an extremely long-lived mechanism for coordinating biological function. This work contributes significantly to our understanding of how quantum biology can speed up, enhance, and optimize behavior in the ``wetware'' environments of living systems. Acknowledging the wide body of research that has been conducted on ultraweak and metabolic photon emissions in the cell, we have incorporated the interaction of neuroprotein tryptophan lattices with the electromagnetic field via the equations of quantum optics, giving us a totally different lens with which to view biology. Such a paradigm shift can greatly enhance our understanding of nature, to visualize biological architectures as chromophore lattices synchronized by long-range interactions, and imbued with unique and specific photophysical properties that are enhanced by collective light-matter interactions governed by the equations of quantum optics. Such a shift reflects a return to understanding, in the (paraphrased) words of Richard Feynman and with the ancients, how external light from the fiery sun causes trees and plants to grow from the carboniferous air; and in parallel symbiosis how oxygen-metabolizing organisms may have evolved their protein architectures to exploit ``internal'' photonic emissions for information processing and to mitigate potentially damaging wavelengths in the cell.

Future work will include performing experiments in order to verify the quantum yield predictions that we have made here. Analogous to prior work \cite{Babcock2023}, where the increased quantum yield in microtubules from tubulin dimers in solution was unambiguously associated with the increased radiative rate due to superradiance, we now have a clear path and approach to experimental validation of collective quantum optical behavior in a wide class of protein polymeric aggregates in solution. We hope that this work will stimulate further experimental efforts in this regard.

\section{\label{sec:methods}Methods}
\subsection{Physical model}
In this work we model tryptophan (Trp), a strongly flourescent amino acid in the ultraviolet (UV) band, as a two-level system \cite{Callis1997Trp1La1Lb} with transition energy $e_0\approx 280 \text{ nm} = 3.57 \times 10^4 \text{ cm}^{-1}$ and decay rate $\gamma\approx2.73 \times 10^{-3}\text{ cm}^{-1}$ \cite{craddock2014feasibility, kurian2017oxidative}. Trp has a large transition dipole moment of $\sim6.0$ debye. We use a non-Hermitian Hamiltonian to describe the interaction of a $N$-dimensional Trp network with the electromagnetic field \cite{Grad1988Hamiltonian, spano1989superradiance, Bienaime2013Hamiltonian, Giusteri2015Hamiltonian, Akkermans2008Hamiltonian}

\begin{equation}
    H_{\text{eff}} = H_0 + \Delta - \frac{i}{2}G
\end{equation}
where
\begin{align}
    &H_0 = \sum_{n=0}^{N-1} \hbar \omega_0 |n\rangle\langle n|\label{eq:non-hermitian-hamiltonian-H0}\\
    &\Delta=\sum_{n\neq m}^N \Delta_{nm}|n\rangle\langle m|\label{eq:non-hermitian-hamiltonian-Delta}\\
    &G = \sum_{n=0}^{N-1} \gamma|n\rangle\langle n| + \sum_{n\neq m}^N G_{nm} |n\rangle\langle m|\label{eq:non-hermitian-hamiltonian-G}
\end{align}
\begin{align}
    \Delta_{nm} = &\frac{3\gamma}{4}\Bigg[\bigg( -\frac{\cos(\alpha_{nm})}{\alpha_{nm}} + \frac{\sin(\alpha_{nm})}{{\alpha_{nm}}^2} + \frac{\cos(\alpha_{nm})}{{\alpha_{nm}}^3} \bigg) \hat{\mu}_n \cdot \hat{\mu}_m \nonumber\\
    &-\bigg(-\frac{\cos(\alpha_{nm})}{\alpha_{nm}} + 3\frac{\sin(\alpha_{nm})}{{\alpha_{nm}}^2} + 3\frac{\cos(\alpha_{nm})}{{\alpha_{nm}}^3} \bigg)(\hat{\mu}_n \cdot \hat{r}_{nm})(\hat{\mu}_m \cdot \hat{r}_{nm})\Bigg]\label{eq:non-hermitian-hamiltonian-Delta-full}
\end{align}
\begin{align}
    G_{nm} = &\frac{3\gamma}{2}\Bigg[\bigg(\frac{\sin(\alpha_{nm})}{\alpha_{nm}} + \frac{\cos(\alpha_{nm})}{{\alpha_{nm}}^2} - \frac{\sin(\alpha_{nm})}{{\alpha_{nm}}^3} \bigg) \hat{\mu}_n \cdot \hat{\mu}_m \nonumber\\
    &-\bigg(\frac{\sin(\alpha_{nm})}{\alpha_{nm}} + 3\frac{\cos(\alpha_{nm})}{{\alpha_{nm}}^2} - 3\frac{\sin(\alpha_{nm})}{{\alpha_{nm}}^3} \bigg)(\hat{\mu}_n \cdot \hat{r}_{nm})(\hat{\mu}_m \cdot \hat{r}_{nm})\Bigg]\label{eq:non-hermitian-hamiltonian-G-full}
\end{align}
where $\alpha_{nm} \equiv k_0 r_{nm}$. The constants $k_0$ and $\omega_0$ are defined in terms of $e_0$ by $k_0 = 2\pi e_0 \times 10^{-8}$ and $\omega_0 = 2\pi e_0 c / n_r  \times 10^{-8}$ where $n_r \approx \sqrt{\varepsilon_r}$ is the refractive index assuming that relative permeability is $1$. The vector $\hat{r}_{nm}$ is the unit vector pointing from the $n^\text{th}$ site to the $m^\text{th}$ site in physical space, and $r_{nm}$ is the distance between the $n^\text{th}$ and $m^\text{th}$ sites. The unit vector $\hat{\mu}_n$ is the transition dipole moment of the $n^\text{th}$ site.
\subsection{Quantum Yield}
The quantum yield (QY) is a dimensionless number from 0 to 1 that is defined to be the ratio of the number of photons emitted to the number of photons absorbed. Equivalently, it can be written as
\begin{equation}
    \label{eq:quantum_yield_basic_def}
    \QY \equiv \frac{\Gamma}{\Gamma+\Gamma_{nr}}
\end{equation}
where \(\Gamma\) is the collective radiative decay rate, and \(\Gamma_{nr}\) is the non-radiative decay rate. It ranges from 0 to 1. If the QY is close to 1, that means that most photons that get absorbed get re-emitted, and if the QY is close to 0, that means that most photons that get absorbed don't get re-emitted. Trp is known to have a significant absorption-emission Stokes shift \cite{Babcock2023}, so photons will be re-emitted at a lower energy than that of their absorption. Therefore, if a Trp network has a high QY, it can be inferred that the Trp network will act in a photoprotective role against high-energy UV photons: absorbing them and then ``downconverting" (red-shifting) them to a lower energy.

Since biological structures exist in a warm and wet environment, we take the thermal average of our quantities. Firstly, the partition function \(Z\) is given by
\begin{equation}
    \label{eq:partition_funct}
    Z = \sum_{j=0}^{N-1} \exp(-\beta E_j)
\end{equation}
where \(\beta\equiv(k_B T)^{-1}\).
The thermal average of the decay rate is then
\begin{equation}
    \label{eq:thermal_avg_of_decay_rate}
    \langle \Gamma \rangle_{th} = \frac{1}{Z} \sum_{j=0}^{N-1} \Gamma_j \exp(-\beta E_j).
\end{equation}
This allows us to calculate the thermal average of the QY
\begin{equation}
    \label{eq:thermal_avg_of_QY}
    \langle \QY \rangle_{th} = \frac{\langle \Gamma \rangle_{th}}{\langle \Gamma \rangle_{th} + \langle \Gamma_{nr} \rangle_{th}}.
\end{equation}
We take \(\langle \Gamma_{nr} \rangle_{th}=\gamma_{nr}\) in this case, assuming conservatively that there is no reduction in the non-radiative decay rate leading to an increase in the QY. This represents the assumption that Trp network formation in protein introduces no change in the non-radiative decay channels, as compared with Trp alone in solution. Recent experimental evidence \cite{Babcock2023} indicates that Trp network formation in tubulin actually \textit{increases} the non-radiative decay rate, suggesting that the protein environment competes with superradiant enhancements to modulate the observed QY.

\subsection{Biological structures}
We model three sets of biological structures in this paper: microtubules, actin filaments, and amyloid fibrils. Python scripts that implement all the following procedures and generate all PDB files for structures of a given length can be found in a GitHub repository link in the Data Availability Statement, which will be made available after publication.

\subsubsection{Microtubules}
We construct models of microtubules of varying length from the tubulin dimer stored in the Protein Data Bank (PDB) entry 1JFF \cite{lowe2001refined} as per the methods given in Appendix A of \cite{Celardo2019} and Section S3 in \cite{Babcock2023}. We briefly summarize the procedure here.

Many identical 1JFF tubulin dimers are laid next to one another to form a left-handed helical microtubule structure with a diameter of 22.4 nm. The initial orientation of one tubulin dimer is such that the $\alpha$ and $\beta$ chains lie both along the protofilament direction. Let this be the x-axis. Then, each tubulin dimer is acted on by the following initial operations: (1) rotated by -55.38$^\circ$ about its longitudinal axis, (2) rotated by 11.7$^\circ$ about the $\beta$-tubulin Trp346 CD2 atom, and (3) translated by 11.2 nm in the y-direction and 0.3 nm in the z-direction. After applying operations (1)-(3) to each tubulin dimer, a set of operations is applied $N$ times to the $N$th dimer to form a single spiral (one spiral consists of 13 tubulin dimers, so $N\leq 13$): (4) rotation of 27.69$^\circ$ about the x-axis and (5) translation of 0.9 nm in the x-direction. This generates a 13-dimer spiral. To create microtubules with multiple spirals, each spiral is translated multiples of 80 nm in the x-direction from the initial spiral. This procedure creates microtubules with a radius of $\sim 11.2$ nm (from the microtubule longitudinal axis to the tubulin dimer center-of-mass), approximately intermediate between the outer (cytoplasm-surface) radius of $\sim 13.5$ nm and the inner (lumen-surface) radius of $\sim 9.5$ nm.

After creating a microtubule, the positions and transition dipole moments of the 8 atoms in every Trp molecule are extracted. The position of a Trp molecule is given by the midpoint of the positions of the CD2 and CE2 carbon atoms in it. The transition dipole moment of a Trp molecule is taken as the well-known $^1L_a$ transition of Trp \cite{Callis1997Trp1La1Lb, Schenkl2005}, which is the vector pointing 46.2$^\circ$ above the axis joining the midpoint between the CD2 and CE2 carbons and carbon CD1, in the plane of the indole ring (i.e. towards nitrogen NE1).

\subsubsection{Actin filaments and actin filament bundles}
We construct models of actin filaments of varying length from the PDB entry 6BNO as per the description of the structure given in \cite{gurel2017cryo}. The procedure to generate actin filaments of a specified length is summarized below.

Many identical 6BNO bare actin filaments are laid next to one another to form a right-handed helical filament. Let the logitudinal direction of the filament be defined as the x-axis. Then, the following operations are applied $N$ times to the $N$th filament: (1) translation of 22.488 nm in the x-direction and (2) rotation by -253.2$^\circ$ about its own axis (the x-axis). This creates actin filaments consisting of any number of 6BNO bare filaments. Extraction of Trp positions and transition dipole moments are the same as for microtubules.

We also take many actin filaments constructed in this way, and pack them together in concentric hexagons, forming actin bundles. This creates bundles containing $3N^2-3N+1$ filaments, where $N$ is the number of concentric hexagons, including the center. We create and study 7-filament ($N=2$) and 19-filament ($N=3$) bundles.

\subsubsection{Amyloid fibrils}
We construct models of human (mouse) amyloid fibrils of varying length from the PDB entries 6MST (6DSO), respectively, as per the descriptions of the structures given in \cite{liberta2019cryo}. The procedure to generate amyloid fibrils of a specified length is summarized below. For both human and mouse amyloid, each subunit is characterized by a six-strand $\beta$-sheet structure, where each $\beta$ strand is joined to its neighbor via hydrogen bonds. The entire fibril is also known as a $\beta$ helix. 

Many identical 6MST (6DSO) amyloid fibrils are laid next to one another to form a right(left)-handed helical fibril. Before applying the necessary operations, we find that a preliminary translation is required for each 6MST (6DSO) fibril: -14.0474 (-14.1715) nm in the x-direction, -14.0376 (-14.1595) nm in the y-direction, and -14.0039 (-11.8823) nm in the z-direction. This moves the center of mass of the molecule to the origin, so that subsequent rotations are performed along the axis parallel to the would-be fibril direction that passes through the center of mass. Let this axis be defined as the x-axis. Then, the following operations are applied $N$ times to the $N$th fibril for 6MST: (1) translation of 2.88 nm in the x-direction and (2) rotation by 9.24$^\circ$ about its own axis (the x-axis). For 6DSO, the translation is the same\footnote{In \cite{liberta2019cryo}, the translations for 6DSO and 6MST are different. We chose to make this modification after our own analysis of the structure.}, but the rotation is -6.90$^\circ$ about its own axis. This creates actin filaments consisting of any number of amyloid fibrils. Extraction of Trp positions and dipole moments are the same as for microtubules.

\section*{Conflict of Interest Statement}

The authors declare that the research was conducted in the absence of any commercial or financial relationships that could be construed as a potential conflict of interest.

\section*{Author Contributions}

HP: Data curation, Formal analysis, Investigation, Software, Validation, Visualization, Writing – original draft, Writing – review \& editing; NSB: Data curation, Investigation, Software, Validation, Visualization; PK: Conceptualization, Funding acquisition, Methodology, Project administration, Resources, Supervision, Visualization, Writing – original draft, Writing – review \& editing.

\section*{Funding}
This work was supported in parts by the Alfred P. Sloan Foundation, Guy Foundation (UK), and Chaikin-Wile Foundation.

\section*{Acknowledgments}
This research used resources of the Argonne Leadership Computing Facility, a U.S. Department of Energy (DOE) Office of Science user facility at Argonne National Laboratory and is based on research supported by the U.S. DOE Office of Science-Advanced Scientific Computing Research Program, under Contract No. DE-AC02-06CH11357. Portions of the manuscript were drafted and discussed during PK's 2023 residency as a Fellow of the Kavli Institute for Theoretical Physics (KITP), which was supported in part by grant NSF PHY-1748958 to KITP.  

\section*{Supplemental Information}
In the Supplemental Information, we present plots of the eigenvectors obtained by diagonalizing the non-Hermitian Hamiltonian \eqref{eq:non-Hermitian-Hamiltonian-simplified-form} for the Trp network of mechanically deformed single microtubules. Each eigenvector is projected onto the site basis, and the color on each Trp represents the probability that it is excited. We would like to thank Marco Agostino Deriu for providing us the atomistic structures of these deformed microtubules for the various vibrational and torsional modes presented.

\bibliographystyle{unsrt}
\bibliography{test}

\clearpage

%%%%%%%%%%%%%%%%%%%%%%%%
\setcounter{figure}{0}
\setcounter{table}{0}
\renewcommand{\thefigure}{S\arabic{figure}}
\renewcommand{\thetable}{S\arabic{table}}

\clearpage
\section*{ }
\begin{minipage}{\textwidth}
    \LARGE
    \centering
    Supplemental Information
    \vspace{10mm}
\end{minipage}

\section*{Supplementary Tables and Figures}
\subsection*{Figures}
\begin{figure}[h!]
\begin{center}
$\,\,\,7$ \includegraphics[width=0.31\linewidth]{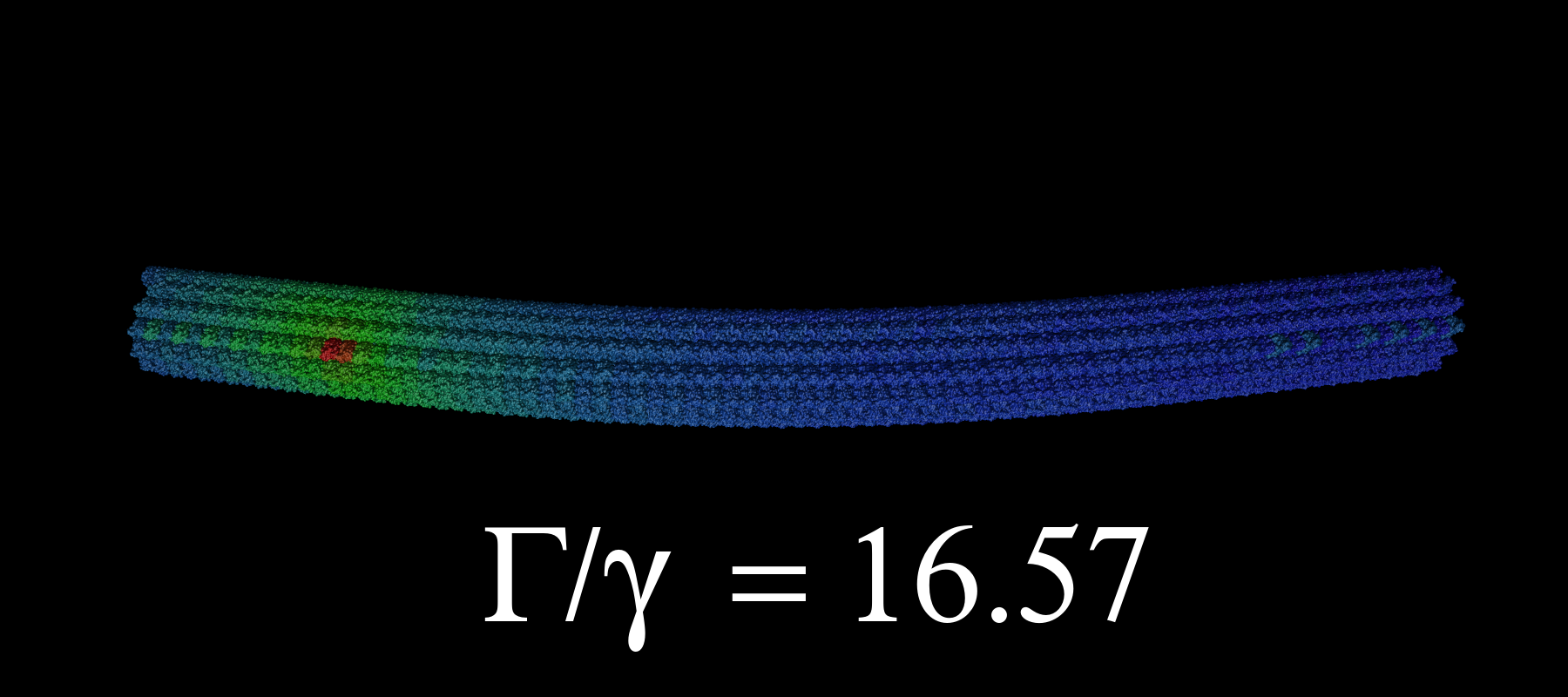}
\includegraphics[width=0.31\linewidth]{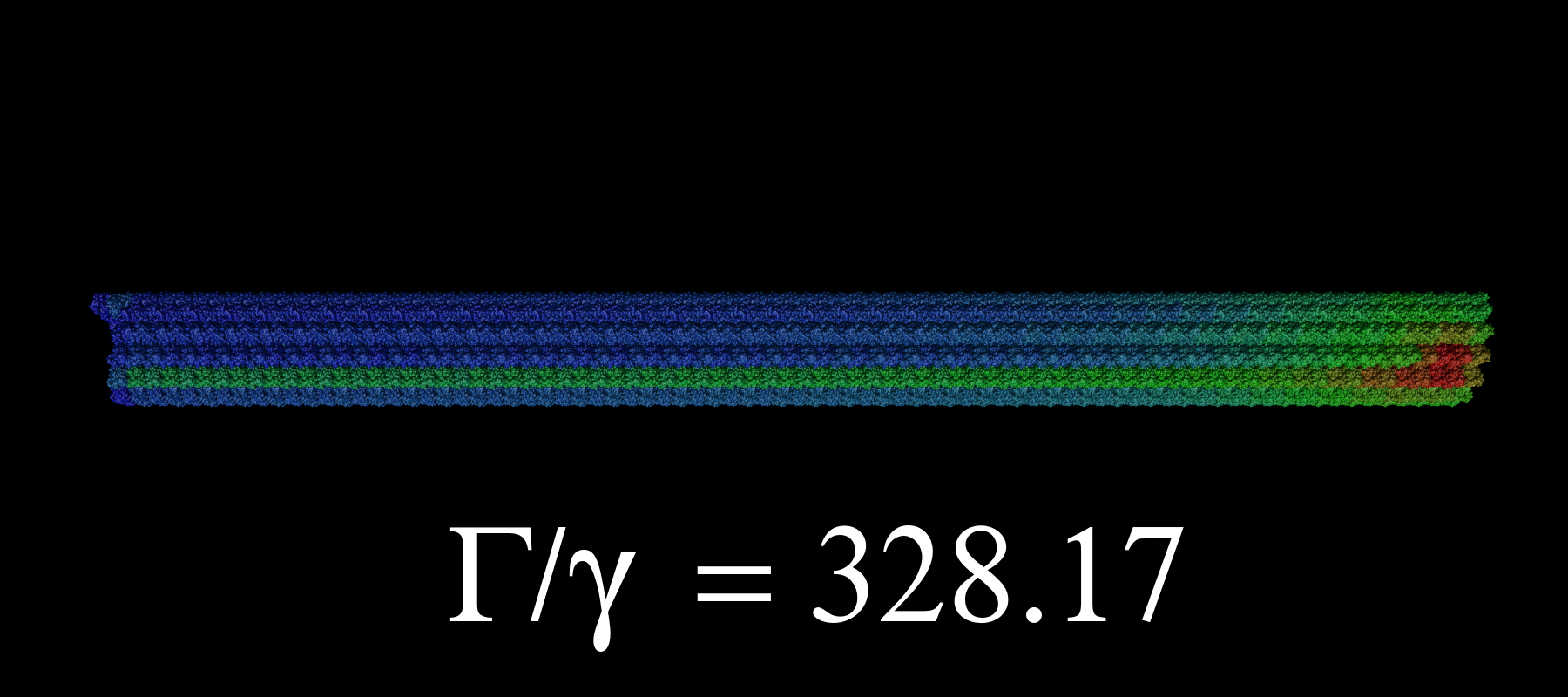}
\includegraphics[width=0.31\linewidth]{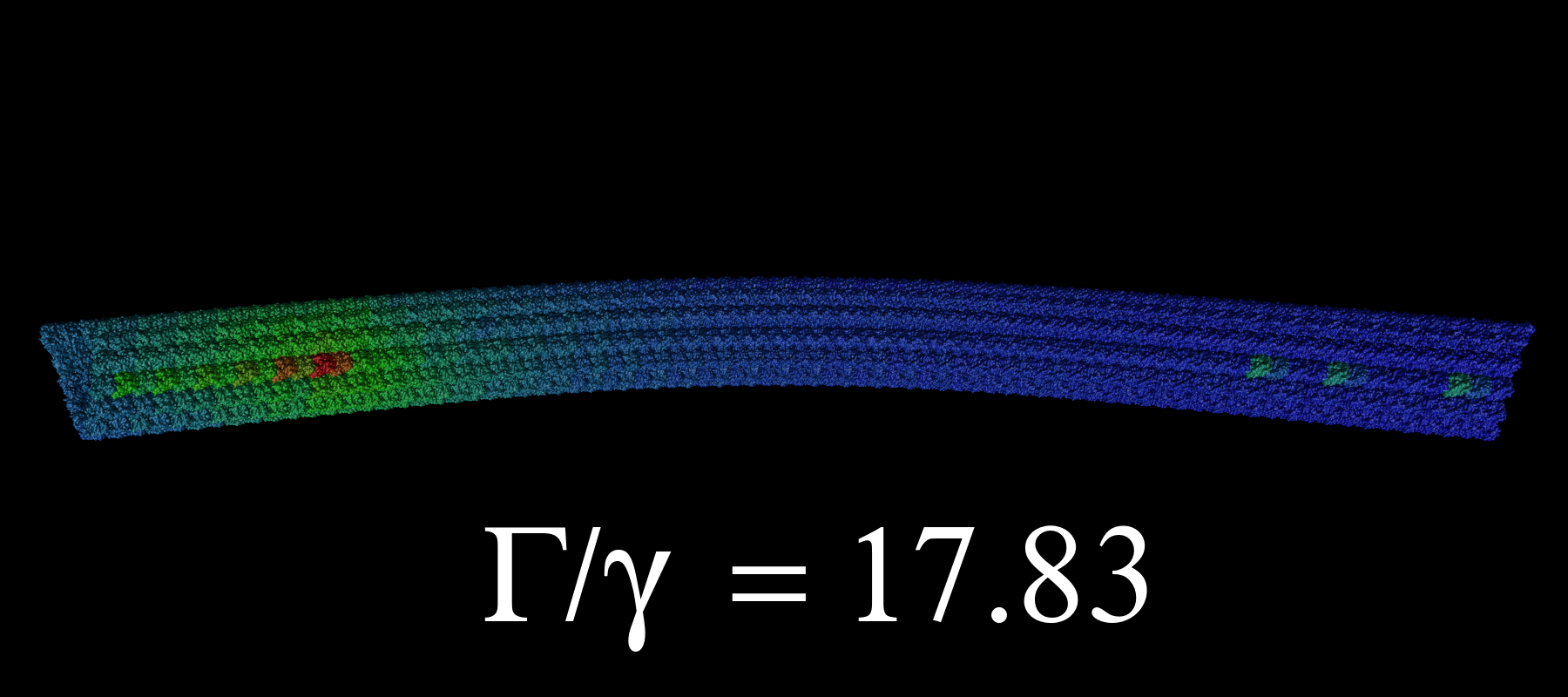} \\
$\,\,\,8$ \includegraphics[width=0.31\linewidth]{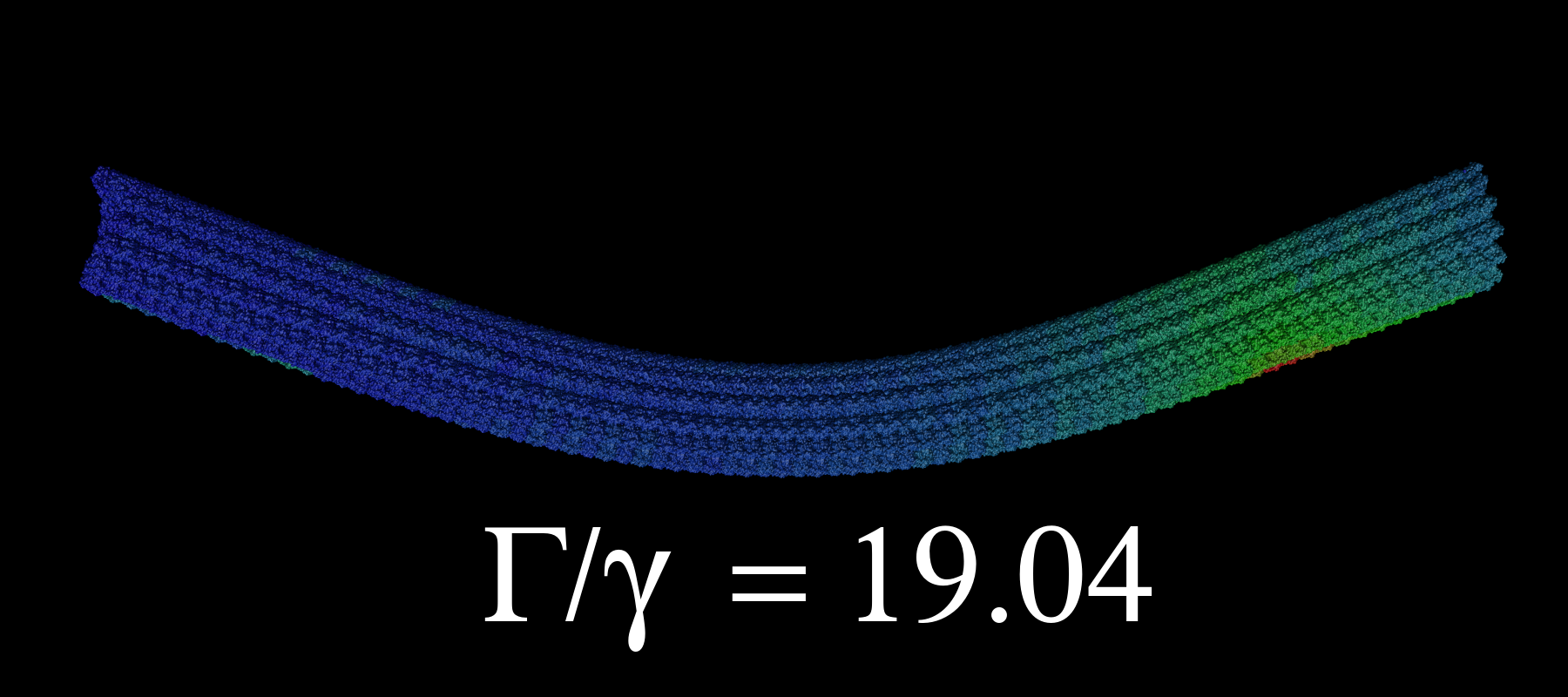}
\includegraphics[width=0.31\linewidth]{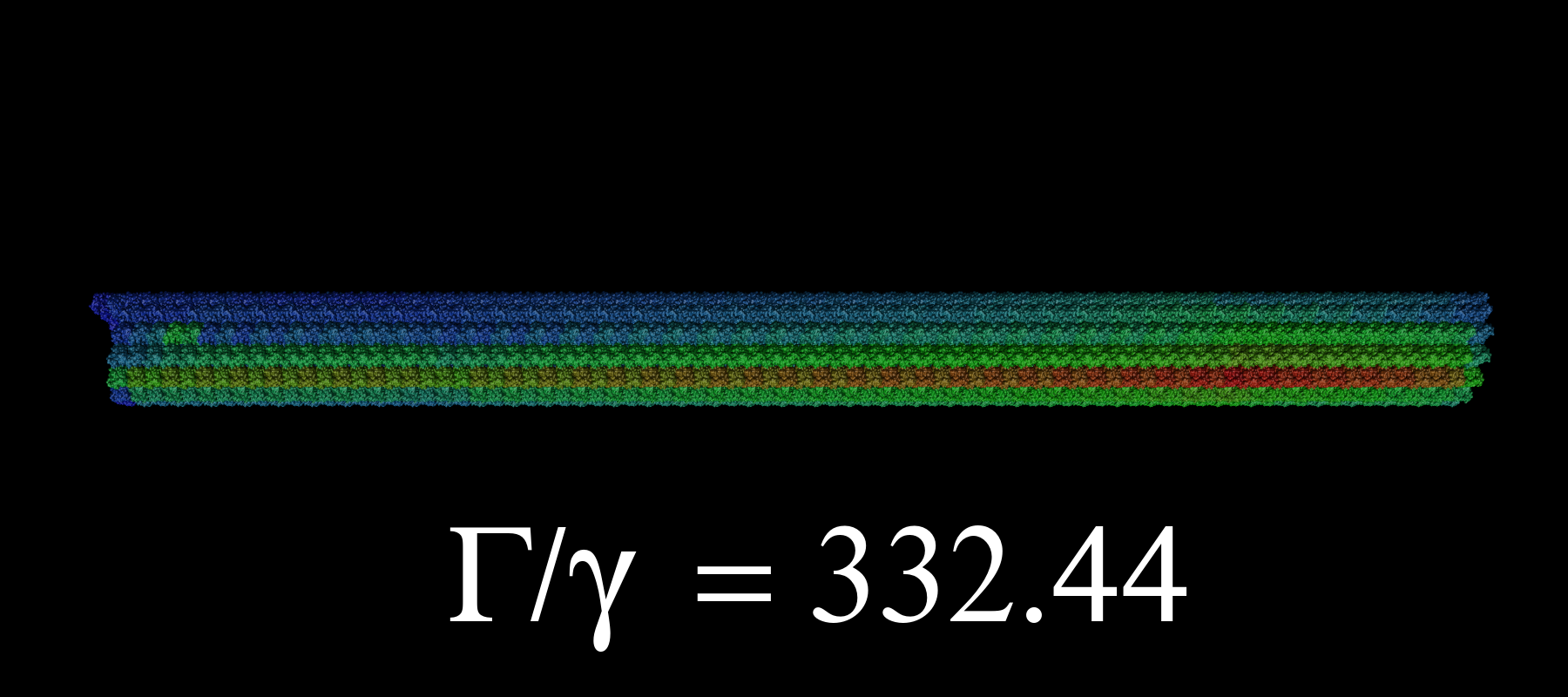}
\includegraphics[width=0.31\linewidth]{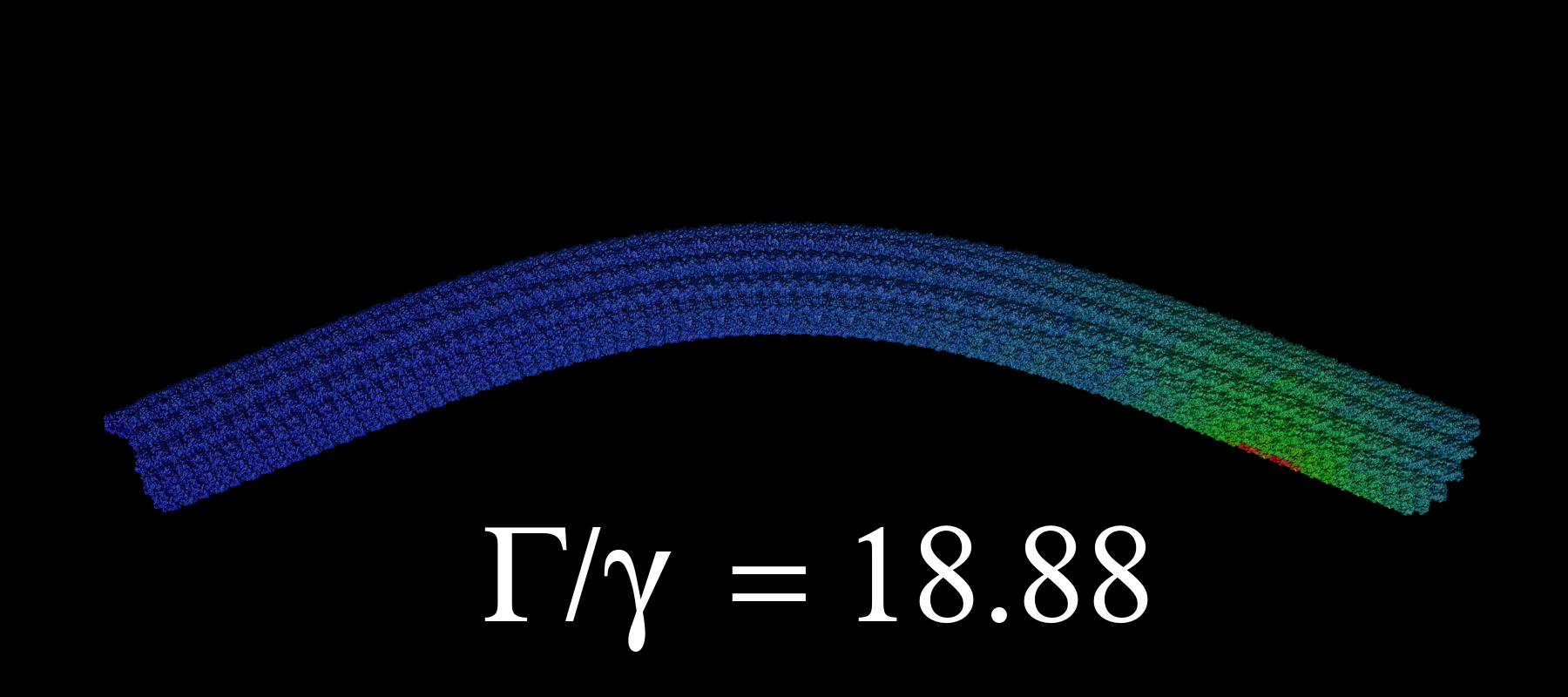} \\
$\,\,\,9$ \includegraphics[width=0.31\linewidth]{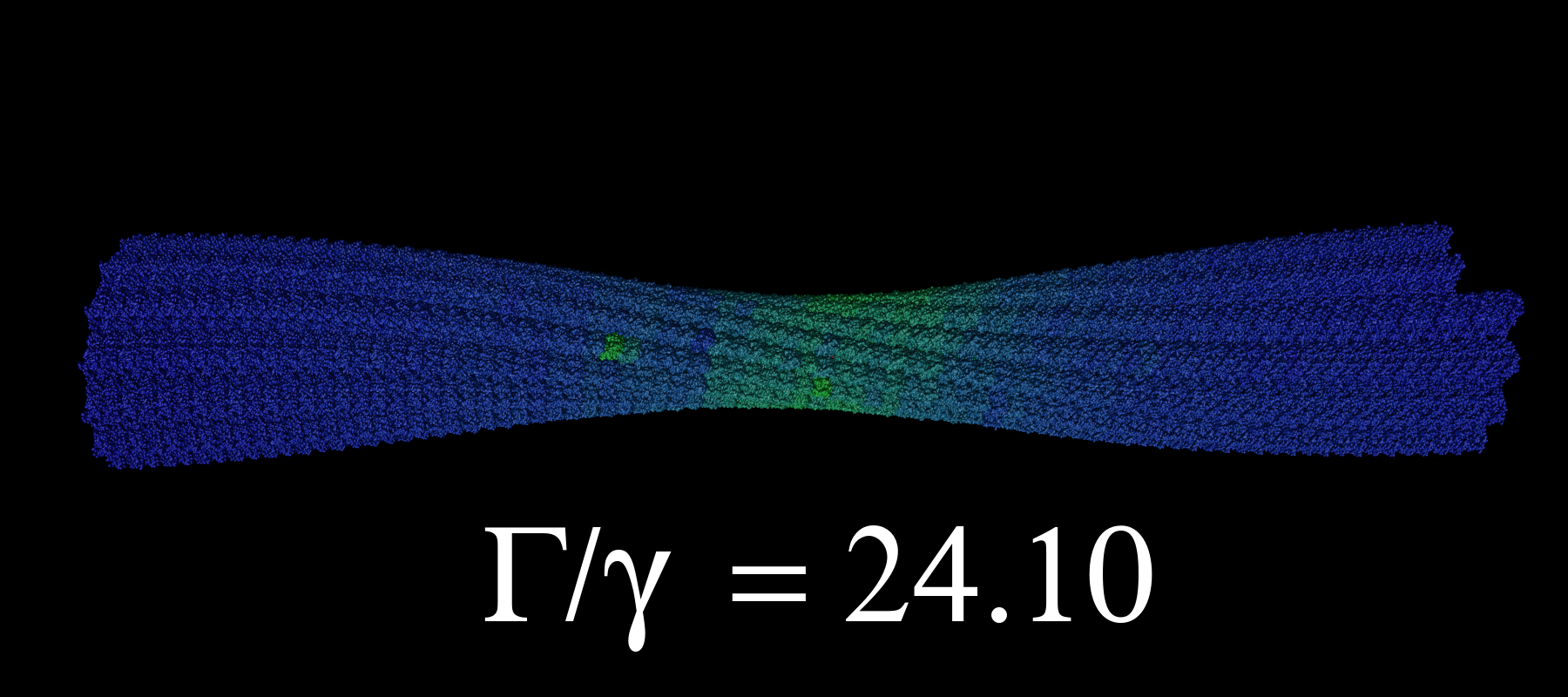}
\includegraphics[width=0.31\linewidth]{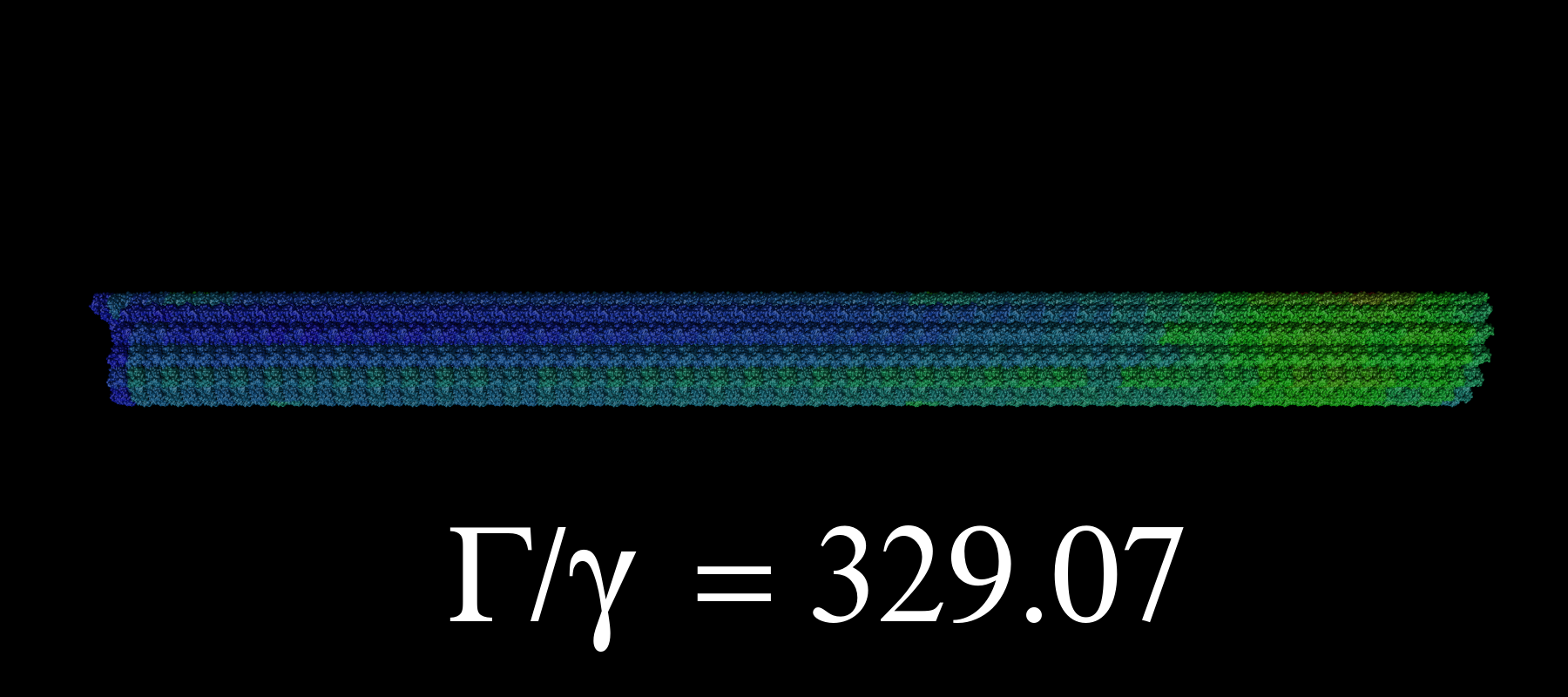}
\includegraphics[width=0.31\linewidth]{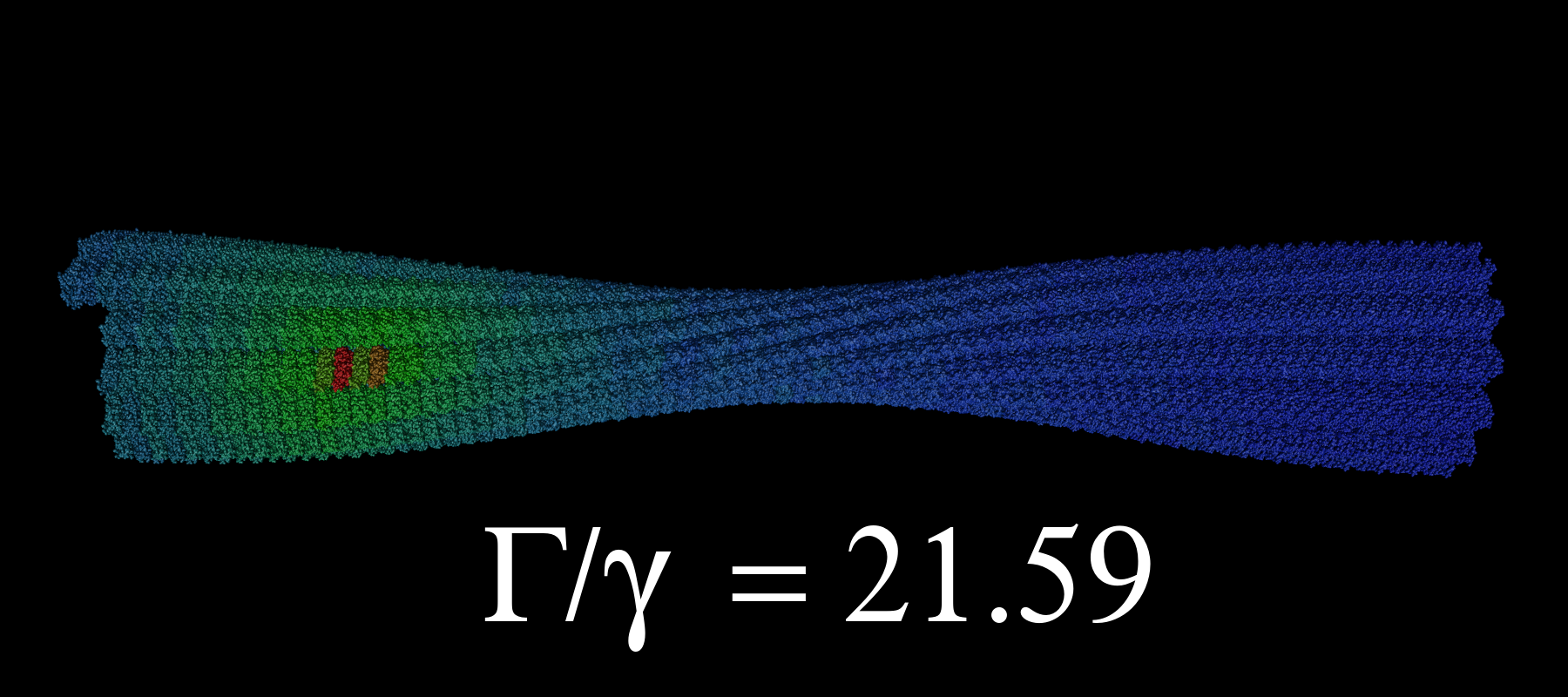} \\
$10$ \includegraphics[width=0.31\linewidth]{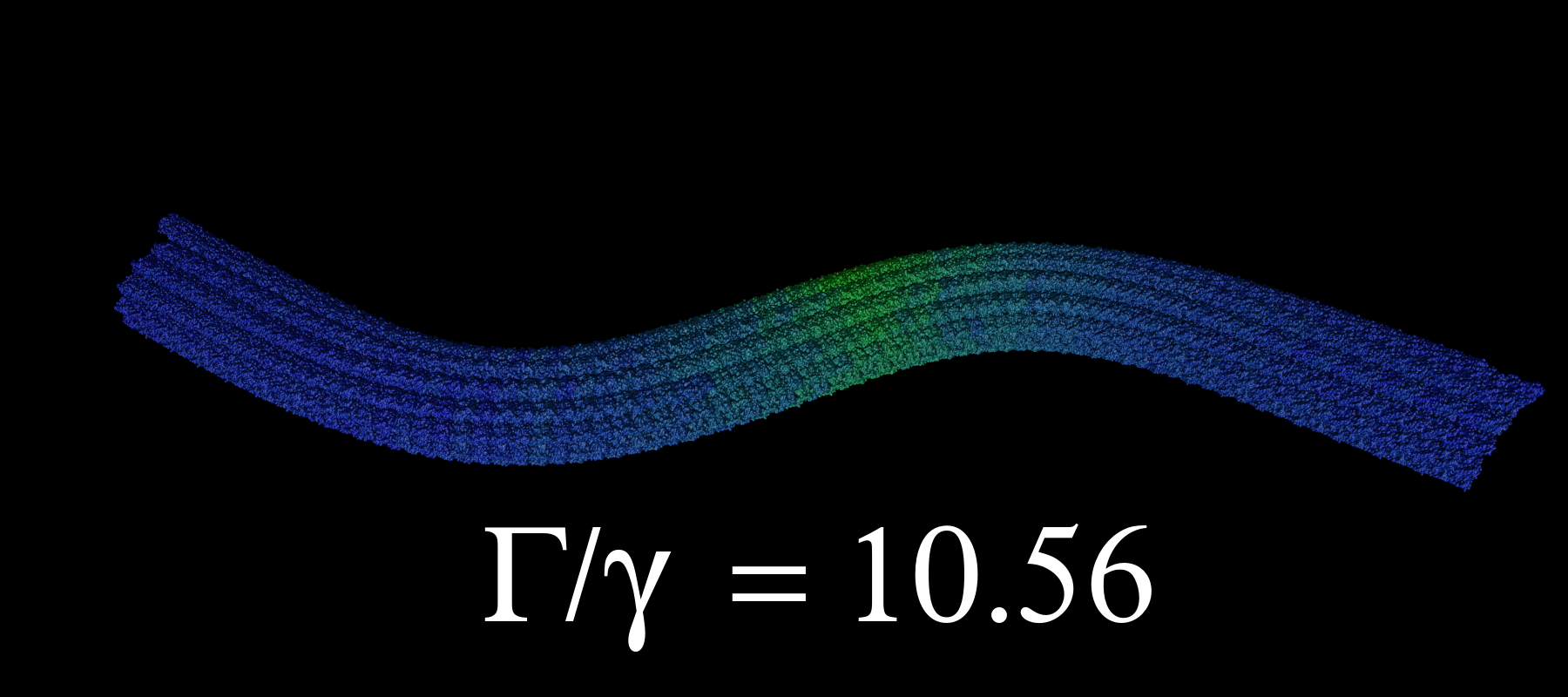}
\includegraphics[width=0.31\linewidth]{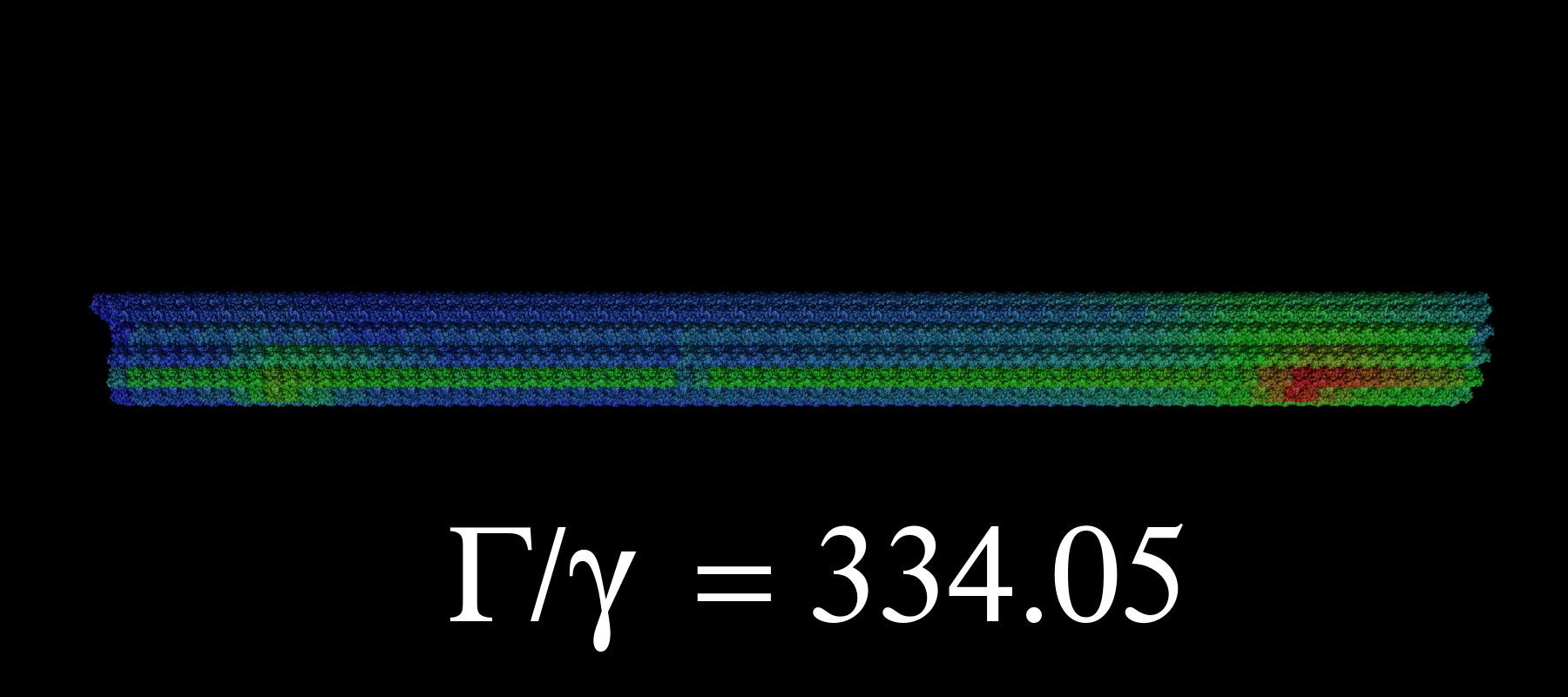}
\includegraphics[width=0.31\linewidth]{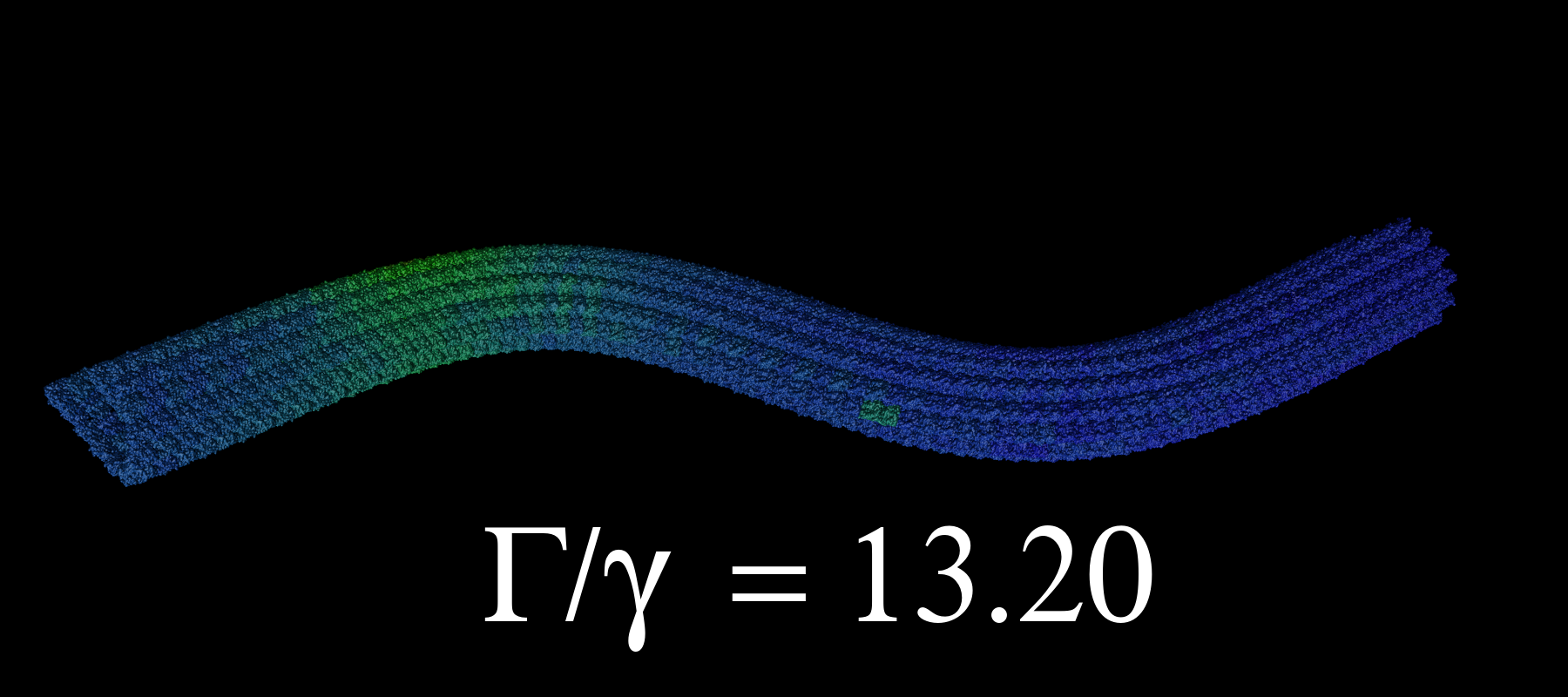} \\
$11$ \includegraphics[width=0.31\linewidth]{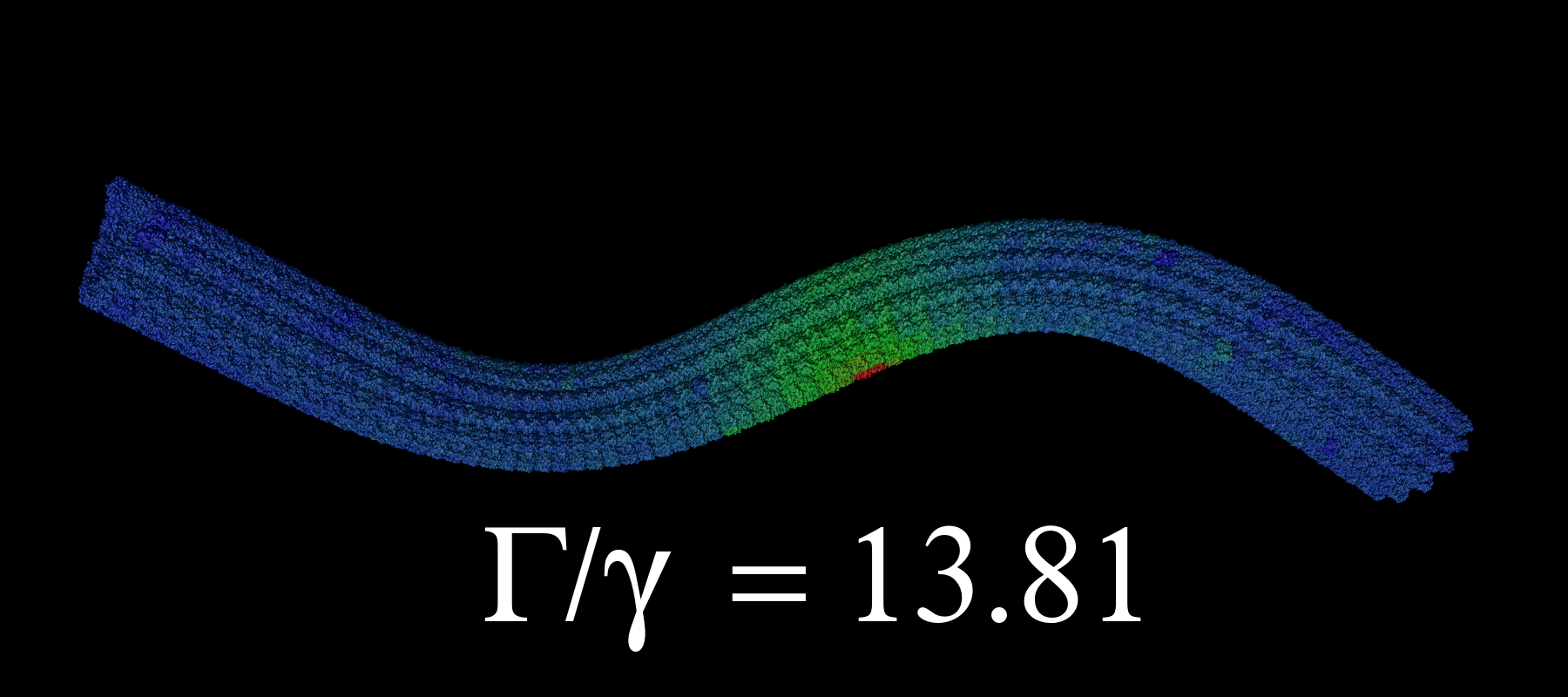}
\includegraphics[width=0.31\linewidth]{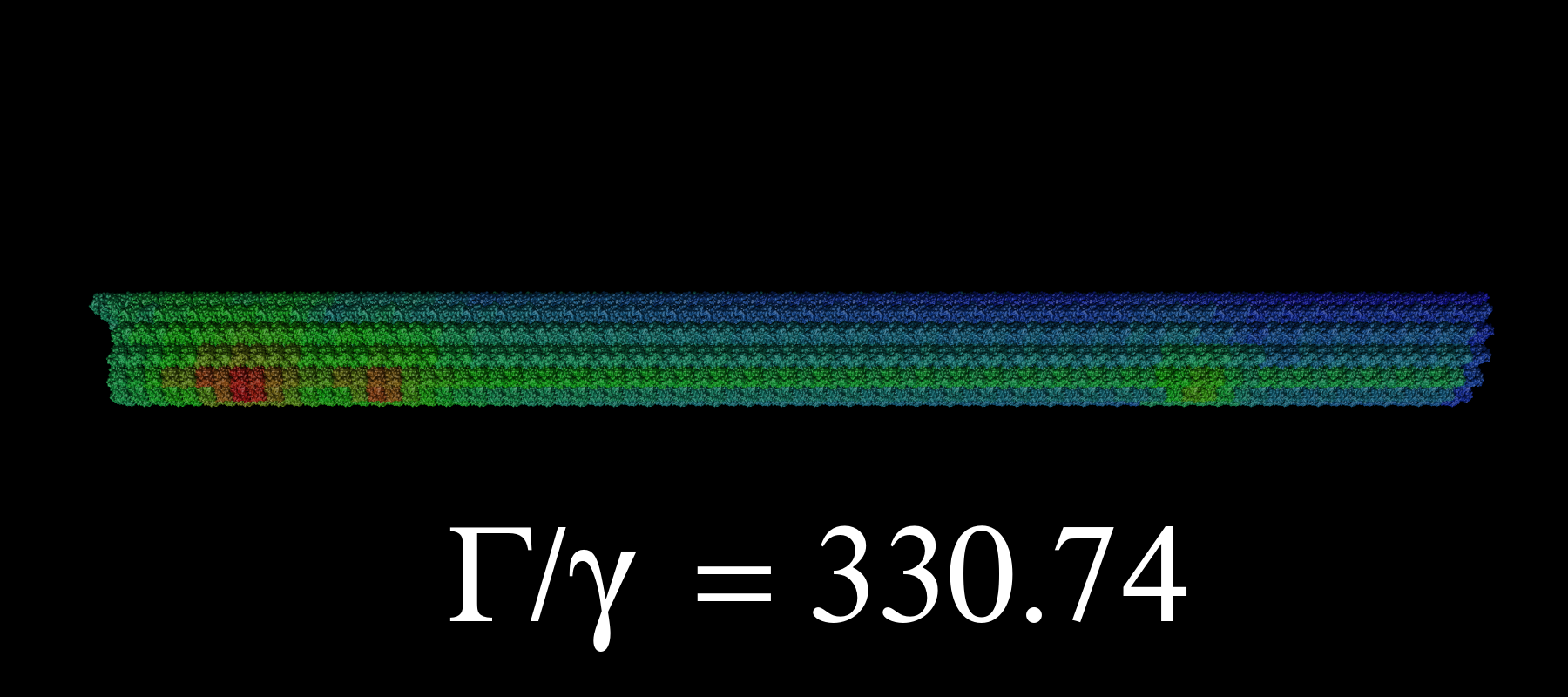}
\includegraphics[width=0.31\linewidth]{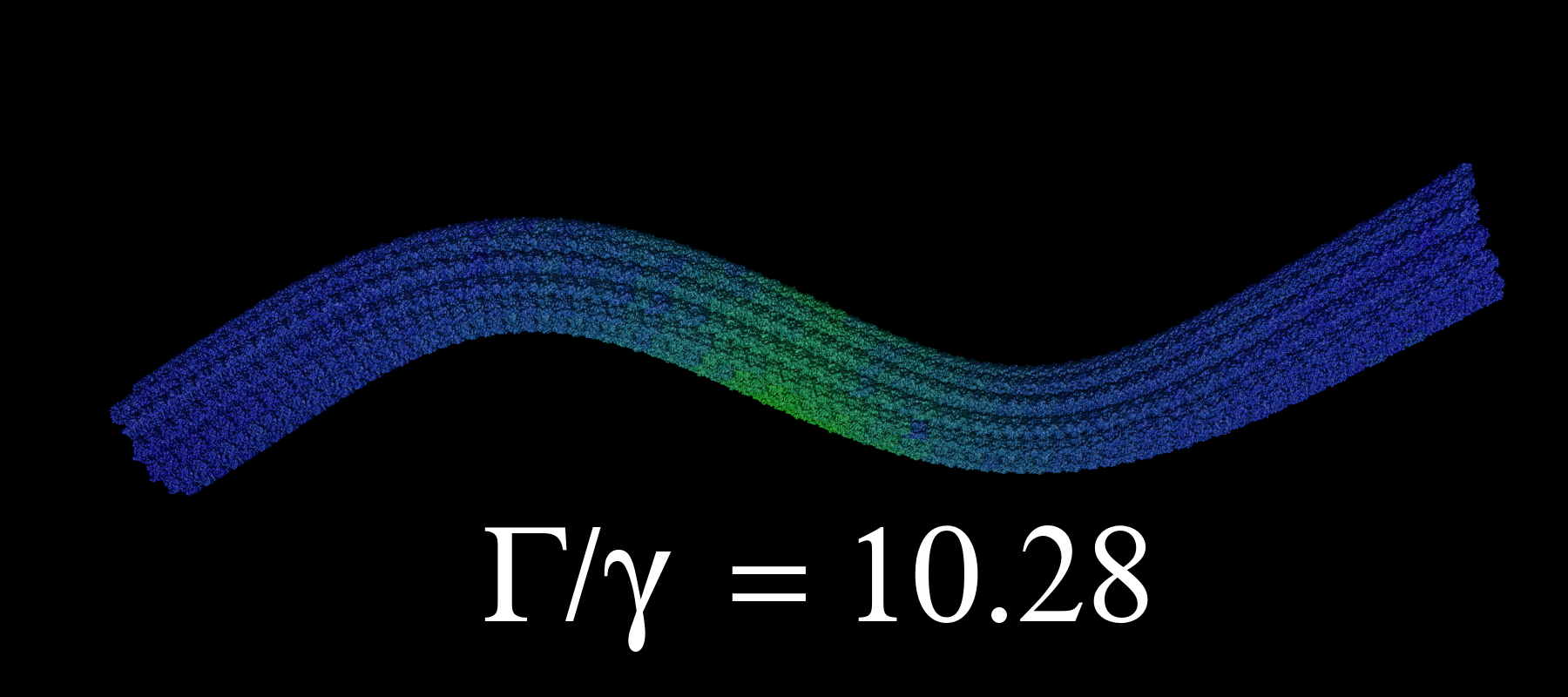} \\
$12$ \includegraphics[width=0.31\linewidth]{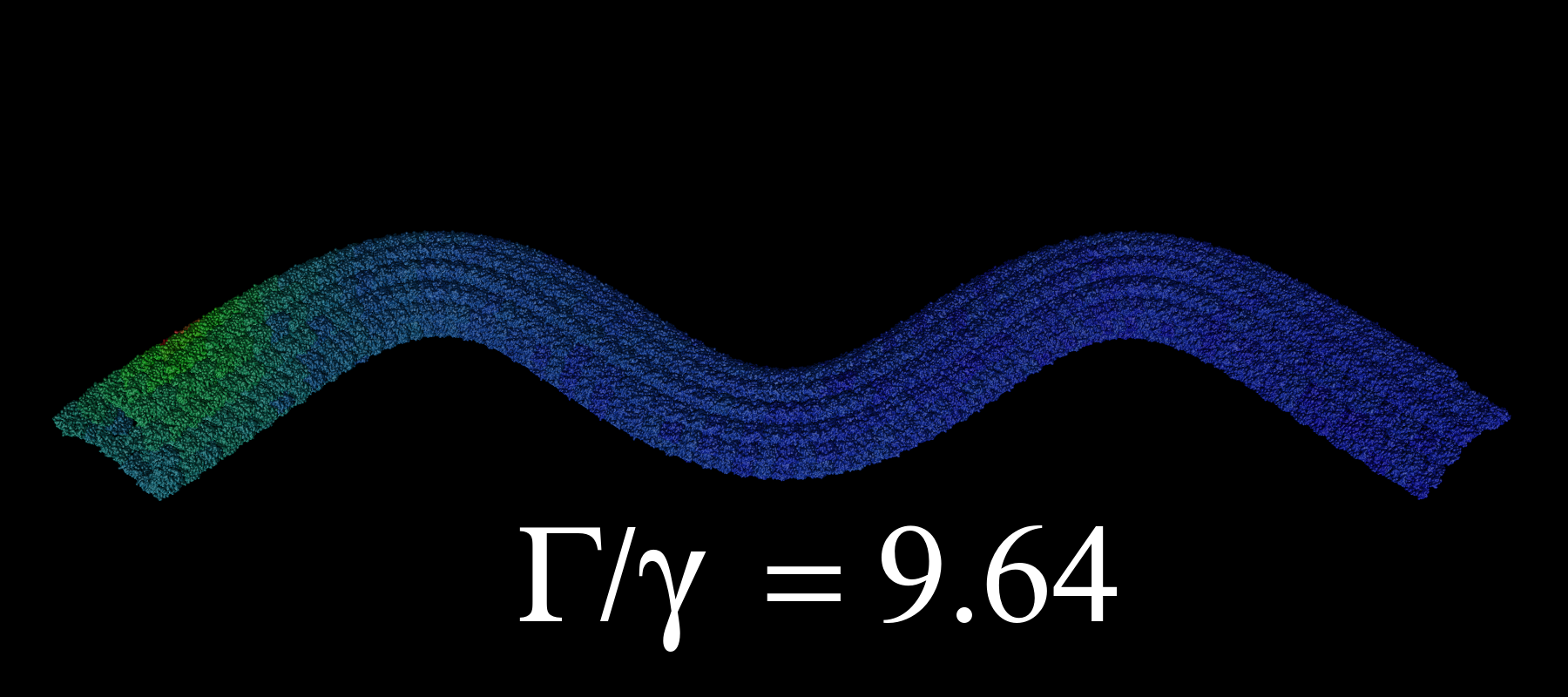}
\includegraphics[width=0.31\linewidth]{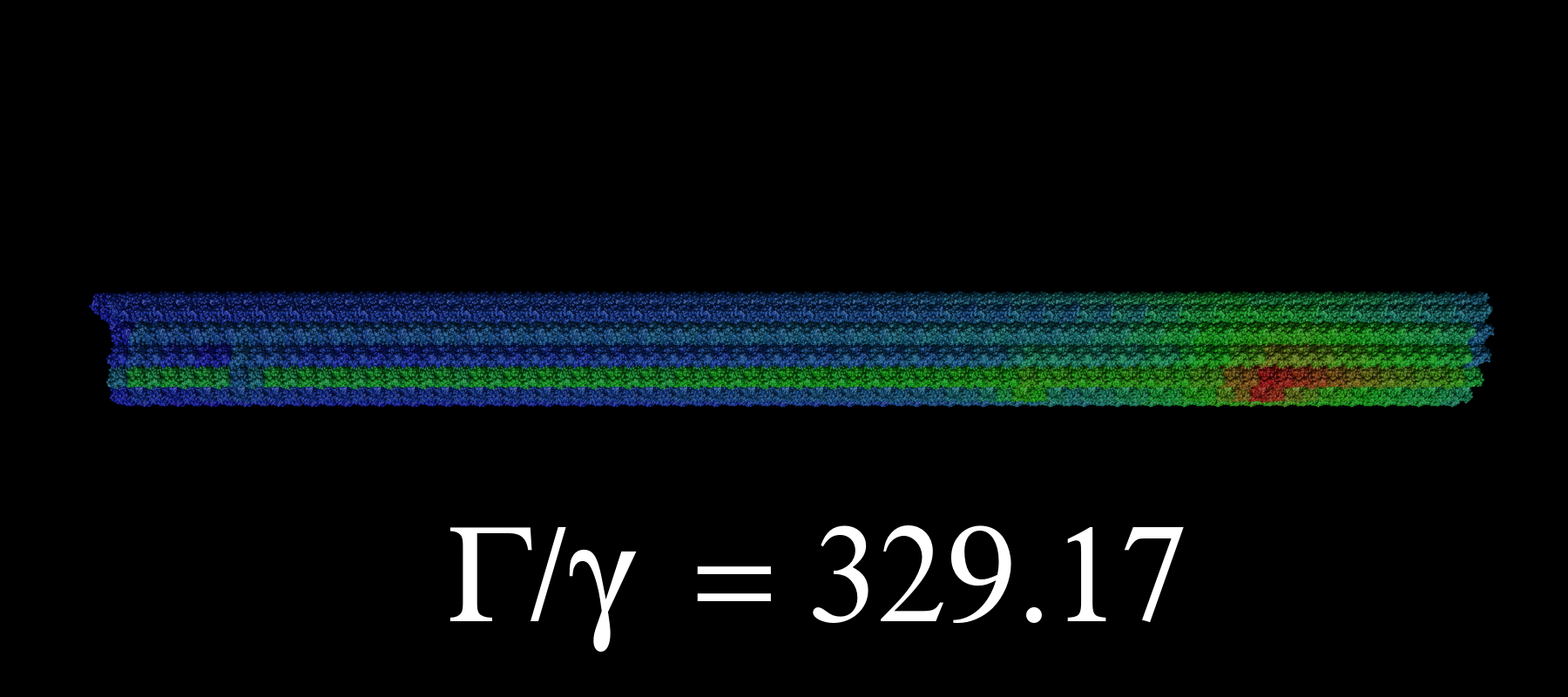}
\includegraphics[width=0.31\linewidth]{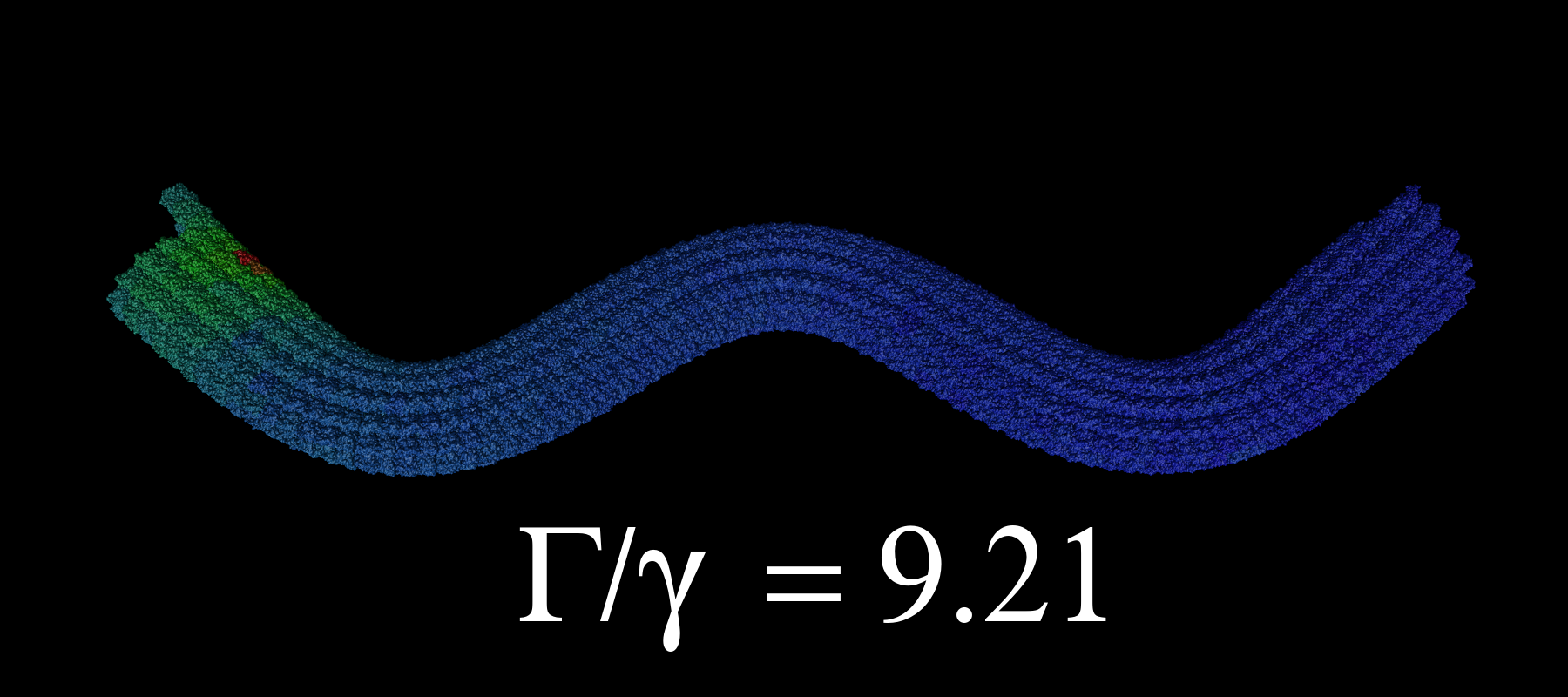} \\
$13$ \includegraphics[width=0.31\linewidth]{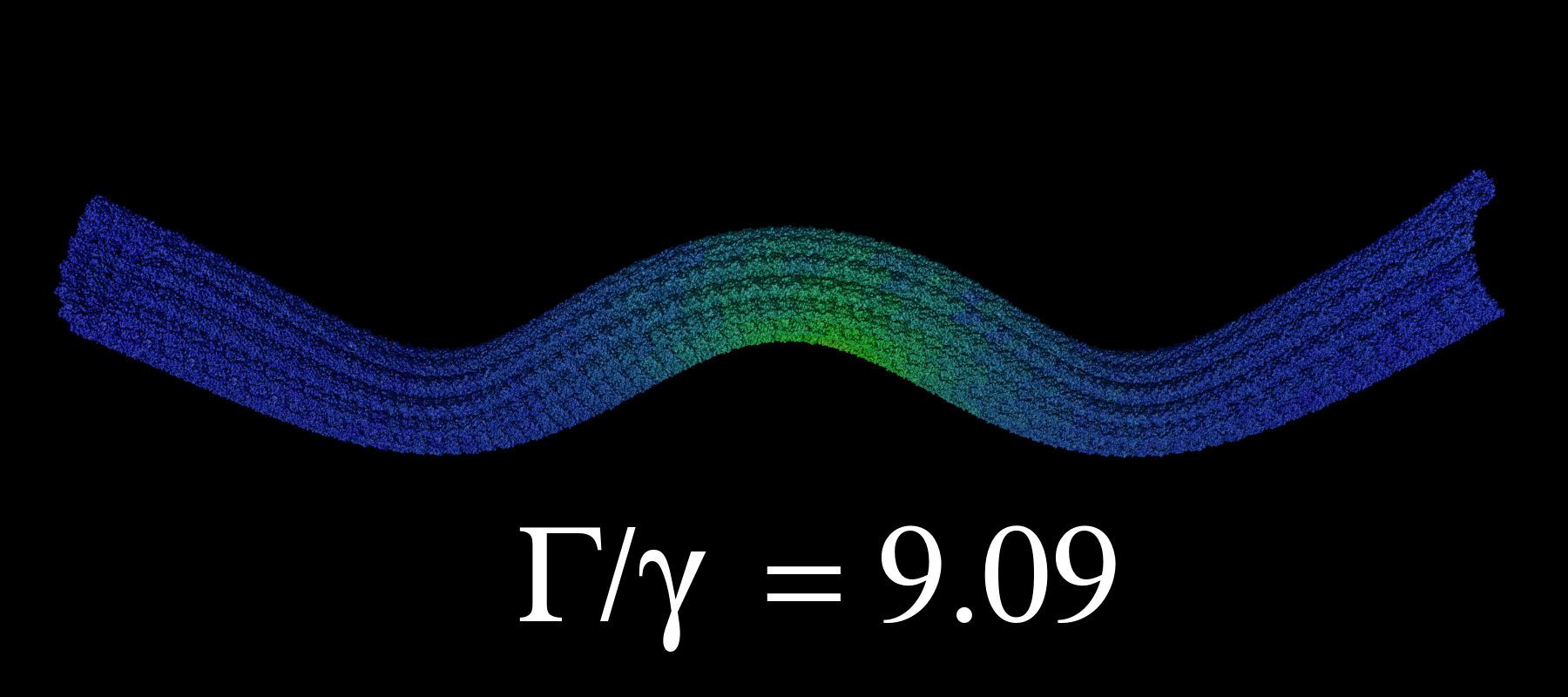}
\includegraphics[width=0.31\linewidth]{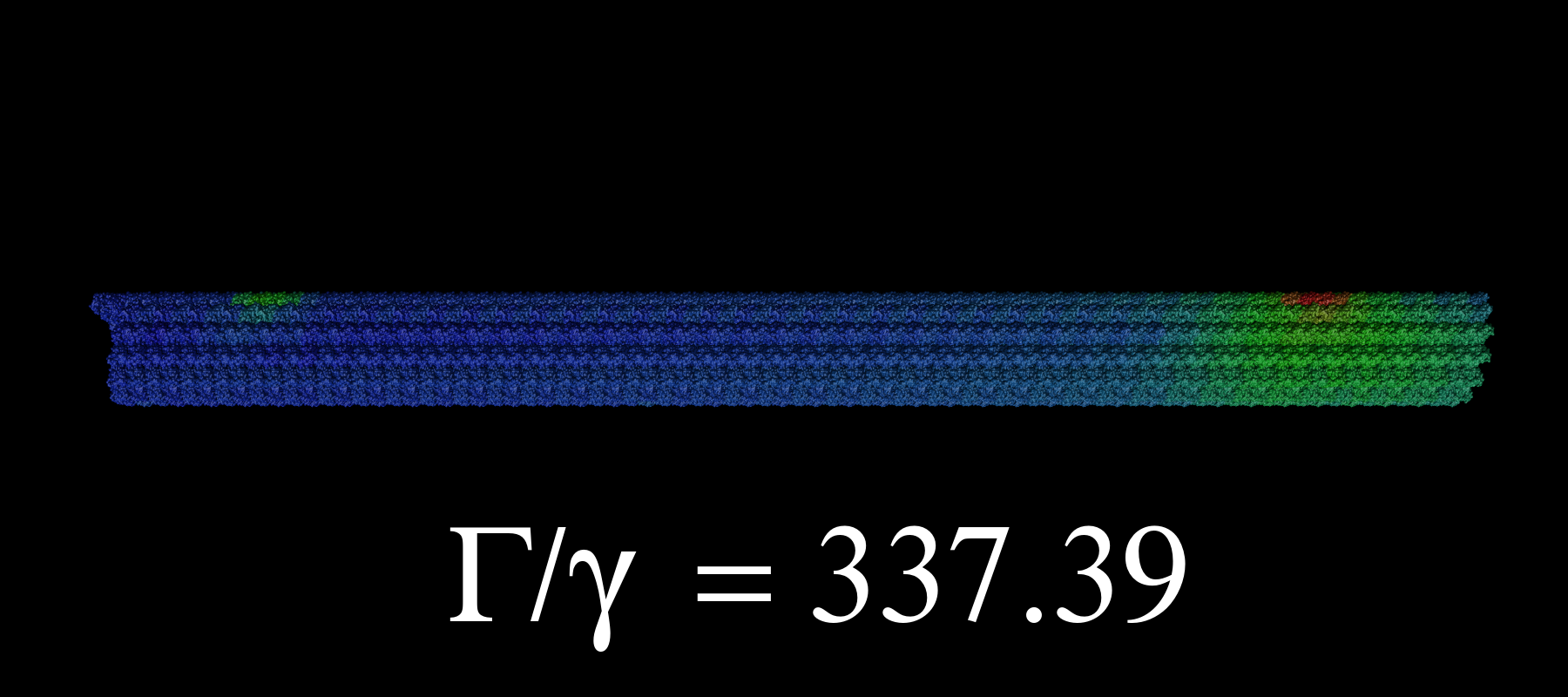}
\includegraphics[width=0.31\linewidth]{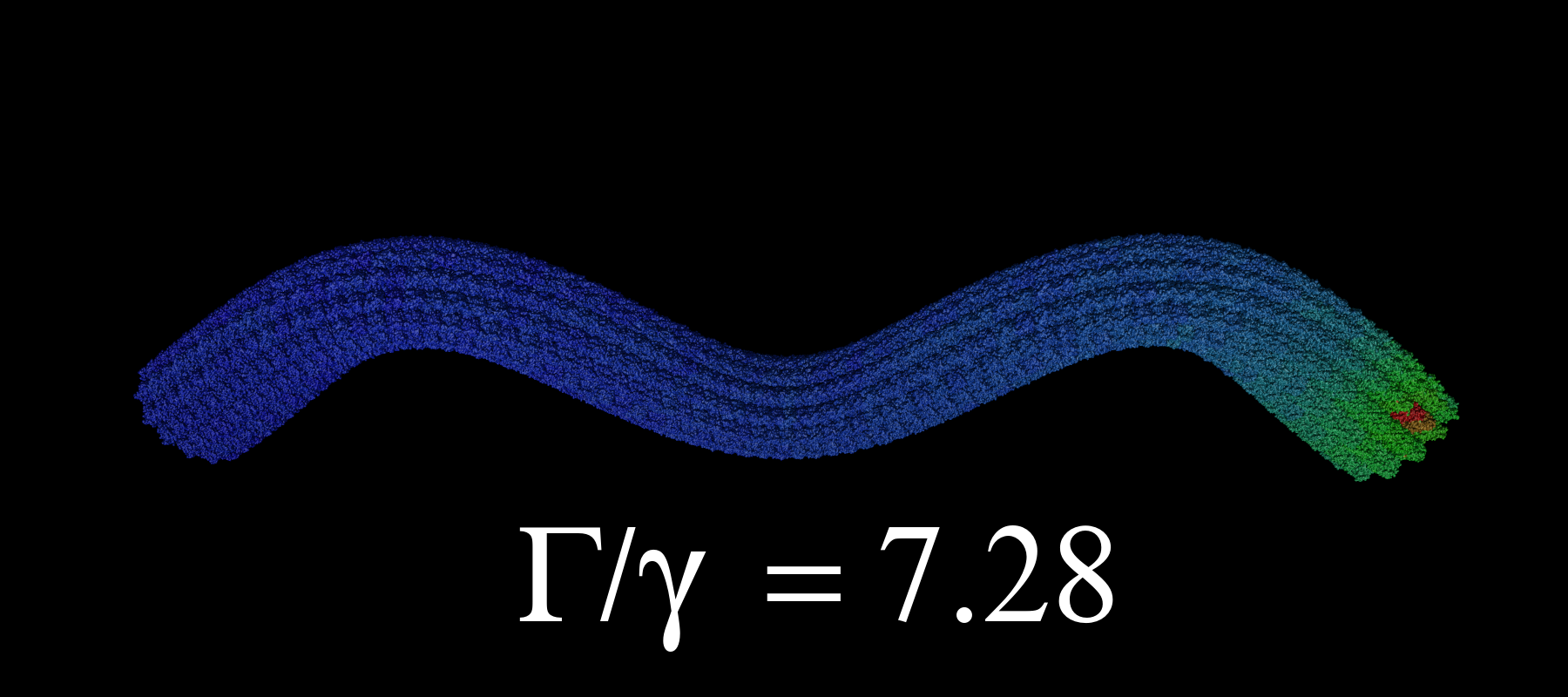} \\
\end{center} %\vspace*{-4mm}
\caption{Visualizations of color-coded probability maps showing exciton occupations for low-lying excitonic energy states in the Trp site basis, labeled by maxima of superradiant enhancement factors $\Gamma/\gamma$ of Trp networks for deformed (left and right columns) and undeformed (middle column) microtubules, realized during half a period of each mechanical mode. Atomistic simulations of vibrational motions were realized using the normal mode analyses of entire microtubules obtained from ~\cite{havelka2017deformation}. Each row displays three snapshots of microtubule conformations for each of the vibrational modes 7-13 (e.g.,~see Fig.~3 of~\cite{havelka2017deformation}). Rigid modes 1-6 are not shown because they do not involve deformations. 
}\phantomsection
\label{fig:Color7-14}
\end{figure}

\begin{figure}[h!]
\begin{center}
$14$ \includegraphics[width=0.31\linewidth]{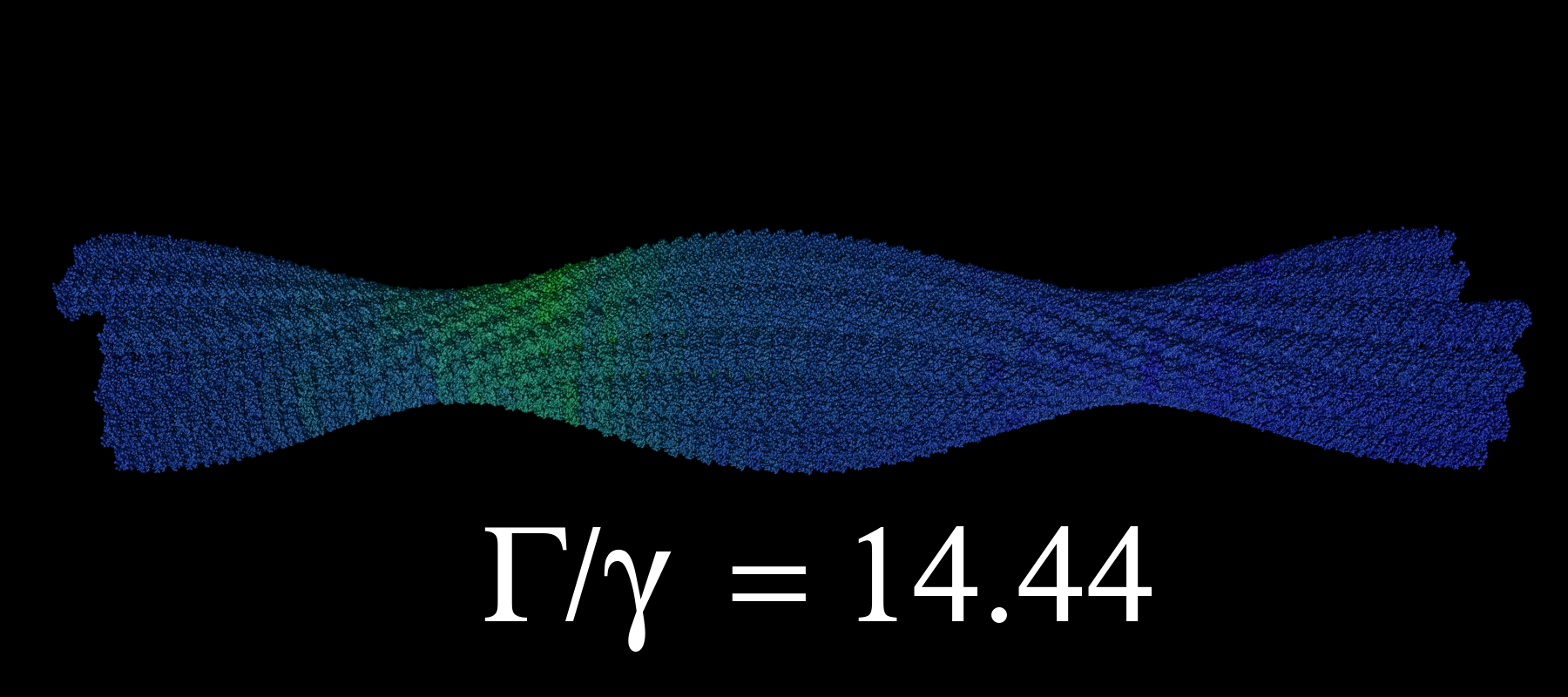}
\includegraphics[width=0.31\linewidth]{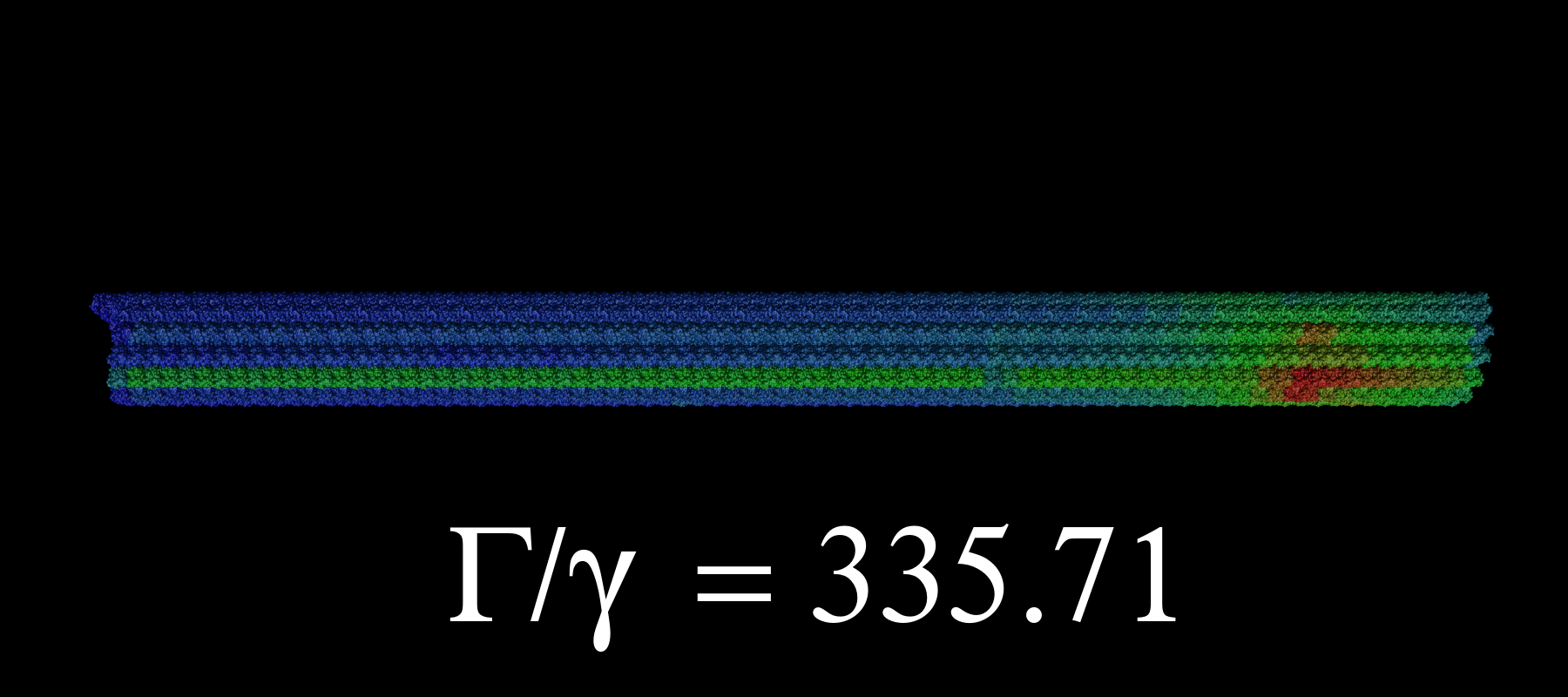}
\includegraphics[width=0.31\linewidth]{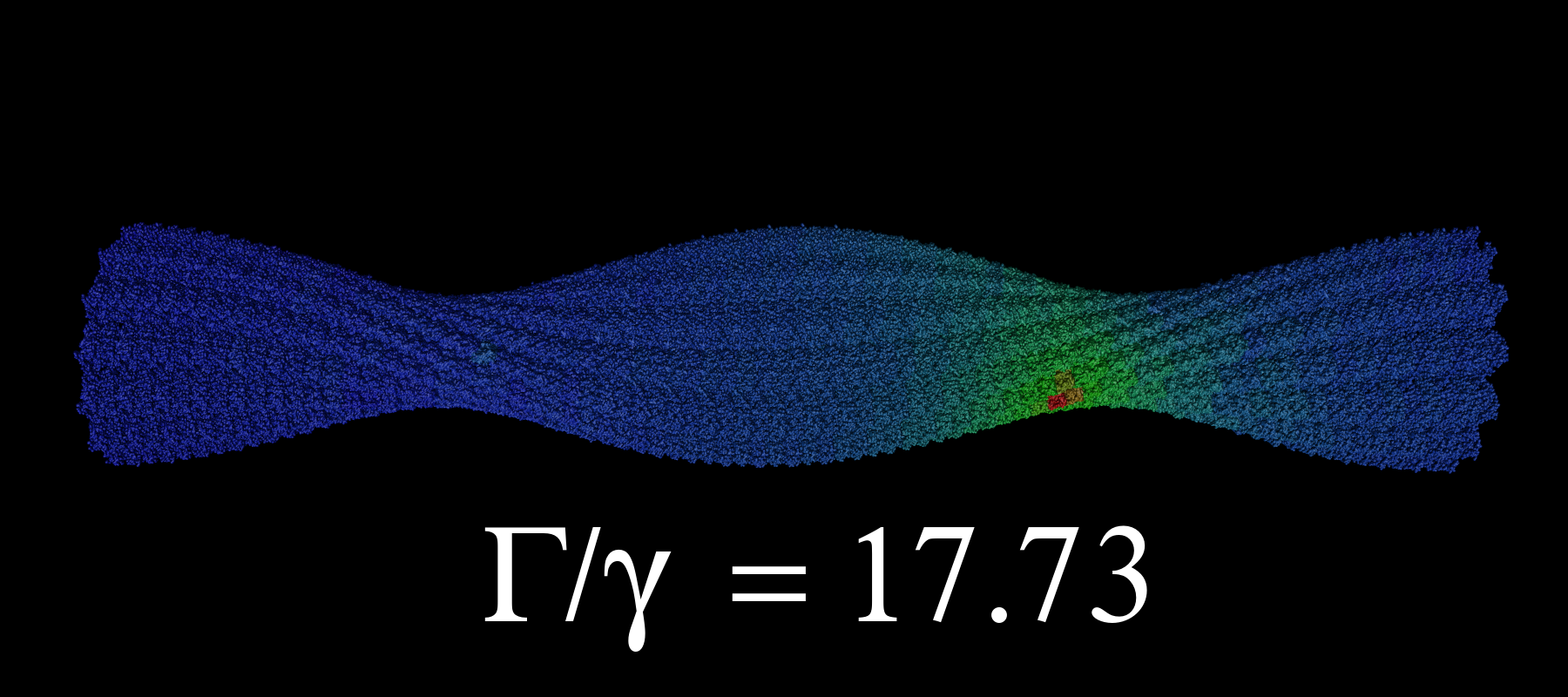} \\
$15$ \includegraphics[width=0.31\linewidth]{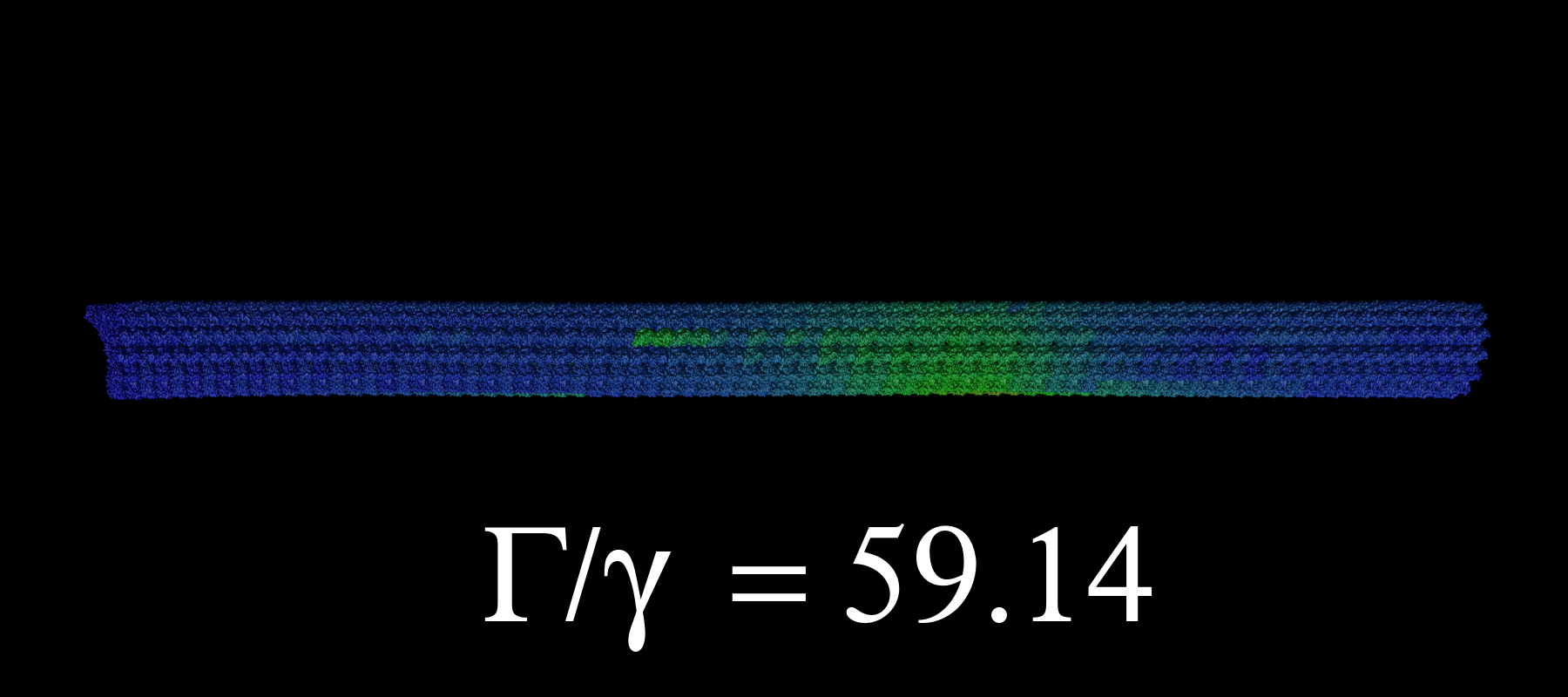}
\includegraphics[width=0.31\linewidth]{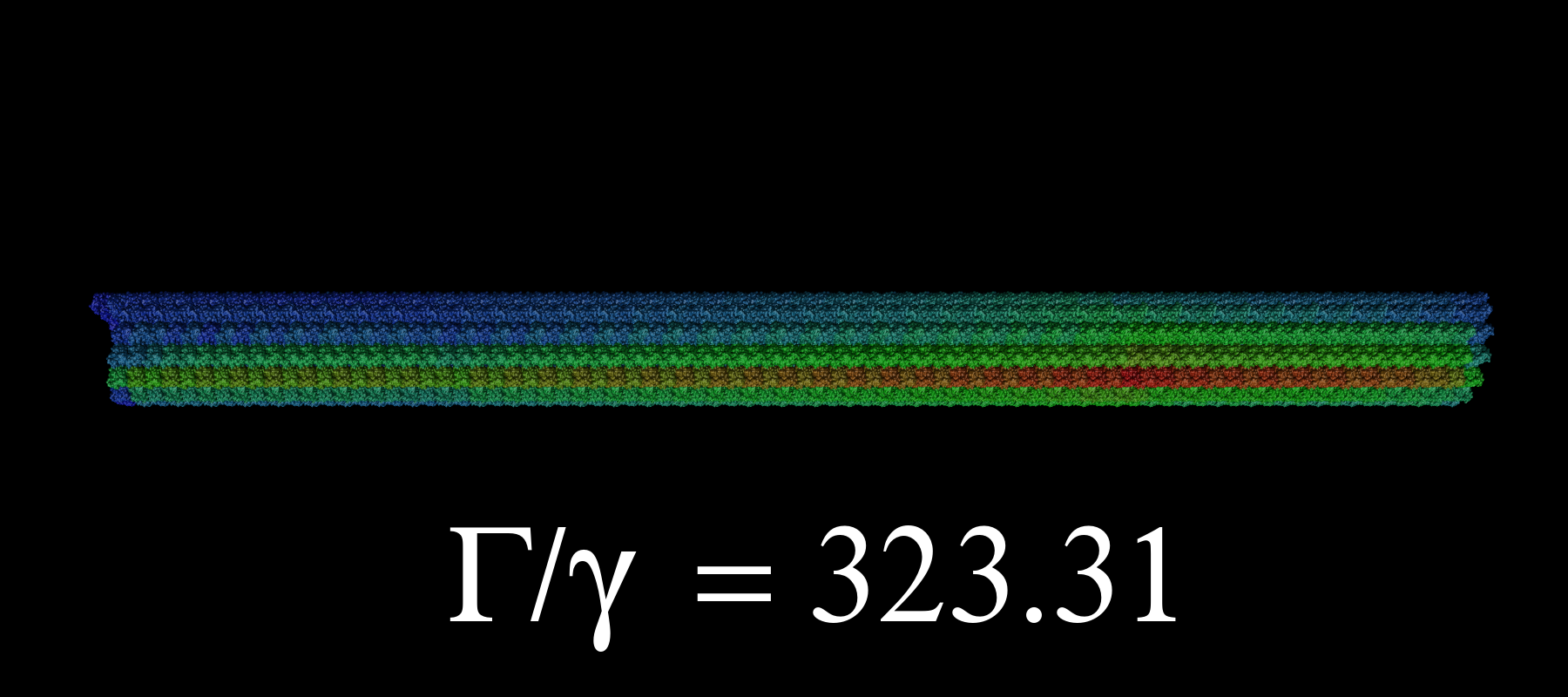}
\includegraphics[width=0.31\linewidth]{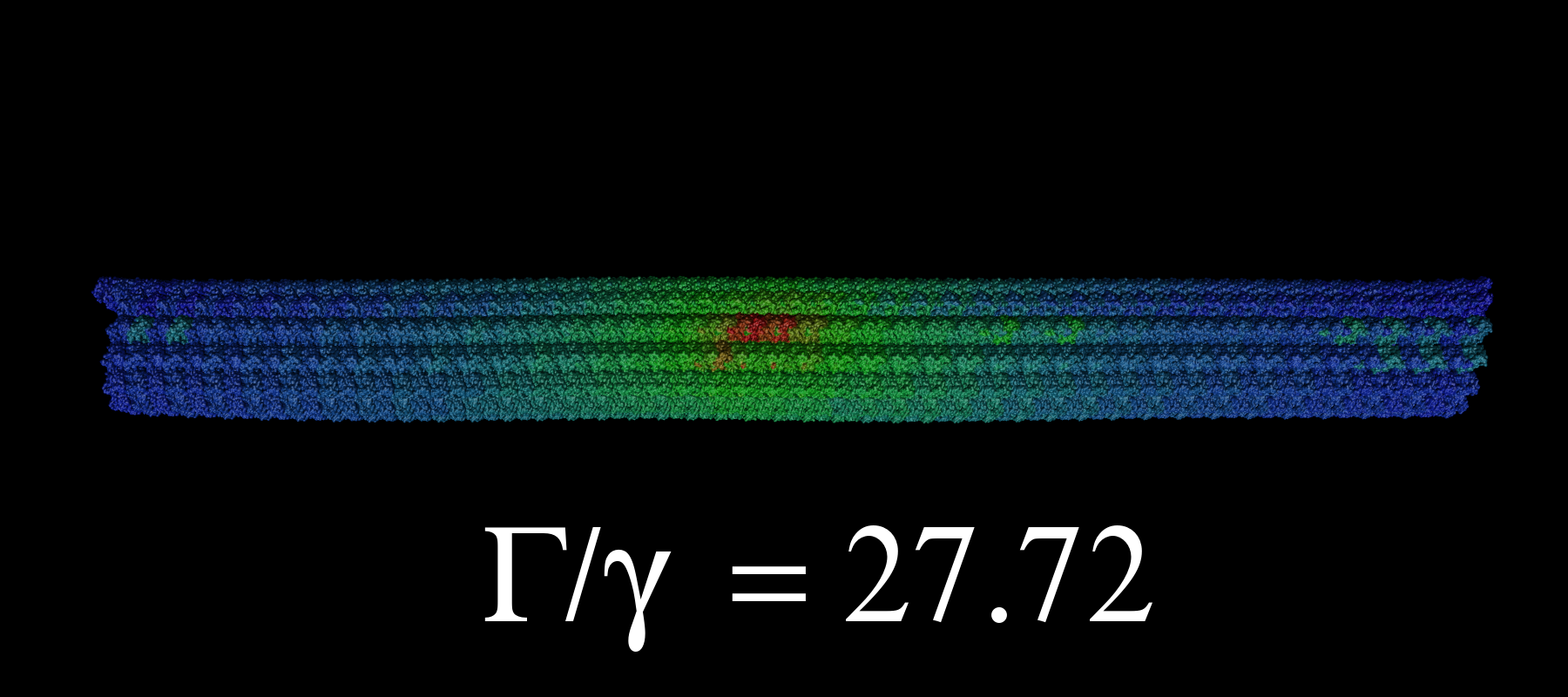} \\
$16$ \includegraphics[width=0.31\linewidth]{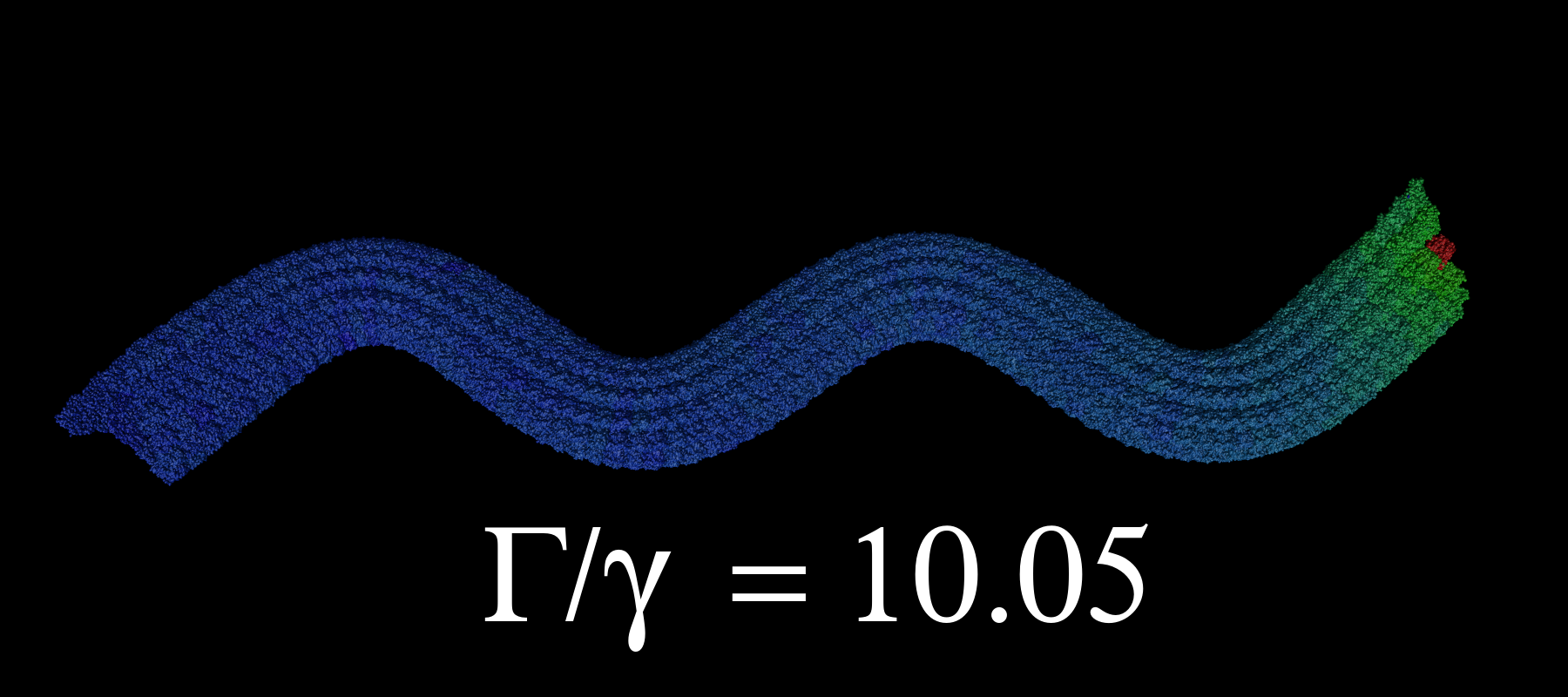}
\includegraphics[width=0.31\linewidth]{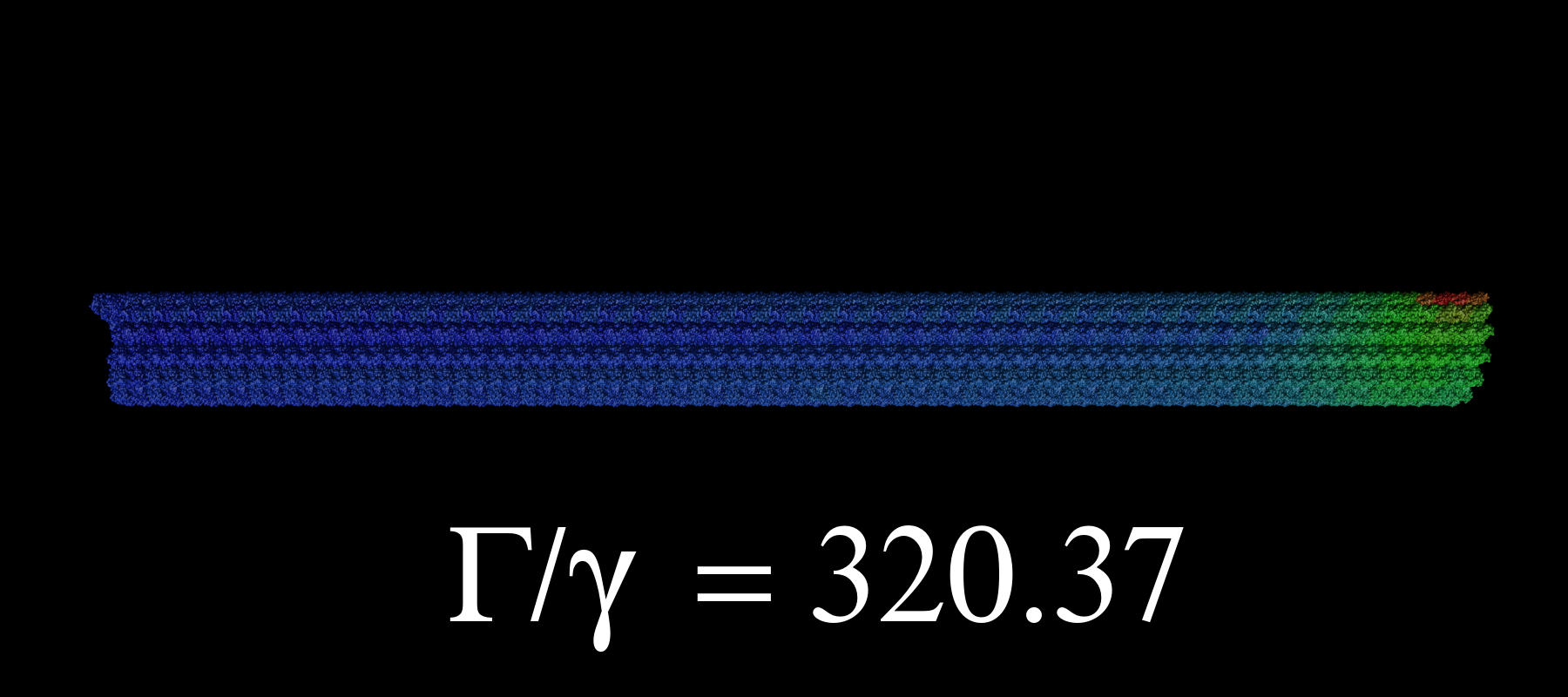}
\includegraphics[width=0.31\linewidth]{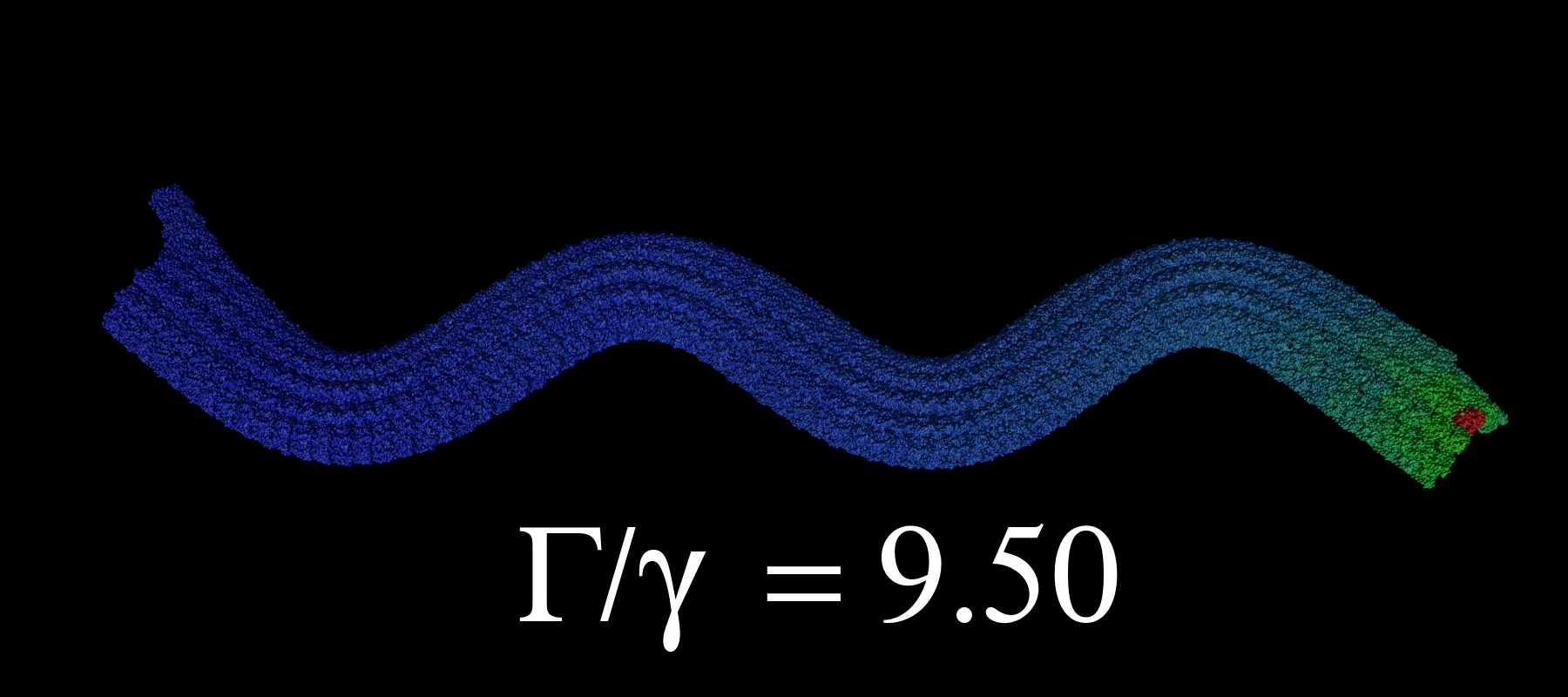} \\
$17$ \includegraphics[width=0.31\linewidth]{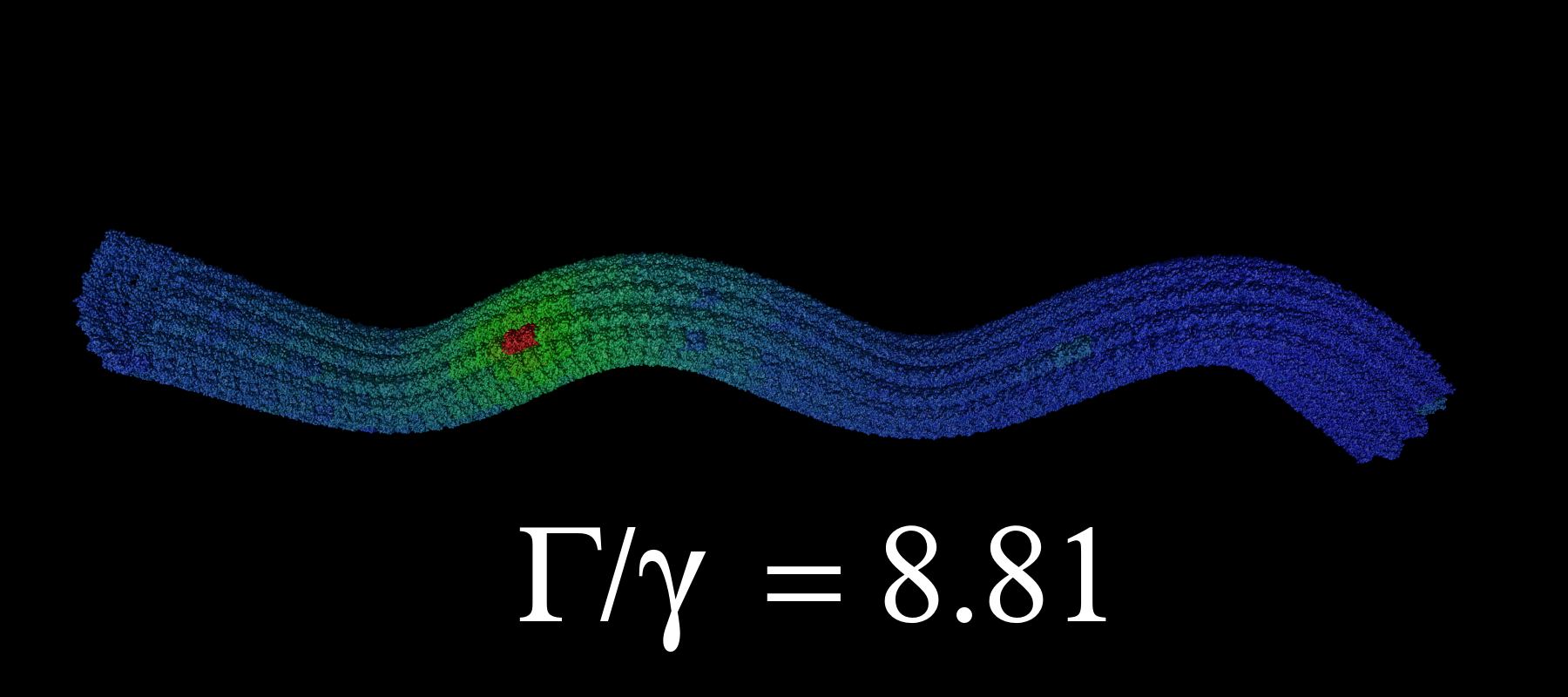}
\includegraphics[width=0.31\linewidth]{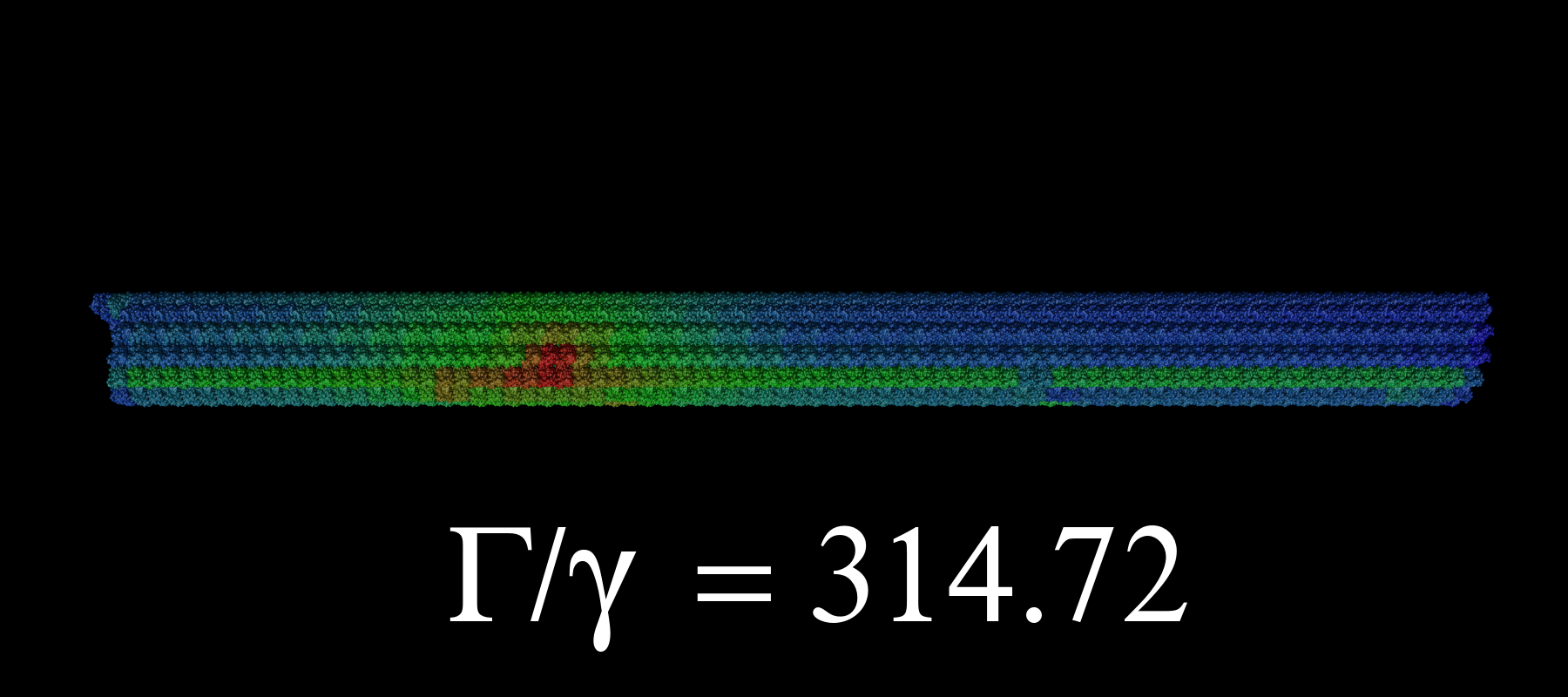}
\includegraphics[width=0.31\linewidth]{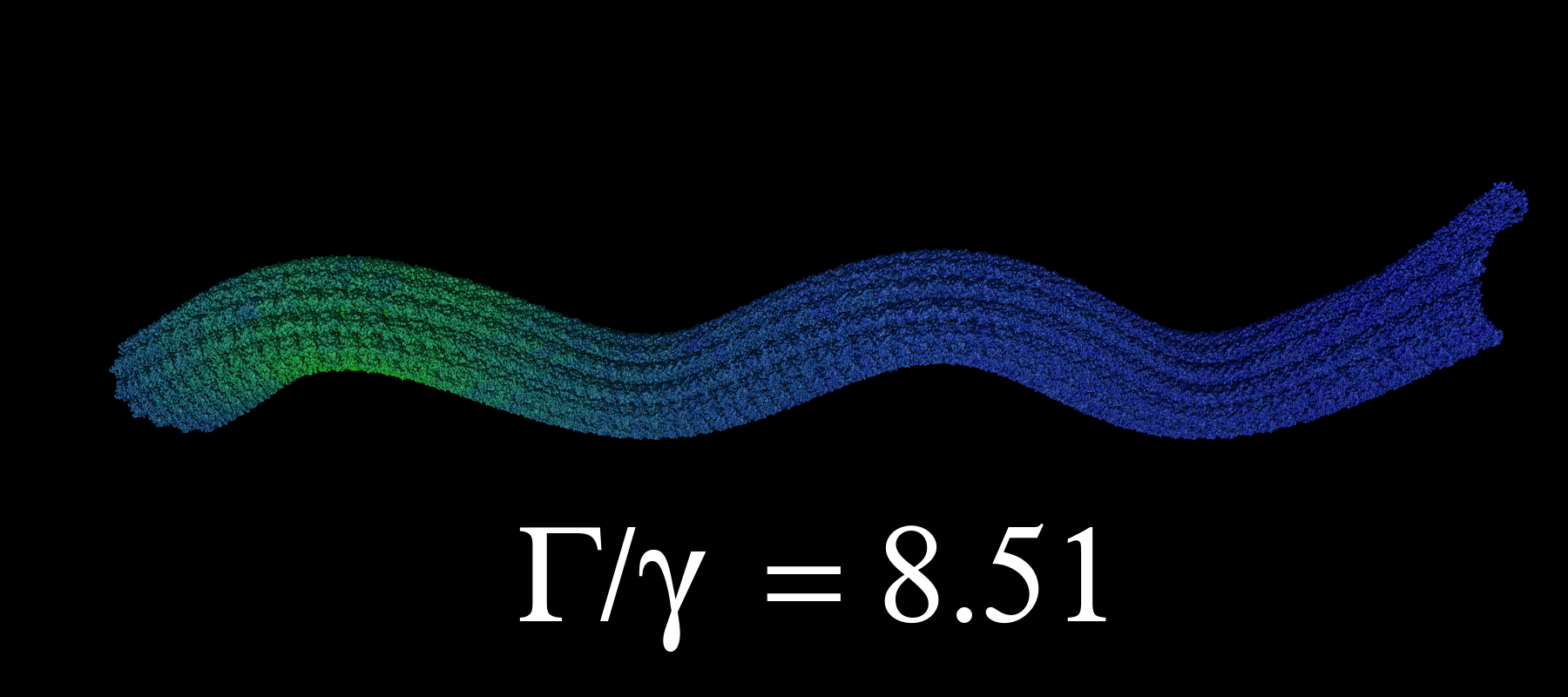} \\
$18$ \includegraphics[width=0.31\linewidth]{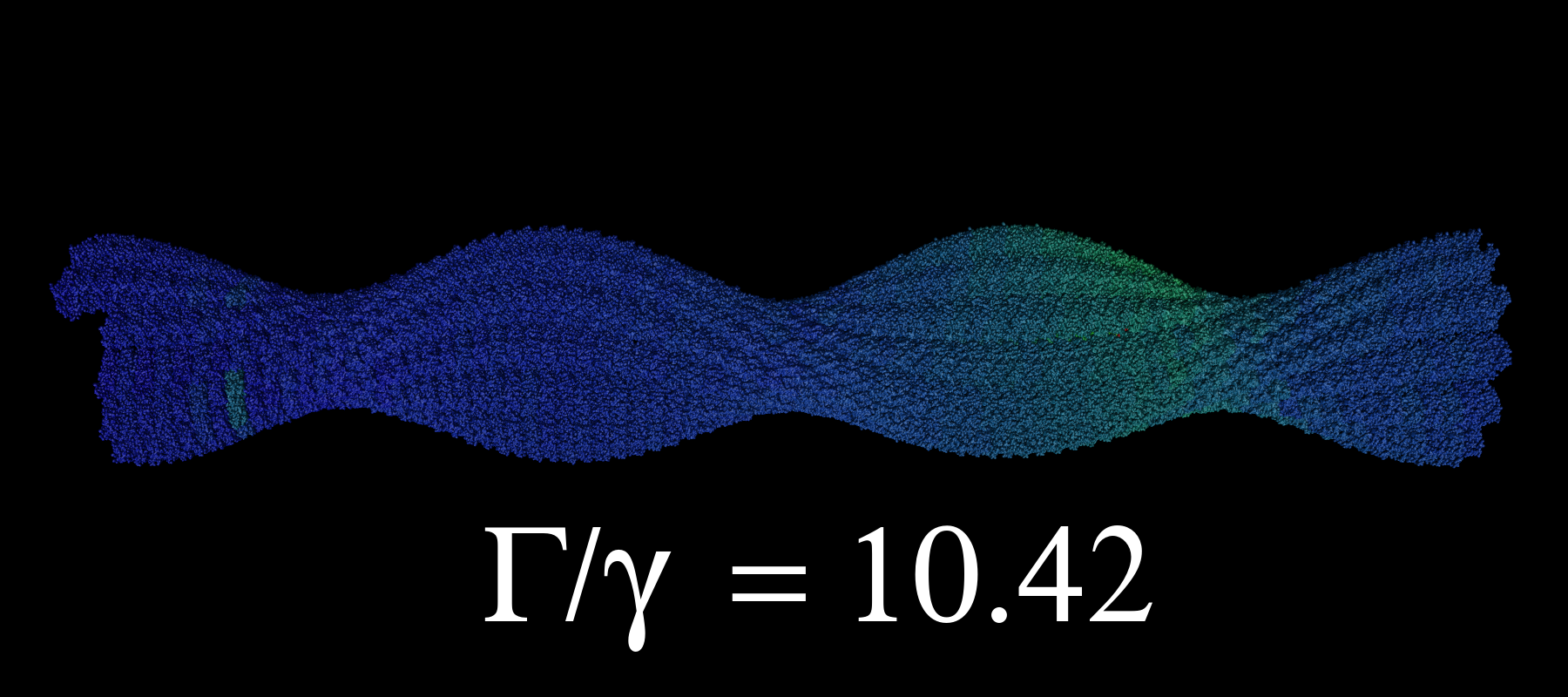}
\includegraphics[width=0.31\linewidth]{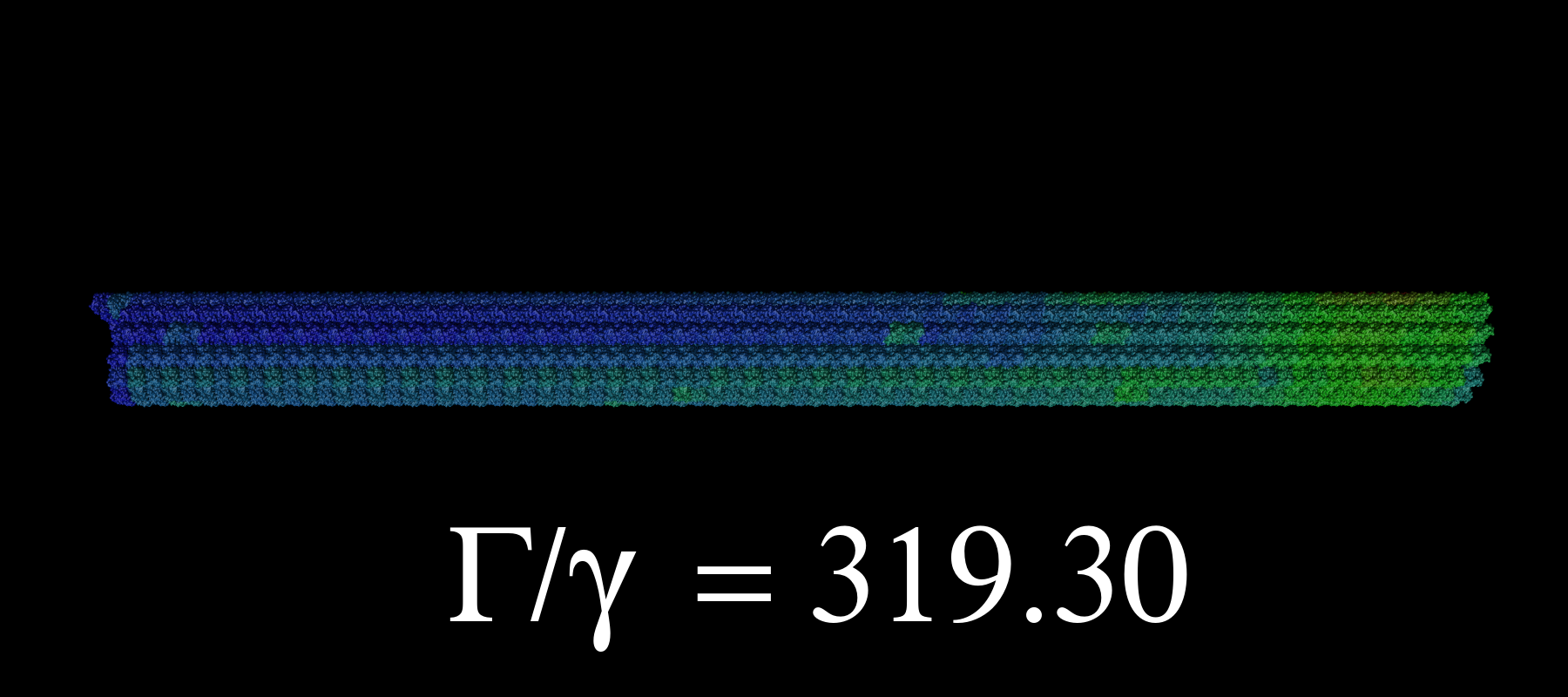}
\includegraphics[width=0.31\linewidth]{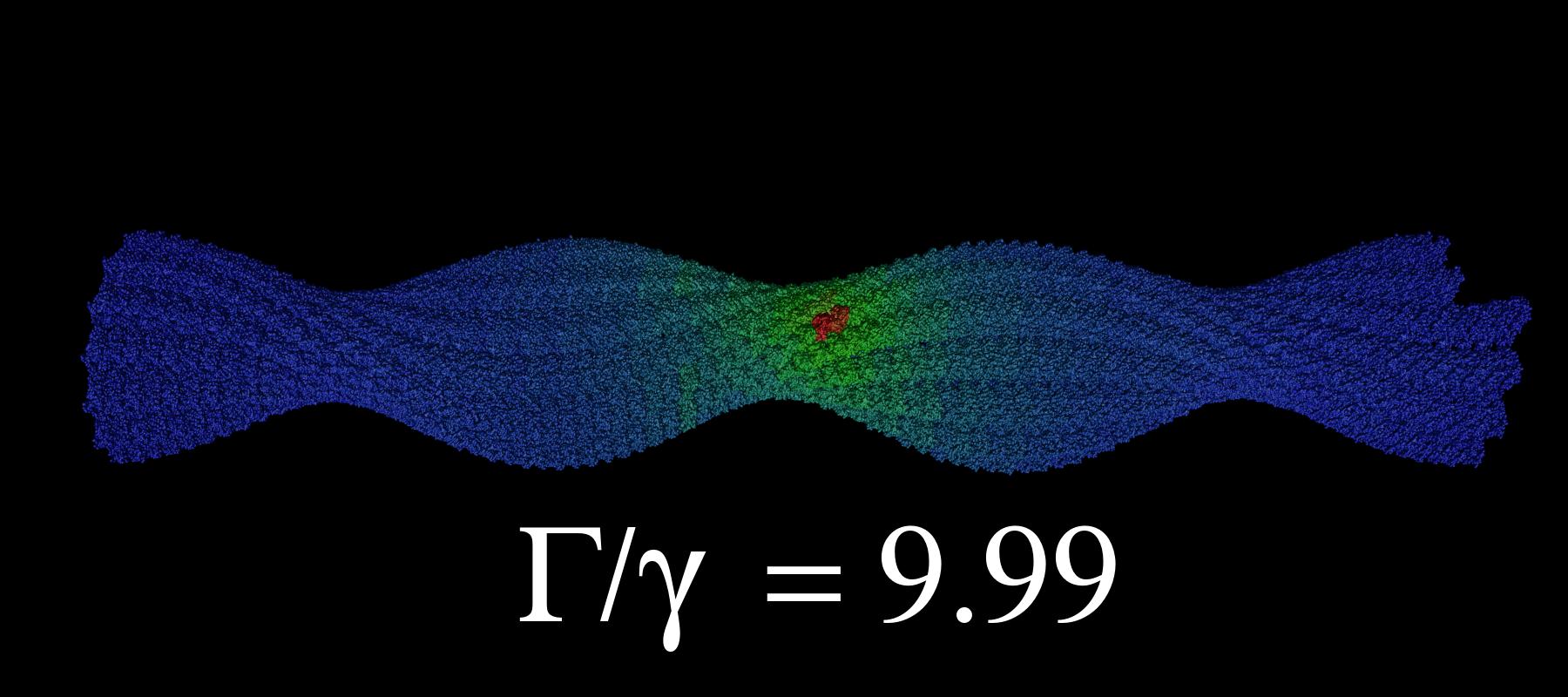} \\
$19$ \includegraphics[width=0.31\linewidth]{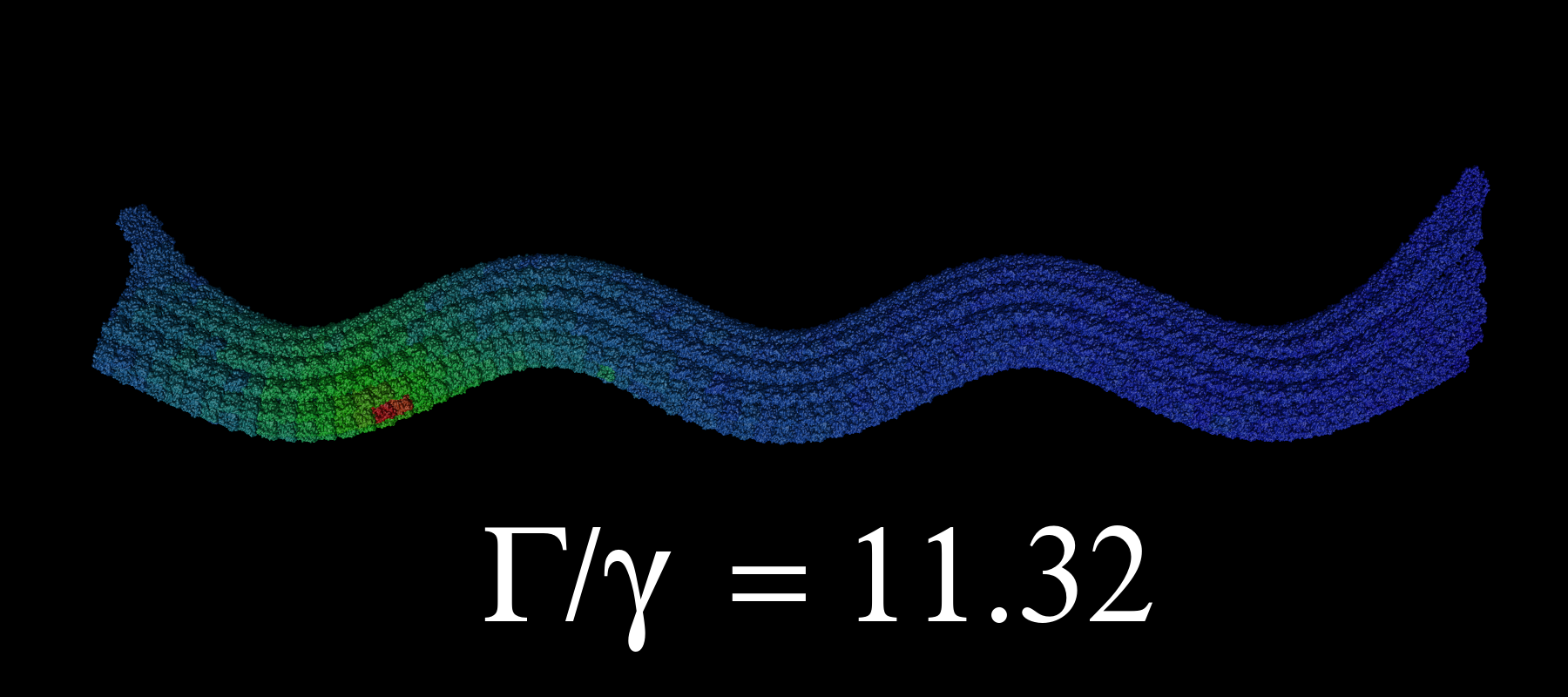}
\includegraphics[width=0.31\linewidth]{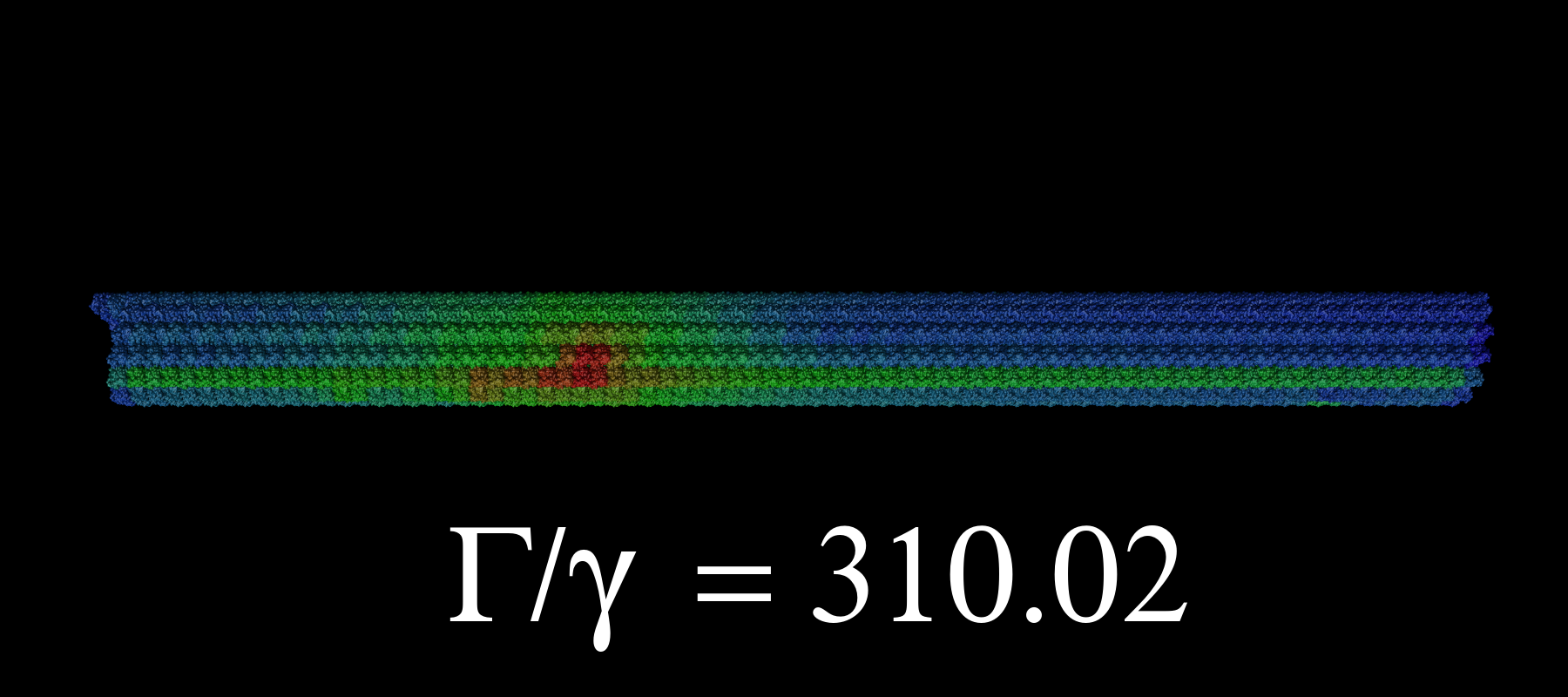}
\includegraphics[width=0.31\linewidth]{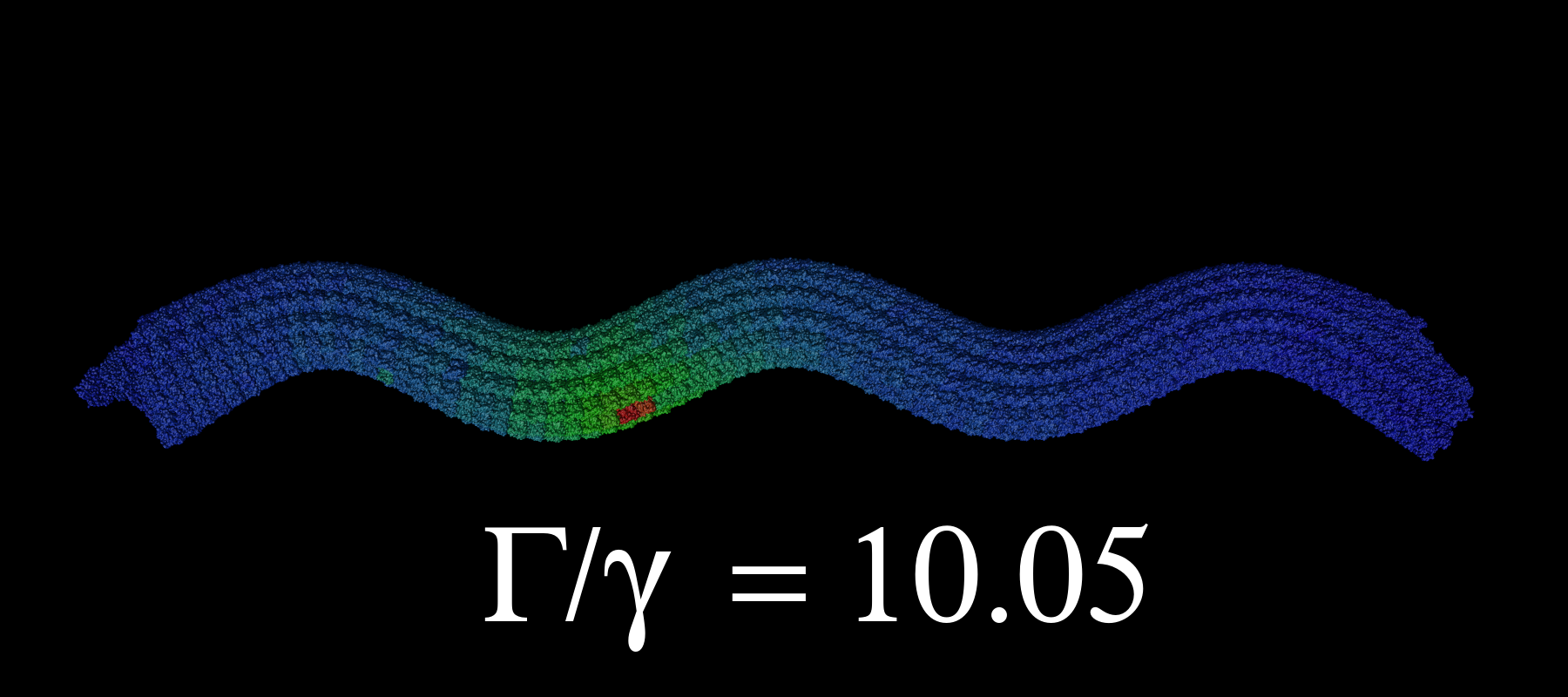} \\
$20$ \includegraphics[width=0.31\linewidth]{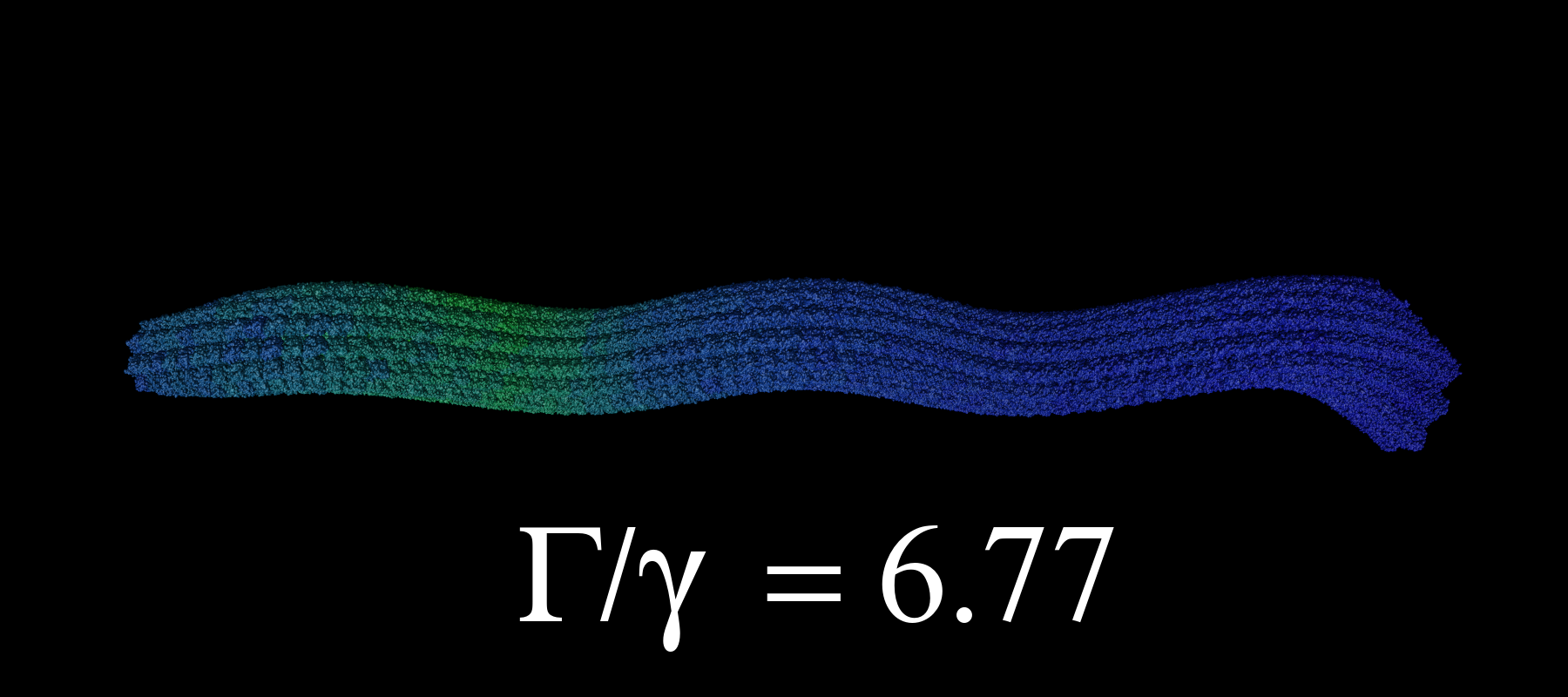}
\includegraphics[width=0.31\linewidth]{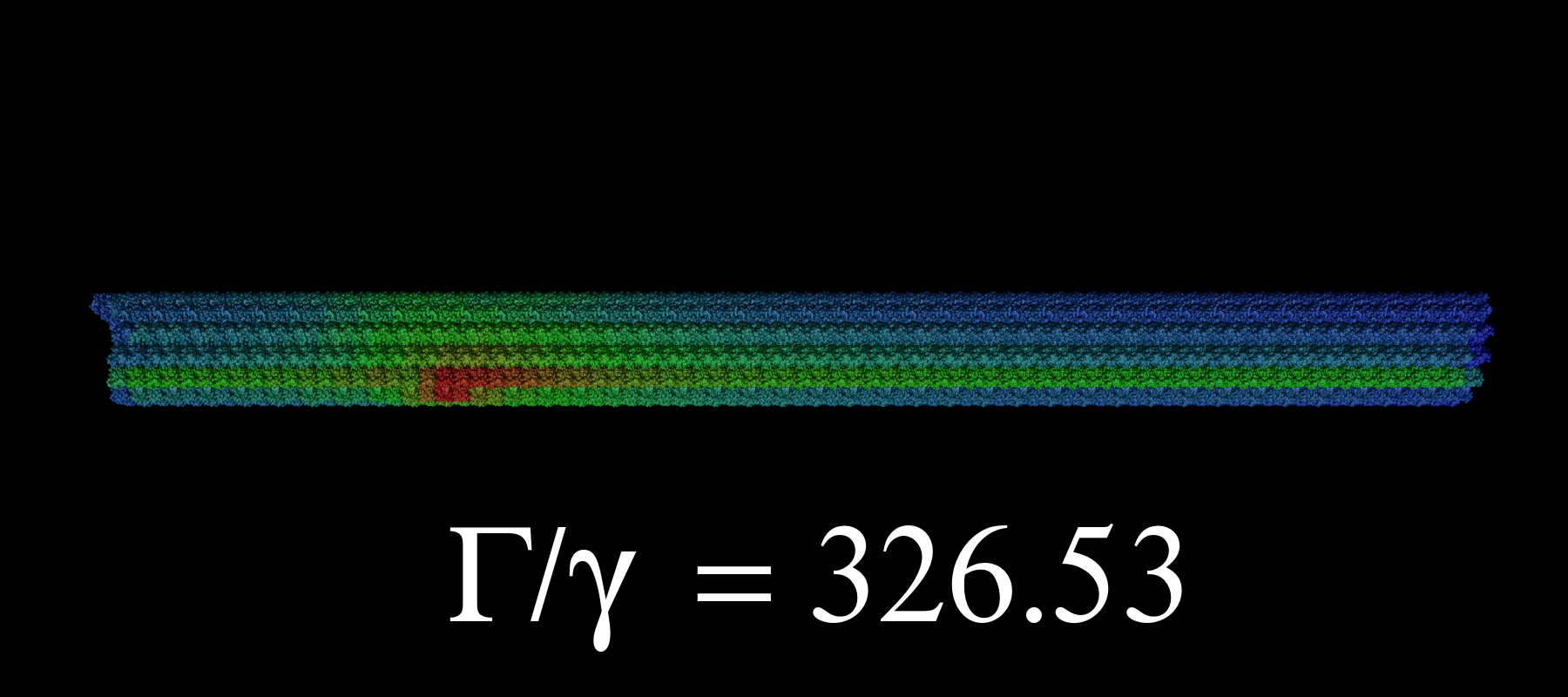}
\includegraphics[width=0.31\linewidth]{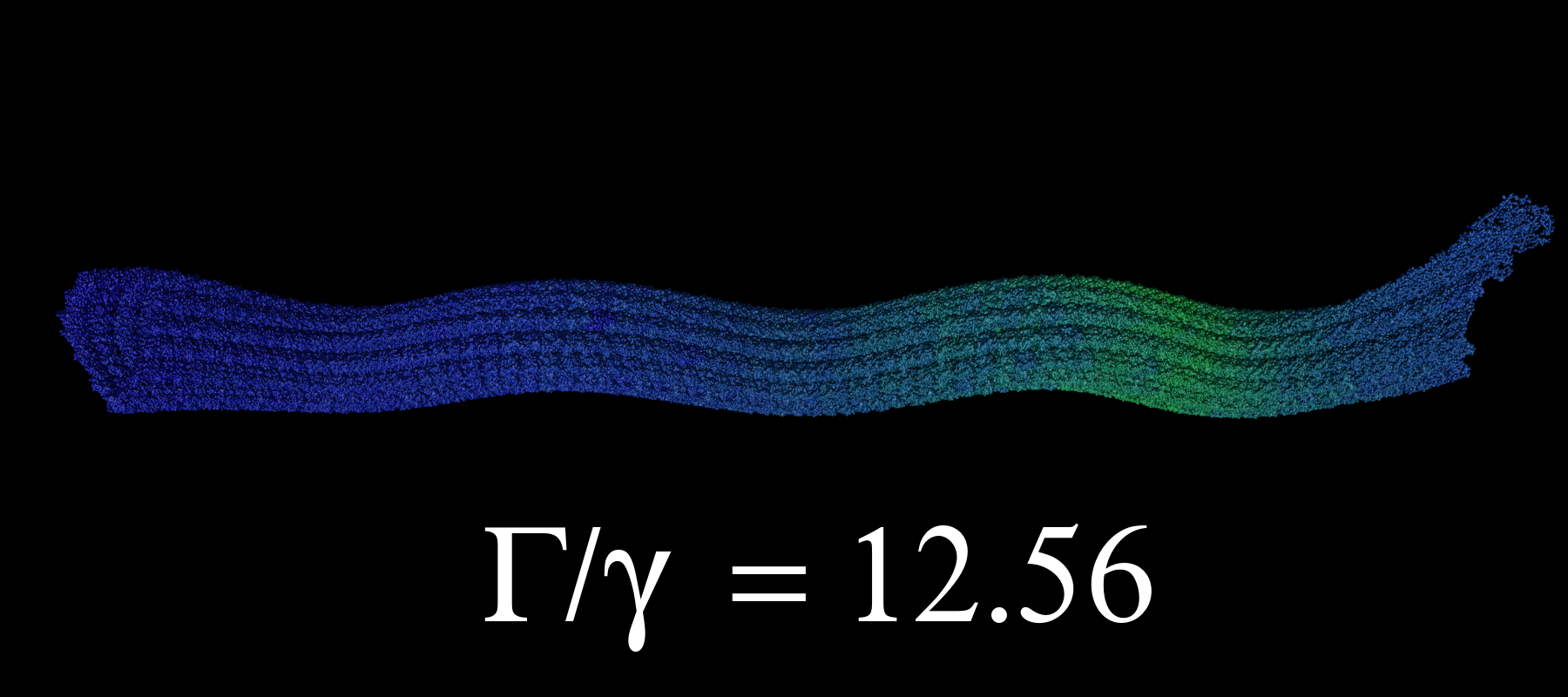} \\
\end{center}
\caption{Visualizations of color-coded probability maps showing exciton occupations for low-lying excitonic energy states, labeled by maxima of superradiant enhancement factors $\Gamma/\gamma$ of Trp networks for deformed (left and right panels) and undeformed (middle panel) microtubules. Atomistic simulations of vibrational motions were realized using the normal mode analyses of entire microtubules obtained from ~\cite{havelka2017deformation}. Each row displays three snapshots of microtubule conformations for each of the vibrational modes 14-20 (e.g.,~see Fig.~3 of~\cite{havelka2017deformation}).}\phantomsection
\label{fig:Color15-20} 
\end{figure}

We present data that show how mechanical motions in microtubules affect the superradiant states they can support. In Figs. \ref{fig:Color7-14} and \ref{fig:Color15-20}, we display visualizations of superradiant eigenstates obtained by diagonalizing the non-Hermitian Hamiltonian in Eq. 1 in the main text. Each eigenstate is near the lowest excitonic state of the system but is not necessarily the maximally superradiant state. The projection of the eigenstate onto the tryptophan (Trp) site basis gives the probabilities of each Trp being in an excited state. Either monomer chain ($\alpha$ or $\beta$) of each tubulin dimer is colored based on the average of the probability of each of the four Trp molecules in that chain being excited. Red implies a higher probability, and blue implies a lower probability. Highly superradiant states in the middle column are delocalized across the microtubule (i.e., a more uniform probability distribution), as reflected in the smaller $P_\text{max}$ values shown. (The $P_\text{min}$ values for all three columns are all numerically zero, to one part in a trillion.) Each row in Figs. \ref{fig:Color7-14} and \ref{fig:Color15-20} represents a different (mechanical) vibrational mode, which can be roughly categorized as stretching, bending, torsional, and breathing modes. The three panels in each row represent snapshots of each specific mode at different instances in time. The left and right columns display each mode at their extreme amplitudes of vibration, and the middle column displays each mode at its zero amplitude.

\subsection*{Tables}
Table \ref{table:heatmap_ranges} describes how the heatmap colors used for each panel in Figs. \ref{fig:Color7-14} and \ref{fig:Color15-20} correspond to probability distributions across each microtubule.
\begin{table}[tbhp!]
    \centering \setlength{\extrarowheight}{3pt}
    \begin{tabular}{|c|cccccc|}
    \hline
    \vspace{-5.35mm}\\
    \multirow{2}{*}{Mode} & \multicolumn{2}{c}{Left Panel} & \multicolumn{2}{c}{Middle Panel} & \multicolumn{2}{c|}{Right Panel}\\\vspace{0mm}\\[-2.8ex]
    \cline{2-7}
    & $P_\text{max}$ & $P_\text{min}$ & $P_\text{max}$ & $P_\text{min}$ & $P_\text{max}$ & $P_\text{min}$ \\
    \hline
    7 & 0.484 & $1.73 \times 10^{-17}$ & 0.222 & $5.74 \times 10^{-15}$ & 0.492 & $8.95 \times 10^{-16}$\\
    8 & 0.414 & $1.30 \times 10^{-16}$ & 0.109 & $1.43 \times 10^{-13}$ & 0.424 & $5.69 \times 10^{-17}$\\
    9 & 0.499 & $1.01 \times 10^{-14}$ & 0.218 & $6.89 \times 10^{-14}$ & 0.498 & $6.89 \times 10^{-16}$\\
    10 & 0.483 & $1.24 \times 10^{-16}$ & 0.261 & $6.78 \times 10^{-16}$ & 0.478 & $3.70 \times 10^{-17}$\\
    11 & 0.428 & $3.64 \times 10^{-16}$ & 0.289 & $1.29 \times 10^{-14}$ & 0.482 & $7.22 \times 10^{-16}$\\
    12 & 0.499 & $4.07 \times 10^{-19}$ & 0.290 & $1.68 \times 10^{-14}$ & 0.499 & $6.94 \times 10^{-17}$\\
    13 & 0.500 & $2.09 \times 10^{-16}$ & 0.394 & $1.05 \times 10^{-15}$ & 0.499 & $1.98 \times 10^{-17}$\\
    14 & 0.500 & $1.19 \times 10^{-16}$ & 0.261 & $2.22 \times 10^{-14}$ & 0.500 & $5.63 \times 10^{-18}$\\
    15 & 0.481 & $5.55 \times 10^{-17}$ & 0.126 & $5.74 \times 10^{-14}$ & 0.358 & $7.09 \times 10^{-14}$\\
    16 & 0.500 & $4.13 \times 10^{-18}$ & 0.395 & $2.92 \times 10^{-15}$ & 0.500 & $2.72 \times 10^{-17}$\\
    17 & 0.492 & $6.22 \times 10^{-17}$ & 0.220 & $4.06 \times 10^{-15}$ & 0.500 & $2.46 \times 10^{-17}$\\
    18 & 0.498 & $1.79 \times 10^{-16}$ & 0.303 & $2.54 \times 10^{-14}$ & 0.490 & $1.01 \times 10^{-15}$\\
    19 & 0.491 & $3.39 \times 10^{-16}$ & 0.214 & $1.745 \times 10^{-15}$ & 0.492 & $5.10 \times 10^{-16}$\\
    20 & 0.491 & $2.38 \times 10^{-16}$ & 0.194 & $2.97 \times 10^{-14}$ & 0.496 & $1.33 \times 10^{-16}$\\

    \hline
    \end{tabular}
    \caption{Heatmap ranges for each panel of Figs. \ref{fig:Color7-14} and \ref{fig:Color15-20}. $P_\text{max}$ and $P_\text{min}$ are the maximum and minimum excitonic occupation probabilities, respectively, which range from 0 to 1. $P_\text{max}$ is associated with red, $P_\text{min}$ is associated with blue, and intermediate probabilities are associated with green in Figs. \ref{fig:Color7-14} and \ref{fig:Color15-20}.}\phantomsection
\label{table:heatmap_ranges}
\end{table}

\end{document}